\documentclass[numbers]{spml_template}

\usepackage[utf8]{inputenc}
\usepackage{enumitem}
\usepackage{amsmath}
\usepackage{amssymb}
\usepackage{amsfonts}
\usepackage{nicefrac}
\usepackage{url}
\usepackage{xspace}
\usepackage[version=4]{mhchem}
\usepackage{placeins}
\usepackage{colortbl}
\usepackage{makecell}
\usepackage{tikz}
\usepackage{overpic}
\usepackage{xparse}
\usepackage{wrapfig}
\usepackage{needspace}
\usepackage{fontawesome}

\definecolor{MyBlue2}{HTML}{5384EC}
\definecolor{rowhl}{HTML}{DDF1FF}
\definecolor{densitygray}{HTML}{E6E6E6}
\definecolor{errorred}{HTML}{F8CACA}

\DeclareRobustCommand{\grayhl}[1]{%
  \begingroup
  \setlength{\fboxsep}{2pt}%
  \colorbox{densitygray}{#1}%
  \endgroup
}

\DeclareRobustCommand{\redhl}[1]{%
  \begingroup
  \setlength{\fboxsep}{2pt}%
  \colorbox{errorred}{#1}%
  \endgroup
}

\newcommand{\method}{MADField\xspace}
\newcommand{\methodfull}{Multi-fidelity Amortized Density Field\xspace}
\newcommand{\methodcdft}{MADField-cDFT\xspace}
\newcommand{\methodgcmc}{MADField-GCMC\xspace}
\newcommand{\methodscalar}{MADField-Scalar\xspace}
\newcommand{\crefpanel}[2]{\hyperref[#1]{\cref*{#1}#2}}

\crefname{figure}{Figure}{Figures}
\Crefname{figure}{Figure}{Figures}
\crefname{table}{Table}{Tables}
\Crefname{table}{Table}{Tables}
\crefname{section}{Section}{Sections}
\Crefname{section}{Section}{Sections}
\crefname{subsection}{Section}{Sections}
\Crefname{subsection}{Section}{Sections}
\crefname{subsubsection}{Section}{Sections}
\Crefname{subsubsection}{Section}{Sections}
\crefname{equation}{Equation}{Equations}
\Crefname{equation}{Equation}{Equations}
\crefname{appendix}{Appendix}{Appendices}
\Crefname{appendix}{Appendix}{Appendices}

\newcommand{\ptz}{\phantom{0}}

\title{MADField: Multi-fidelity Amortized Density Field\\[-0.05em]for Adsorption in Nanoporous Materials}

\author[1]{Yoonho Kim}
\author[1]{Seongsu Kim}
\author[1,\dagger]{Sungsoo Ahn}
\author[1,\dagger]{Honghui Kim}
\affiliation[1]{KAIST Graduate School of AI}
\contribution[\dagger]{Corresponding authors}
\correspondence{\email{yoonho2001@kaist.ac.kr}, \email{sungsoo.ahn@kaist.ac.kr}, \email{ildmb96@kaist.ac.kr}}
\metadata[Code]{TBA}
\metadata[Data]{TBA}

\abstract{%
High-throughput computational screening of nanoporous materials for gas storage and separation requires fast and accurate characterization of adsorption equilibrium. Particle-based grand canonical Monte Carlo (GCMC) and density-based classical density functional theory (cDFT) provide simulation-based estimates of gas uptake and adsorbate density fields, but their speed--accuracy tradeoff remains insufficient for large-scale screening. In this work, we address this gap with \methodfull{} (\method{}), which reframes adsorption prediction as equilibrium density-field estimation. \method{} learns from two complementary fidelities, combining broad and scalable cDFT density supervision with higher-fidelity GCMC density labels, and recovers gas uptake by integrating the predicted density field. \method{} improves uptake accuracy over the strongest baselines by $6.0\times$ for cDFT and $15.4\times$ for GCMC, and its predicted fields accelerate cDFT solvers with $2.0\times$ fewer iteration steps while recovering $42\%$ of cases that fail under standard settings. Finally, we evaluate \method{} for conventional \ce{CH4} working capacity screening on the 270k-structure ARC-MOF database. Within this space of extremely rare high-capacity targets (167 in total), the model achieves $56\times$ higher average precision than the strongest baseline and accelerates inference by five orders of magnitude compared to GCMC. By prioritizing the \method{} rankings, selecting the top $1.7\%$ of candidates recovers $95\%$ of all targets, while the top $6\%$ ensures $100\%$ recall.
}

\begin{document}

\maketitle

\section{Introduction}

High-throughput computational screening is central to discovering nanoporous materials for gas storage and separation~\citep{wilmer2012large}. 
Metal--organic frameworks (MOFs) and amorphous carbons (ACs), for example, have been widely screened for methane storage~\citep{mason2014evaluating} and carbon capture~\citep{lin2012silico,charalambous2024holistic}. 
The primary screening observable is gas uptake, i.e., the amount adsorbed at a given pressure and temperature. 
Computing uptake, however, requires characterizing adsorption equilibrium: the distribution of gas configurations inside a complex porous material under a specified thermodynamic condition.

Adsorption equilibrium can be modeled through two complementary views. 
\textit{Grand canonical Monte Carlo (GCMC)} provides a \textit{particle-based view}, sampling gas molecules in the grand canonical ensemble through insertion, deletion, and roto-translational moves as illustrated in \crefpanel{fig:concept}{(a)}~\citep{adams1975grand}. 
As it estimates ensemble averages from particle configurations under a chosen force field, GCMC is widely used as a reference for adsorption screening~\citep{dubbeldam2016raspa, ran2024raspa3}. 
However, long Markov chains with millions of Monte Carlo steps limit its use in high-throughput settings. 
\textit{Classical density functional theory (cDFT)} provides a complementary \textit{density-based view}, replacing stochastic sampling with optimization of a grand-potential functional over the one-body density field as illustrated in \crefpanel{fig:concept}{(b)}~\citep{evans1979nature,roth2010fundamental}. 
Recent work shows that cDFT can reproduce GCMC-level adsorption trends in MOFs at lower cost~\citep{dufour2025classical,thiele2026efficient}. 
Nevertheless, cDFT relies on an approximate free-energy functional, and each material--adsorbate--condition still requires an iterative solve.

Existing machine learning approaches do not fully exploit this particle--density structure. 
Most methods regress scalar uptake from structural, textural, or energetic descriptors~\citep{sarikas2024gas,chen2022interpretable}, but discard where gas adsorbs inside the material and leave a non-trivial accuracy gap relative to simulation references. 
A smaller line of work predicts adsorbate density fields~\citep{sun2024understanding,burner2026rapid}, but existing models either produce normalized probability-like fields that integrate to unity, or require the uptake itself as an auxiliary input. 
Thus, they cannot independently recover the equilibrium adsorbate density.


\newif\ifpaneldebug

\tikzset{
  panel debug/.style={draw=red, line width=0.5pt},
  panel nodebug/.style={},
}

\NewDocumentCommand{\panellabel}{O{0pt 0pt 0pt 0pt} m m}{%
  \begin{tikzpicture}
    \node[anchor=north west, inner sep=0,
          \ifpaneldebug panel debug\else panel nodebug\fi] (img) 
      {\includegraphics[width=\linewidth, trim=#1, clip]{#2}};
    \node[anchor=north west, inner sep=2pt, font=\footnotesize] at (img.north west) {#3};
  \end{tikzpicture}%
}

\begin{figure}[t]
    \centering
    \begin{minipage}[t]{0.622\linewidth}
        \panellabel[-13pt 4pt 9pt 3pt]{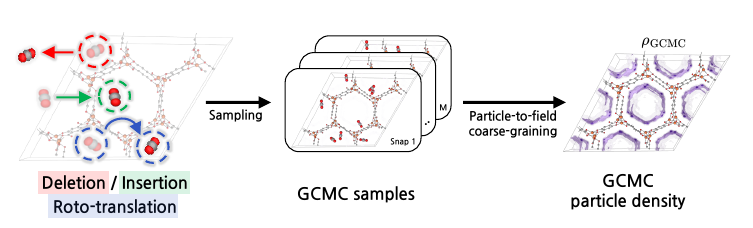}{(a)}
    \end{minipage}
    \hfill
    \begin{minipage}[t]{0.358\linewidth}
        \panellabel[0pt -3pt 9pt -10pt]{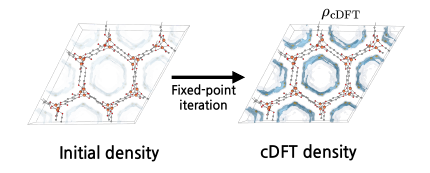}{(b)}
    \end{minipage}
    \vspace{4pt}
    
    \begin{minipage}[t]{\linewidth}
        \panellabel[-13pt 5pt 9pt 0pt]{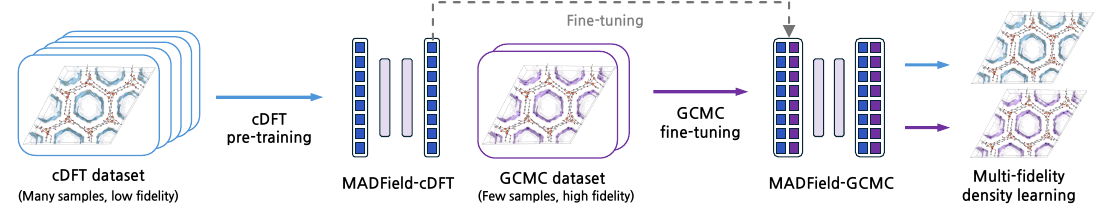}{(c)}
    \end{minipage}
    
\caption{
\textbf{Multi-fidelity adsorbate density learning from cDFT and GCMC.}
\textbf{(a)} GCMC samples a particle ensemble through deletion, insertion, and roto-translation moves, and particle-to-field coarse-graining converts the ensemble into $\rho_{\mathrm{GCMC}}$.
\textbf{(b)} cDFT solves for $\rho_{\mathrm{cDFT}}$ by fixed-point iteration from an initial density.
\textbf{(c)} MADField is pre-trained on broad cDFT labels and fine-tuned on sparse GCMC labels, combining cDFT coverage with GCMC fidelity.
}
\label{fig:concept}
\end{figure}

\paragraph{Contribution.}
In this work, we introduce \methodfull{} (\method{}) for adsorption simulation in nanoporous materials. Given a porous framework, adsorbate, and thermodynamic condition, \method{} predicts the unnormalized equilibrium adsorbate density; scalar uptake is then recovered by integration rather than by an unconstrained regression head. This density-level formulation provides a unified target for screening: it predicts gas uptake, preserves spatial adsorption structure, and can initialize iterative cDFT solvers.

The key challenge is that density supervision is available at different fidelity-cost regimes. 
cDFT provides relatively inexpensive density fields but inherits the approximation error of its free-energy functional, while GCMC provides costly particle-sampling references that can be coarse-grained into density. 
We therefore pre-train \method{} on large-scale cDFT density data and adapt it to sparse GCMC-derived density fields, learning a broad density prior from cDFT and correcting the remaining fidelity gap toward particle-simulation references as illustrated in \crefpanel{fig:concept}{(c)}. 

Our main contributions:
\begin{itemize}[leftmargin=*, itemsep=4pt, topsep=2pt]
    \item \textbf{Equilibrium density as a unified target.}
    We formulate adsorption prediction as unnormalized equilibrium adsorbate density field modeling. 
    \method{} predicts the 3D equilibrium density in physical units, so uptake $N$ is obtained by integration while spatial adsorption structure is retained.

    \item \textbf{Multi-fidelity density learning.}
    We bridge cDFT and GCMC by pre-training \method{} on cDFT density fields across nine non-polar adsorbates and fine-tuning it to a GCMC dataset that is $14.7\times$ smaller. 
    This cDFT-to-GCMC transfer outperforms GCMC-only training, especially on out-of-distribution materials, including amorphous carbons (ACs), polymers of intrinsic microporosity (PIMs), hyper-cross-linked polymers (HCPs), and kerogens.

    \item \textbf{Density-based acceleration of cDFT.}
    Since \method{} predicts density fields in physical units, its output can serve as a high-quality initial density for the original cDFT fixed-point solver. 
    Such initialization allows solving $42\%$ of cases where the standard initialization fails and achieves a $2.0\times$ reduction in the number of iterations for the solver to converge.


\item \textbf{Comprehensive adsorption benchmark.}
    We benchmark \method{} on both approximation fidelity and downstream screening, covering uptake and density prediction across cDFT and GCMC labels over nine adsorbates and diverse material classes.
    On MOFs, \method{} reduces uptake MAE over the strongest baselines by $6.0\times$ for cDFT labels and $15.4\times$ for GCMC labels.
    On out-of-distribution disordered materials, it also reduces uptake MAE by $4.9\times$ and $3.5\times$ on cDFT and GCMC labels, respectively.
    For database-scale screening across all $270{,}583$ ARC-MOF frameworks, \method{} achieves a $56\times$ higher average precision for conventional \ce{CH4} working-capacity retrieval than the strongest learned baseline, while accelerating inference by five orders of magnitude relative to GCMC. By prioritizing its rankings, the top $1.7\%$ of candidates recovers $95\%$ of the $167$ rare high-capacity targets and the top $6\%$ recovers all of them.
    We will release the cDFT benchmark dataset of $280{,}000$ calculations, produced with $3{,}600$ H200 GPU hours.
\end{itemize}
\section{Background and related work}
\label{sec:background}

This section introduces adsorption equilibrium and its particle- and density-based formulations (\cref{sec:adsorption_equilibrium}), followed by related work on machine learning for adsorption prediction (\cref{sec:ml_related}). 

\subsection{Adsorption equilibrium as particle and density fields}
\label{sec:adsorption_equilibrium}

Adsorption is the process by which guest gas molecules accumulate on the surfaces and within the pores of a host porous material.
For a rigid host framework $\mathcal{M}$, guest species $s$, temperature $T$, and pressure $P$, the central field-level object is the equilibrium one-body number density $\rho_{\mathrm{eq}}(\mathbf{r})$, defined as the expected number of adsorbate molecule centers per unit volume at position $\mathbf{r}$ in the periodic unit cell $\mathcal{V}$. 
\textit{Scalar uptake} $N$ and \textit{adsorption isotherm} $\mathcal{I}_{T}$ are obtained from this field:
\begin{equation}
    N(P,T;\mathcal{M},s)
    =
    \int_{\mathcal{V}}
    \rho_{\mathrm{eq}}(\mathbf{r})\,d\mathbf{r},
    \qquad
    \mathcal{I}_{T}(\mathcal{M},s)
    =
    \{N(P,T;\mathcal{M},s):P\!\in\!\mathcal{P}\}.
    \label{eq:loading_integral}
\end{equation}
Scalar uptake is the primary quantity used to rank porous materials in high-throughput screening, since practical applications often require maximizing the amount of gas stored or captured at a target pressure and temperature. The adsorption isotherm further characterizes how uptake changes with pressure, enabling evaluation of process-relevant quantities such as working capacity between adsorption and desorption pressures, pressure-dependent selectivity, and operating-window robustness.

\paragraph{Particle view: grand canonical Monte Carlo.}
GCMC simulates adsorption by sampling many-body particle configurations in the grand canonical ($\mu VT$) ensemble~\citep{adams1975grand}. 
For adsorbate species $s$, a configuration with $n$ molecules at positions $\mathbf{r}_{1:n}=(\mathbf{r}_1,\dots,\mathbf{r}_n)$ in a porous material has energy
\begin{equation}
    U_n(\mathbf{r}_{1:n};\mathcal{M},s)
    =
    \sum_{i=1}^{n} V_{\mathrm{ext}}(\mathbf{r}_i;\mathcal{M},s)
    +
    \sum_{i<j} u_{ss}(\mathbf{r}_i,\mathbf{r}_j),
    \label{eq:gcmc_energy}
\end{equation}
where $V_{\mathrm{ext}}$ is the framework--fluid potential and $u_{ss}$ is the fluid--fluid pair interaction.
At chemical potential $\mu=\mu(P,T;s)$, a Markov chain proposes insertion, deletion, translation, and rotation moves, accepted by a Metropolis criterion so that the chain samples the $\mu VT$ distribution over variable-size configurations.
Adsorption observables are then computed as ensemble averages, \textit{i.e.}, uptake is the expected particle count, and the one-body density field is obtained by coarse-graining the particles:
\begin{equation}
    N(P,T;\mathcal{M},s)
    =
    \langle n \rangle_{\mu,V,T},
    \qquad
    \rho_{\mathrm{eq}}(\mathbf{r})
    =
    \left\langle
    \sum_{i=1}^{n}
    \delta(\mathbf{r}-\mathbf{r}_i)
    \right\rangle_{\mu,V,T}.
    \label{eq:gcmc_observables}
\end{equation}
Integrating this density recovers the uptake consistent with \cref{eq:loading_integral}. In practice, GCMC with classical force fields is a standard simulation tool for large-scale MOF screening, with studies evaluating over $10^5$ of structures for gas storage and separation~\citep{wilmer2012large,colon2014high,simon2015materials}.

\paragraph{Density view: classical density functional theory.}
cDFT reformulates adsorption equilibrium as a variational problem over the same one-body density field~\citep{evans1979nature,roth2010fundamental}.
For a guest species $s$ at fixed $(T,\mu)$, the equilibrium density $\rho_{\mathrm{eq}}$ is defined as the minimizer of the grand-potential functional
\begin{equation}
    \Omega[\rho;\mathcal{M},s]
    =
    F_{\mathrm{id}}[\rho]
    +
    F_{\mathrm{exc}}[\rho;s]
    +
    \int_{\mathcal{V}}
    \rho(\mathbf{r})
    \bigl[
        V_{\mathrm{ext}}(\mathbf{r};\mathcal{M},s)
        -\mu
    \bigr]\,d\mathbf{r},
    \label{eq:grand_potential}
\end{equation}
where $F_{\mathrm{id}}$ is the ideal-gas free energy, $F_{\mathrm{exc}}$ captures non-ideal fluid correlations, and $V_{\mathrm{ext}}$ is the framework--fluid external potential.
Setting $\delta\Omega/\delta\rho = 0$ yields the Euler--Lagrange condition
\begin{equation}
    \rho_{\mathrm{eq}}(\mathbf{r})
    =
    \rho_{\mathrm{bulk}}\,
    \exp\!\Bigl[
        -\beta\,V_{\mathrm{ext}}(\mathbf{r})
        -\beta\,\frac{\delta F_{\mathrm{exc}}}{\delta\rho}\biggr|_{\rho_{\mathrm{eq}}}
        +\beta\,\mu_{\mathrm{exc}}
    \Bigr],
    \label{eq:euler_lagrange}
\end{equation}
where $\beta = 1/k_BT$, $\rho_{\mathrm{bulk}}$ is the bulk fluid density at the given $(T,\mu)$, and $\mu_{\mathrm{exc}}$ is the excess chemical potential of the bulk fluid.
Because $F_{\mathrm{exc}}$ depends nonlinearly on $\rho$, this self-consistent equation is solved by fixed-point iteration on a volumetric grid until convergence.
Uptake is then computed by integrating the converged density field.
Recent work has shown that cDFT can reproduce GCMC-level gas uptake in MOFs at substantially lower computational cost~\citep{dufour2025classical,thiele2026efficient}.
We use the PC-SAFT functional~\citep{sauer2017classical}, which parameterizes each species by segment number $m$, segment diameter $\sigma$, and dispersion energy $\varepsilon$. 
Full derivations of the Euler--Lagrange equation, the Boltzmann reference density, the Picard update used in our solver, and the PC-SAFT functional are given in \cref{app:cdft_gcmc}.

\subsection{Machine learning for adsorption observables}
\label{sec:ml_related}

Machine learning approaches to adsorption have mostly targeted scalar uptake at a given thermodynamic state.
These models differ in representation and backbone. 
Existing models have progressed from building-block and topological embeddings~\citep{lee2021computational} to energy-based representations such as 2D energy histograms~\citep{shi2023two} and voxelized potential-energy fields processed by 3D CNNs~\citep{sarikas2024gas}.
More recent work adopts graph and transformer architectures that combine local atomic environments, global textural features, and transferable pre-training across materials and gas species~\citep{chen2022interpretable,lin2025unified,cui2023direct,kang2023multi,wang2024comprehensive}.

A smaller line of work predicts adsorbate density fields, but its outputs are either conditioned on scalar uptake or normalized, and therefore do not directly determine the uptake. Sun and Siepmann~\citep{sun2024understanding} predict voxel-wise differential contributions to global quadratic isotherm parameters, enabling thermodynamic-state-dependent density reconstruction but requiring global uptake information. Burner et al.~\citep{burner2026rapid} use an equivariant graph network based on DeepDFT~\citep{jorgensen2020deepdft,jorgensen2022equivariant} to predict normalized adsorbate density fields in MOFs. While suitable for binding sites, this approach cannot recover global uptake $N$, since $\rho$ is normalized to integrate to unity.



\section{Method}
\label{sec_method}

\method{} is a multi-fidelity neural density operator for amortized adsorption simulation. 
Given a porous framework $\mathcal M$, an adsorbate species $s$, and a thermodynamic condition $(P,T)$, \method{} predicts the equilibrium adsorbate density field in physical units. 
Gas uptake is obtained by integrating this field, while the same prediction preserves spatial adsorption structure and can serve as an initial density for iterative cDFT solvers.
The training pipeline first learns \methodcdft{} from solver-converged cDFT density fields, then uses its predictions to reinitialize failed cDFT calculations and expand the cDFT label set, and finally adapts the resulting density operator to sparse GCMC-derived density fields to obtain \methodgcmc{}.

\subsection{Multi-fidelity density training}
\label{subsec:multifidelity}

\paragraph{Shared density target.}
\method{} is trained to predict the equilibrium one-body density field $\rho_{\mathrm{eq}}(\mathbf r;\mathcal M,s,P,T)$. 
For \methodcdft{}, this density is supervised by solver-converged PC-SAFT cDFT labels; for \methodgcmc{}, it is supervised by particle-to-grid averaged GCMC labels. 
Thus, both stages learn the same field-level object, while the source of supervision changes. 
The scalar uptake is recovered from the predicted density by integrating over the accessible pore region $\mathcal P$.

\paragraph{cDFT pre-training.}
The first stage trains \methodcdft{} on PC-SAFT cDFT density fields. 
These labels are available at larger scale and across more adsorbates than GCMC-derived density fields, but inherit the approximation and convergence properties of the cDFT solver. 
This stage learns a broad adsorption-density prior that captures how framework geometry, adsorbate identity, and thermodynamic condition shape the equilibrium density field.


\paragraph{cDFT refinement from predicted densities.}
Because \methodcdft{} predicts an unnormalized equilibrium adsorbate density field, its output can be used not only as a one-shot prediction but also as an initial density for the original cDFT solver.
For PC-SAFT cDFT, the equilibrium density is a fixed point of the Picard map, $\rho_{\mathrm{eq}}=\mathcal T_{\boldsymbol{\xi}}[\rho_{\mathrm{eq}}]$, with iterations $\rho^{(n+1)}=\mathcal T_{\boldsymbol{\xi}}[\rho^{(n)}]$, where $\boldsymbol{\xi}$ denotes the PC-SAFT adsorbate parameters. 
In a standard cDFT run, the initial density is the Boltzmann density $\rho_{\mathrm B}$. 
In our refinement procedure, we keep the same solver but replace this initial density with the predicted density, $\rho^{(0)}=\hat\rho_\theta$. 
The subsequent iterations are still performed by the original cDFT solver, and only converged runs are retained. Thus, the refined outputs are solver-converged cDFT densities rather than neural pseudo-labels.

\begin{wrapfigure}[17]{r}{0.42\textwidth}
    \vspace{-12pt}
    \centering
    \includegraphics[
    width=\linewidth,
    trim= 95pt 6pt 0pt 0pt,
    clip
    ]{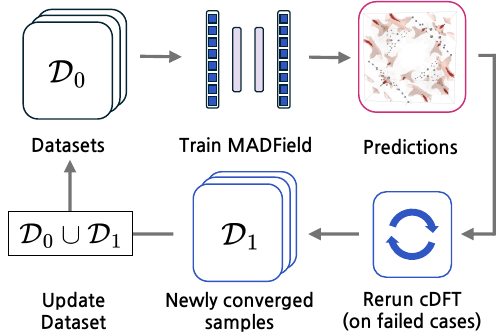}
    \caption{
    \textbf{Bootstrapping cDFT labels.}
        \method{} reinitializes failed cDFT cases with predicted density fields, which are then rerun to convergence.
        Recovered samples are solver-converged cDFT labels, not neural pseudo-labels.
        }
    \label{fig:bootstrap}
\end{wrapfigure}

\paragraph{Label bootstrapping.}
We use this refinement procedure for label bootstrapping, as illustrated in \cref{fig:bootstrap}.
After training \methodcdft{} on converged cDFT calculations, we use its predictions as initial densities for cases that previously failed under the same standard initialization and rerun the original solver. 
When the reinitialized solve converges, the resulting fixed-point density satisfies the same cDFT convergence criteria and is added to the supervised cDFT dataset. 
This expands cDFT supervision without treating neural predictions as ground truth.


\paragraph{GCMC adaptation.}
After cDFT pre-training and bootstrapping, we adapt the resulting \methodcdft{} checkpoint to GCMC-derived density fields.
GCMC provides particle-simulation references under the chosen force field, but field-level GCMC labels are more expensive and therefore sparser.
\methodgcmc{} is obtained by fine-tuning \methodcdft{} on paired GCMC density grids.
This stage uses the cDFT-trained model as an adsorption-density prior and updates it toward particle-simulation fidelity, rather than learning the GCMC density operator from scratch.
Thus, the abundant cDFT labels provide broad geometric and thermodynamic coverage, while the sparse GCMC labels correct the remaining fidelity gap.

\paragraph{Training objective.}
All training stages use the same density-field objective,
\begin{equation}
    \mathcal L
    =
    \mathcal L_{\mathrm{field}}
    +
    \lambda_N \mathcal L_N ,
    \label{eq_main_loss_compact}
\end{equation}
where $\mathcal L_{\mathrm{field}}$ denotes the collection of voxel-wise density-supervision terms.
It includes log-residual supervision relative to the Boltzmann reference, relative density error in the pore region, and auxiliary field losses that emphasize high-density adsorption sites and supervise the coarse branch against a low-pass residual target.
The uptake loss $\mathcal L_N$ matches the integrated uptake of the prediction to that of the target density.
Across cDFT and GCMC stages, the forward architecture, density parameterization, and loss form remain unchanged, only the supervision target and molecular-parameter registry differ.
The full expanded loss, including all density-field terms and weights, is given in \cref{app:training}.


\subsection{Periodic volumetric density operator}
\label{subsec:operator}

\paragraph{Periodic volumetric representation.}
The continuous density fields are represented on a periodic discrete voxel grid for learning and evaluation. 
The operator is designed around three requirements: periodic cell geometry, three-dimensional pore connectivity, and continuous dependence on thermodynamic and molecular conditions. 
The volumetric input contains the reduced external potential and the log-Boltzmann density. 
All boundary-facing operations are periodic: the convolutional stem uses circular padding, shifted-window attention uses circular indexing, and positional information is encoded with periodic Fourier features. We provide implementation details in \cref{app:architecture}.

\paragraph{Boltzmann-residual parameterization.}
To encode the dominant one-body framework response, we use the ideal-gas Boltzmann density $\rho_{\text{Boltz}}(\mathbf r)=\rho_{\mathrm{bulk}}\exp[-\beta V_{\mathrm{ext}}(\mathbf r)]$, where $\beta=(k_BT)^{-1}$. 
Its derivation and relation to the cDFT Euler--Lagrange equation are given in \cref{app:cdft_gcmc}. 
\method{} predicts a gated correction in log-density space:
\begin{equation}
    \log \hat\rho_\theta(\mathbf r)
    =
    \log \rho_{\text{Boltz}}(\mathbf r)
    +
    \sigma\!\left(h_\theta(\mathbf r)\right)
    \delta_\theta(\mathbf r),
    \label{eq_residual_output}
\end{equation}
where $\delta_\theta(\mathbf r)$ is a residual field and $h_\theta(\mathbf r)$ is a gate field. 
This guarantees $\hat\rho_\theta(\mathbf r)>0$ and makes the network learn the many-body correction to the one-body Boltzmann response.

\paragraph{Backbone and conditioning.}
The backbone is a U-shaped 3D Swin Transformer~\citep{liu2021swin}. 
A two-stage convolutional stem downsamples the discrete voxel grid to a token resolution where windowed attention is tractable. 
Shifted windows, skip connections, and the decoder allow adsorption features to propagate across connected pores while keeping memory cost manageable. 
Non-volumetric information is injected through a conditioning vector containing $\log\rho_{\mathrm{bulk}}$, $\log P$, lattice parameters, and a three-scalar molecular descriptor. 
This vector modulates every Swin block through adaptive normalization. 
For \methodcdft{}, the descriptor is the PC-SAFT triplet $(m,\sigma,\varepsilon)$; for \methodgcmc{}, the same descriptor slot is filled with Lennard-Jones parameters from the GCMC references. 
Full details are given in \cref{app:architecture,app:training,app:parameters}.

\FloatBarrier
\section{Experiments}
\label{sec:experiments}

This section presents benchmark results for density field (\cref{sec:density_bench}) and uptake prediction (\cref{sec:uptake_bench}) in MOFs, comparing \method{} with other baseline models.
We further demonstrate transferability to ACs and other disordered porous materials with larger unit cells (\cref{sec:transfer_bench}).
We then evaluate \ce{CH4} working-capacity screening across the full ARC-MOF database, comprising $270{,}583$ frameworks (\cref{sec:wc_screening}).
In addition, we highlight the computational speedup \method{} provides for cDFT solvers (\cref{sec:warmstart}).

\subsection{Setup}
\label{sec:setup}

\paragraph{Datasets.}
We train and evaluate at two fidelities, cDFT and GCMC. Each fidelity provides a dataset for training and in-distribution (ID) testing, paired with a separate out-of-distribution (OOD) benchmark to assess zero-shot transferability.
As part of this work, we release a large-scale cDFT adsorption dataset comprising approximately 280{,}000 PC-SAFT cDFT calculations across 4{,}000 unique MOF frameworks, nine adsorbates, and tens of pressure points per isotherm, requiring a total of 3{,}600 H200 GPU hours to generate. 

We perform cDFT calculations to generate density fields and uptake values, using MOF structures from the QMOF database~\citep{rosen2021machine} for ID (nine adsorbates) and amorphous carbon structures from Gardner et al.~\citep{gardner2023synthetic} for OOD (\ce{CH4}).
For GCMC data, we use \ce{CH4} and \ce{Xe} density fields and uptake values on a subset of the ARC-MOF database from Burner et al.~\citep{burner_2023arcmof, burner_2025_CH4_1bar, burner_2025_CH4_65bar,burner_2025_Xe_1bar} for ID, and \ce{CH4} uptake on diverse disordered materials from Thyagarajan and Sholl~\citep{thyagarajan2020database} for OOD, including polymers of intrinsic microporosity (PIM), hyper-cross-linked polymers (HCP), Kerogen, and ACs.
For the screening task, we further use all $270{,}583$ frameworks of the ARC-MOF database~\citep{burner_2023arcmof}.
Details on data generation and settings are provided in \cref{app:data_pipeline}.

\definecolor{myred}{HTML}{B85450}
\definecolor{mygreen}{HTML}{1B9E77}
\definecolor{myamber}{HTML}{E6A700}

\newcommand{\yes}{\textcolor{mygreen}{\faCheck}}
\newcommand{\no}{\textcolor{myred}{\faTimes}}
\newcommand{\umm}{\textcolor[HTML]{E6A700}{\scalebox{0.85}{\rotatebox[origin=c]{90}{\faPlay}}}}


\begin{wraptable}{r}{0.45\textwidth}
\vspace{-12pt}
\caption{\textbf{Comparison of \method{} with baselines.}
Columns indicate support for uptake prediction, density field prediction, adsorbate conditioning, pressure conditioning, and material-class generalization.
\yes/\no/\umm\ denote full/no/partial support.
\method{} satisfies all five criteria.}
\label{tab:model_comparison}
\centering
\setlength{\tabcolsep}{4pt}
\renewcommand{\arraystretch}{0.95}
\footnotesize
\resizebox{\linewidth}{!}{%
\begin{tabular}{l *{5}{>{\centering\arraybackslash}m{2.2em}}}
\toprule
Model & $N$ & $\rho$ & Ads. & $P$ & Mat. \\
\midrule
RetNet           & \yes & \no   & \no   & \no   & \yes \\
MOFTransformer   & \yes & \no   & \no   & \no   & \no  \\
DeepSorption     & \yes & \no   & \no   & \umm  & \yes \\
IsothermNet      & \yes & \no   & \yes  & \yes  & \yes \\
Uni-MOF          & \yes & \no   & \yes  & \yes  & \no  \\
SorbIIT          & \no  & \umm  & \yes  & \yes  & \yes \\
DeepAPD          & \no  & \umm  & \no   & \no   & \yes \\
\arrayrulecolor{black!40}\midrule\arrayrulecolor{black}
\textbf{MADField (Ours)} & \yes & \yes & \yes & \yes & \yes \\
\bottomrule
\end{tabular}
}
\vspace{-10pt}
\end{wraptable}

\paragraph{Baselines.}
We compare \method{} against several baseline models: RetNet \citep{sarikas2024gas}, DeepSorption \citep{cui2023direct}, MOFTransformer \citep{kang2023multi}, Uni-MOF \citep{wang2024comprehensive}, and IsothermNet \citep{lin2025unified} for scalar uptake prediction, as well as DeepAPD \citep{burner2026rapid} and SorbIIT \citep{sun2024understanding} for density field prediction. Notably, DeepAPD does not support the prediction of unnormalized density fields, and SorbIIT requires gas uptake data as an input to predict them. Accordingly, these two models were excluded from the gas uptake benchmark and only assessed in the density field benchmark. Detailed implementation of each baselines are in \cref{app:baselines}.

We train \method{} on each fidelity dataset to obtain \methodgcmc{} and \methodcdft{}, as described in \cref{subsec:multifidelity}. Baselines are likewise trained separately on each fidelity dataset. For RetNet, MOFTransformer, and DeepSorption, which do not support conditioning on adsorbate and pressure, we additionally train a separate model for each (adsorbate, pressure) combination. The exception is DeepSorption, which predicts uptake at a predefined set of pressure points in a single forward pass and thus requires one model per adsorbate. \cref{tab:model_comparison} summarizes how MADField compares to existing models across five criteria, gas uptake prediction, density field prediction, adsorbate and pressure conditioning, and material-class generalization. It shows that \method{} is the only model satisfying all five.

\paragraph{Metrics.}
We report gas uptake mean absolute error $N_{\mathrm{err}}$ in cm$^3$(STP)/g and Tanimoto coefficient $T_\rho \in [0, 1]$ to evaluate the similarity between predicted and reference densities, where $T_\rho = 0$ indicates non-overlapping densities and $T_\rho = 1$ indicates identical densities. Since DeepAPD predicts only normalized densities (i.e., normalized to unit integral), we evaluate $T_\rho$ on normalized densities for all models.
For the screening task, we evaluate \ce{CH4} working capacity across all $270{,}583$ frameworks of the ARC-MOF database~\citep{burner_2023arcmof}. For each framework $\mathcal{M}$, we define working capacity over the storage pressure swing from the desorption pressure $P_{\mathrm{des}} = 5.8~\mathrm{bar}$ to the adsorption pressure $P_{\mathrm{ads}} = 65~\mathrm{bar}$ at $T = 298~\mathrm{K}$ as
\begin{equation}
\mathrm{WC}(\mathcal{M},\ce{CH4})
=
N(P_{\mathrm{ads}}, T;\, \mathcal{M},\ce{CH4})
-
N(P_{\mathrm{des}}, T;\, \mathcal{M},\ce{CH4})
\label{eq:wc}
\end{equation}
where $N(P,T;\mathcal{M},s)$ is the volumetric uptake defined in \cref{eq:loading_integral}. We treat the $167$ frameworks with GCMC $\mathrm{WC} \ge 200~\mathrm{cm^3/cm^3}$, corresponding to $0.06\%$ of the database, as high-working-capacity targets. Each model ranks the database by predicted WC, and we report Average Precision (AP, the area under the precision--recall curve induced by this ranking) together with the top-$k$ recall of true positives, while fully converged GCMC serves as the upper bound (AP $= 1$, at about $3$~h per framework).
Detailed metric definitions are provided in \cref{app:metrics}.

\subsection{Density fields prediction on MOFs}
\label{sec:density_bench}

\cref{tab:density_bench_updated} reports the density prediction benchmark on QMOF dataset (cDFT density) and a subset of ARC-MOF (GCMC density).
\method{} consistently achieves a $T_\rho$ close to 1 across all adsorbates, with an overall mean $T_\rho$ of 0.996 for cDFT density and 0.965 for GCMC density.
In comparison, DeepAPD follows with mean $T_\rho$ values of 0.943 for cDFT density and 0.891 for GCMC density.
SorbIIT shows lower $T_\rho$, falling significantly short of \method{} performance.
The density prediction benchmark on OOD ACs is provided in \cref{app:additional}, where \method{} outperforms all baselines by a large margin ($T_\rho$: \method{} $0.885$ vs.\ SorbIIT $0.492$ and DeepAPD $0.418$).
\cref{fig:density_prediction_example} visualizes representative \ce{CH4} density predictions on QMOF and AC.
\method{} produces lower error across both cases, while DeepAPD and SorbIIT show substantially larger errors.

\definecolor{cdftbase}{HTML}{E1EBF7}  
\definecolor{gcmcbase}{HTML}{EDCCEB}  

\newcommand{\cdftopacity}{75}   
\newcommand{\gcmcopacity}{50}   

\colorlet{cdfthighlight}{cdftbase!\cdftopacity!white}
\colorlet{gcmchighlight}{gcmcbase!\gcmcopacity!white}

\DeclareRobustCommand{\hlcdft}[1]{%
  \begingroup
  \setlength{\fboxsep}{2pt}%
  \colorbox{cdfthighlight}{#1}%
  \endgroup
}

\DeclareRobustCommand{\hlgcmc}[1]{%
  \begingroup
  \setlength{\fboxsep}{2pt}%
  \colorbox{gcmchighlight}{#1}%
  \endgroup
}

\begin{table}[t]
\caption{\textbf{Density field prediction on MOFs.}
Density prediction quality is measured by Tanimoto similarity ($\uparrow$) between predicted and ground-truth normalized density on \hlcdft{\textbf{cDFT}} densities (nine adsorbates) and on \hlgcmc{\textbf{GCMC}} densities (\ce{CH4}, \ce{Xe}). \textbf{Total} denotes the sample-weighted mean over all evaluated density fields within each fidelity group. The conditioning columns (Cond.) indicate whether adsorbate identity (Ads.) and pressure ($P$) enter the model as conditioning inputs.}
\label{tab:density_bench_updated}
\footnotesize
\centering
\setlength{\tabcolsep}{3.0pt}
\renewcommand{\arraystretch}{1.1}
\resizebox{\textwidth}{!}{%
\begin{tabular}{cc l ccccccccc c cc c}
\toprule
\multicolumn{2}{c}{Cond.} & \multicolumn{1}{c}{\multirow{2}{*}{Model}} 
& \multicolumn{10}{c}{\hlcdft{\textbf{cDFT density}}} 
& \multicolumn{3}{c}{\hlgcmc{\textbf{GCMC density}}} \\
\cmidrule(lr){1-2} \cmidrule(lr){4-13} \cmidrule(lr){14-16}
Ads. & $P$ & & \ce{H2} & \ce{Ar} & \ce{Kr} & \ce{Xe} & \ce{N2} & \ce{CH4} & \ce{CO2} & \ce{C2H6} & \ce{C3H8} & \textbf{Total} & \ce{CH4} & \ce{Xe} & \textbf{Total} \\
\midrule
\yes & \yes & SorbIIT  & 0.881 & 0.749 & 0.693 & 0.736 & 0.812 & 0.714 & 0.908 & 0.819 & 0.823 & 0.805 & 0.730 & 0.690 & 0.717 \\
\no  & \no  & DeepAPD  & 0.968 & 0.955 & 0.942 & 0.905 & 0.958 & 0.947 & 0.929 & 0.902 & 0.849 & 0.943 & 0.904 & 0.863 & 0.891 \\
\yes & \yes & 
\textbf{MADField} & \textbf{1.000} & \textbf{1.000} & \textbf{1.000} & \textbf{0.978} & \textbf{1.000} & \textbf{0.999} & \textbf{0.998} & \textbf{0.991} & \textbf{0.960} & \textbf{0.996} & \textbf{0.982} & \textbf{0.931} & \textbf{0.965} \\
\bottomrule
\end{tabular}
}
\end{table}

\begin{figure}[!htbp]
    \centering
    \setlength{\tabcolsep}{0pt}
    \renewcommand{\arraystretch}{0.95}

    \newcommand{\densfig}[1]{%
        \includegraphics[width=0.175\textwidth,trim=4pt 0pt 4pt 4pt,clip]{figures/fig5_sub/#1}%
    }
    \newcommand{\simtext}[1]{%
        {\tiny #1}%
    }

    {\scriptsize
    \resizebox{\linewidth}{!}{%
    \begin{tabular}{
        @{}
        >{\centering\arraybackslash}m{0.025\textwidth}
        >{\centering\arraybackslash}m{0.175\textwidth}
        >{\centering\arraybackslash}m{0.175\textwidth}
        >{\centering\arraybackslash}m{0.175\textwidth}
        >{\centering\arraybackslash}m{0.175\textwidth}
        >{\centering\arraybackslash}m{0.175\textwidth}
        @{}
    }
        \toprule
        & \textbf{Framework} & \textbf{cDFT} & \textbf{\method{} (Ours)} & \textbf{DeepAPD} & \textbf{SorbIIT} \\
        \midrule

        \multirow{2}{*}{\rotatebox[origin=c]{90}{\textbf{QMOF}}}
        & \densfig{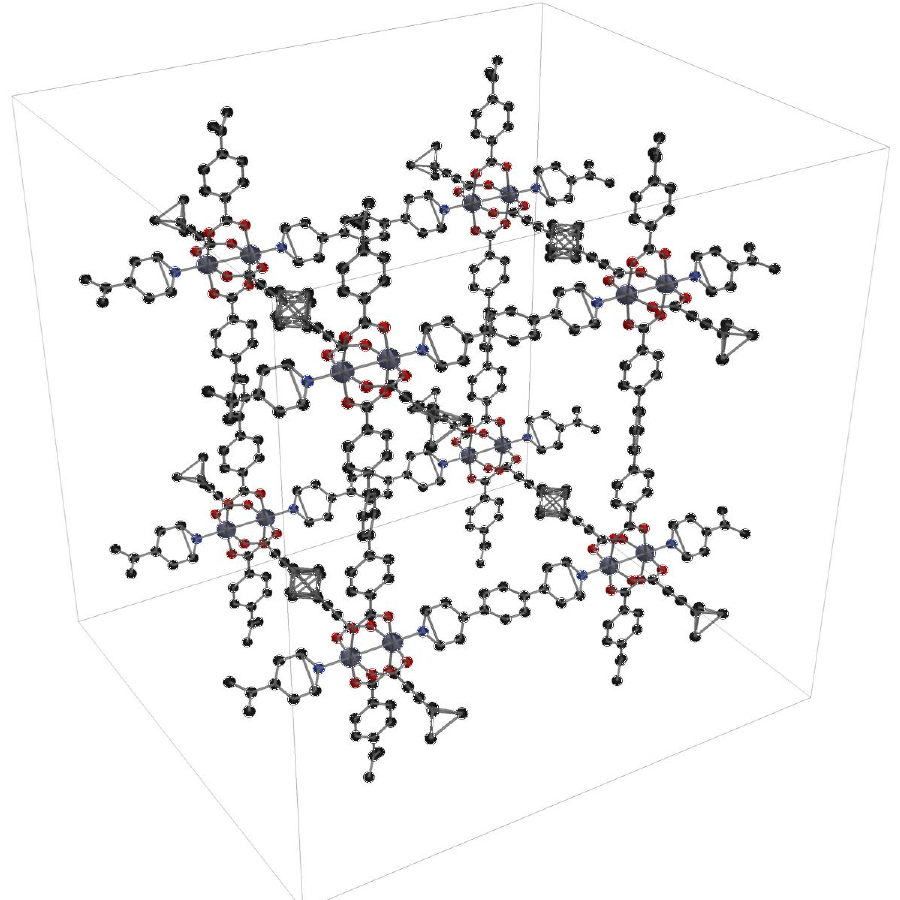}
        & \densfig{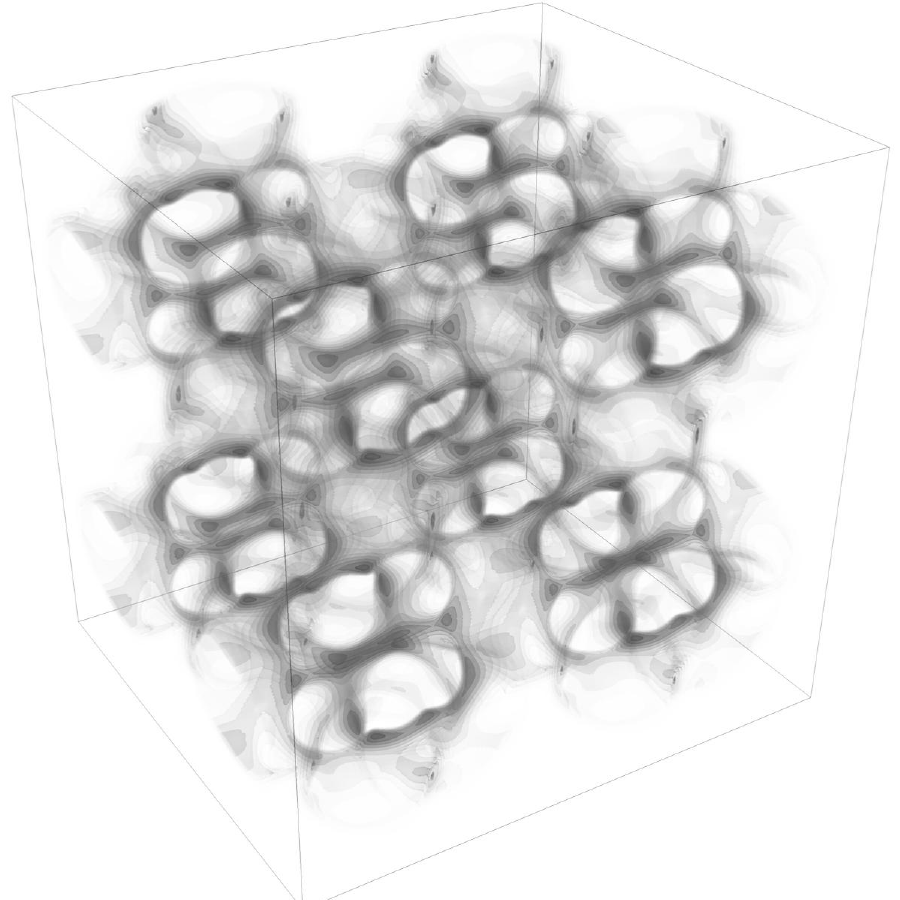}
        & \densfig{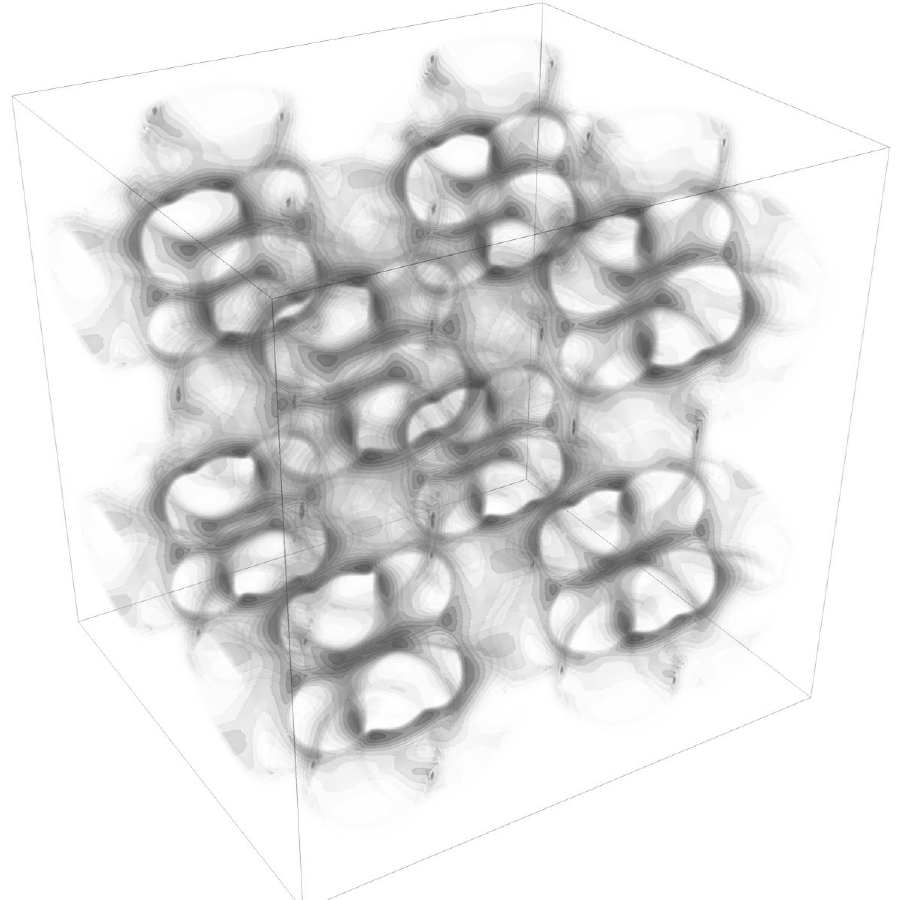}
        & \densfig{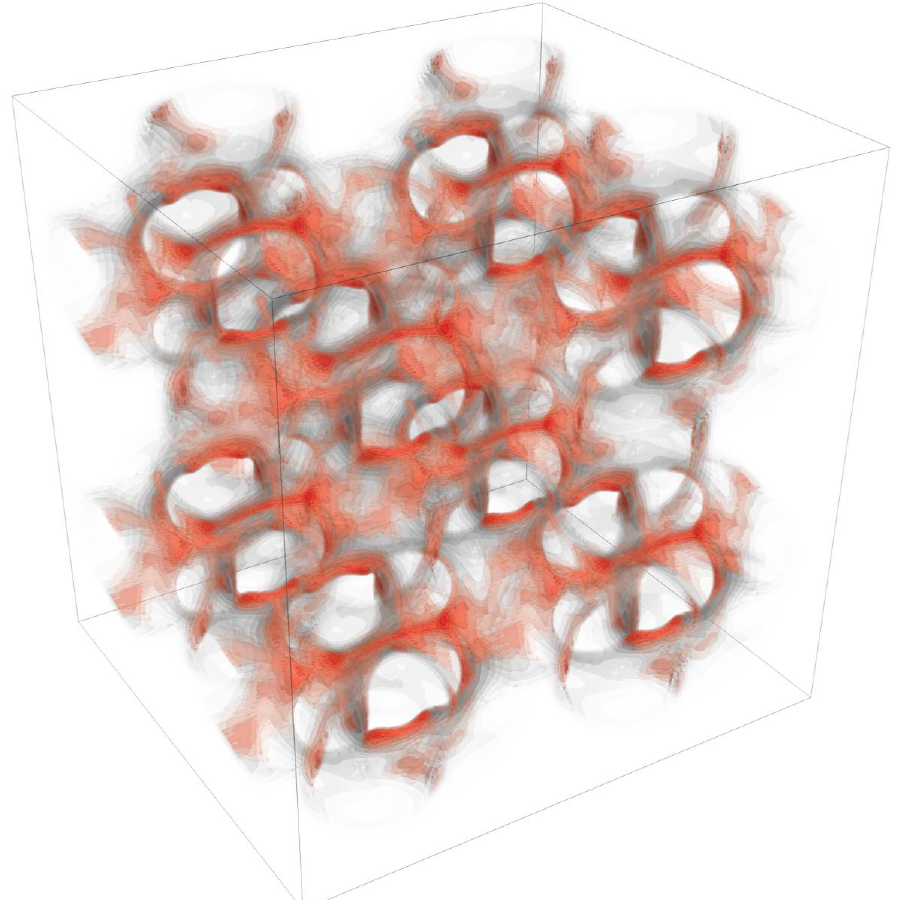}
        & \densfig{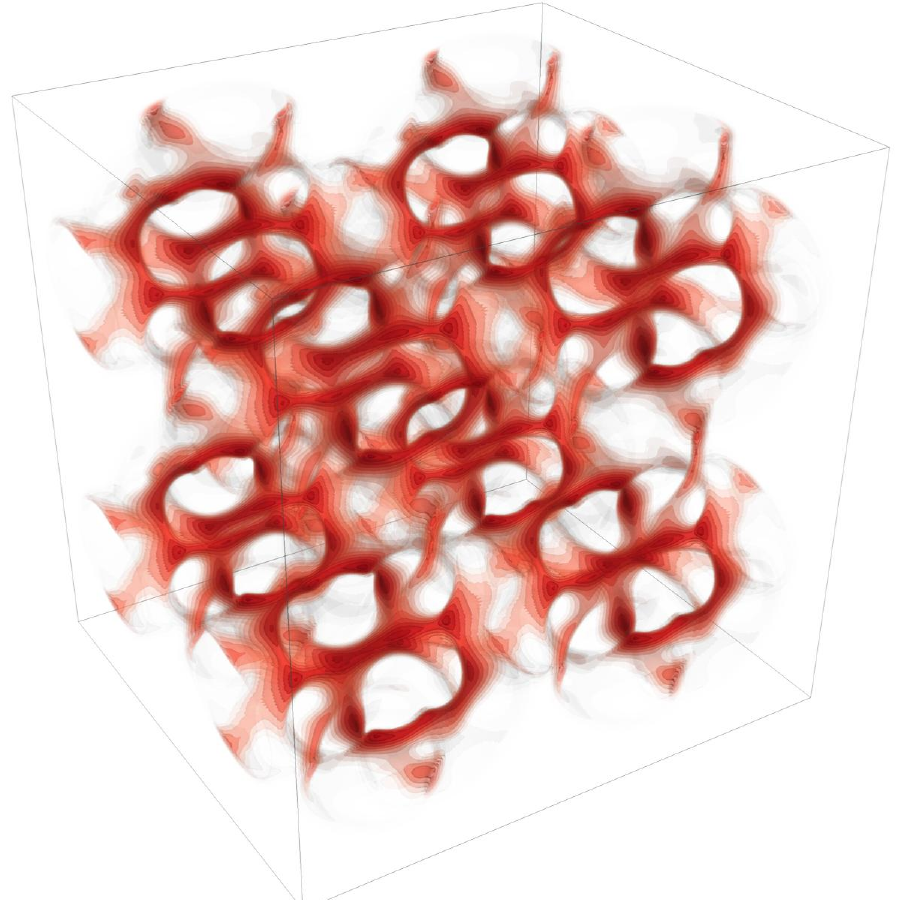} \\
        & 
        & 
        & \simtext{1.000}
        & \simtext{0.921}
        & \simtext{0.671} \\
        
        \arrayrulecolor{black!40}\midrule\arrayrulecolor{black}

        \multirow{2}{*}{\rotatebox[origin=c]{90}{\textbf{AC}}}
        & \densfig{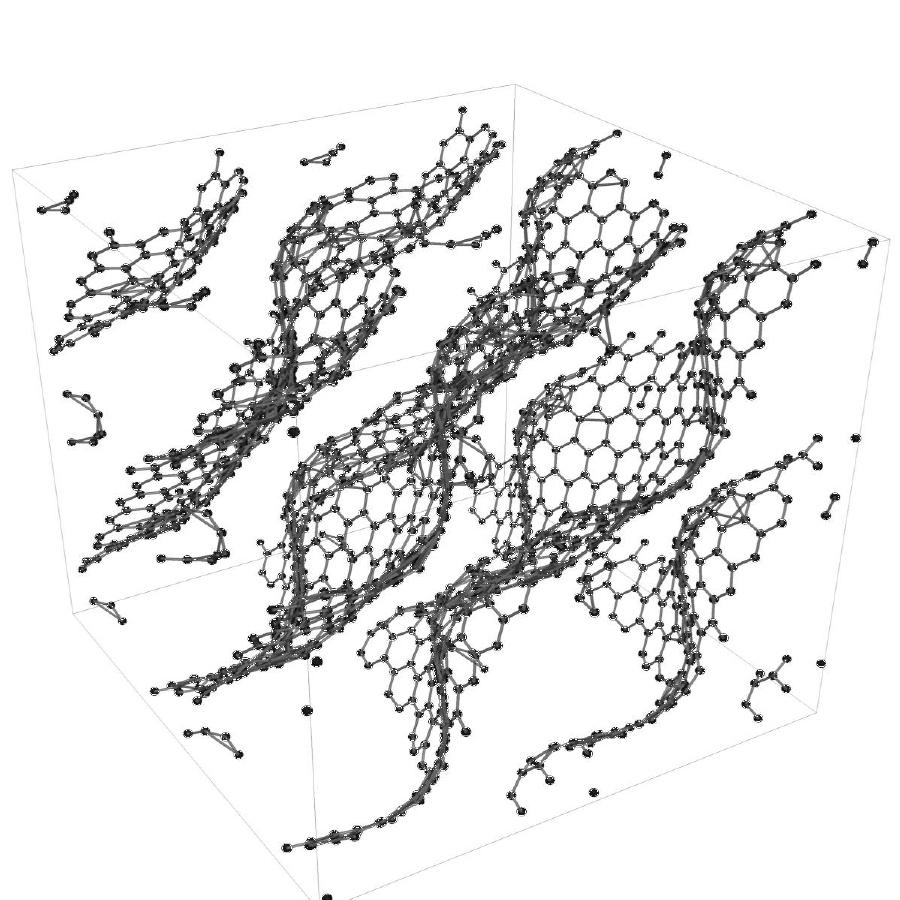}
        & \densfig{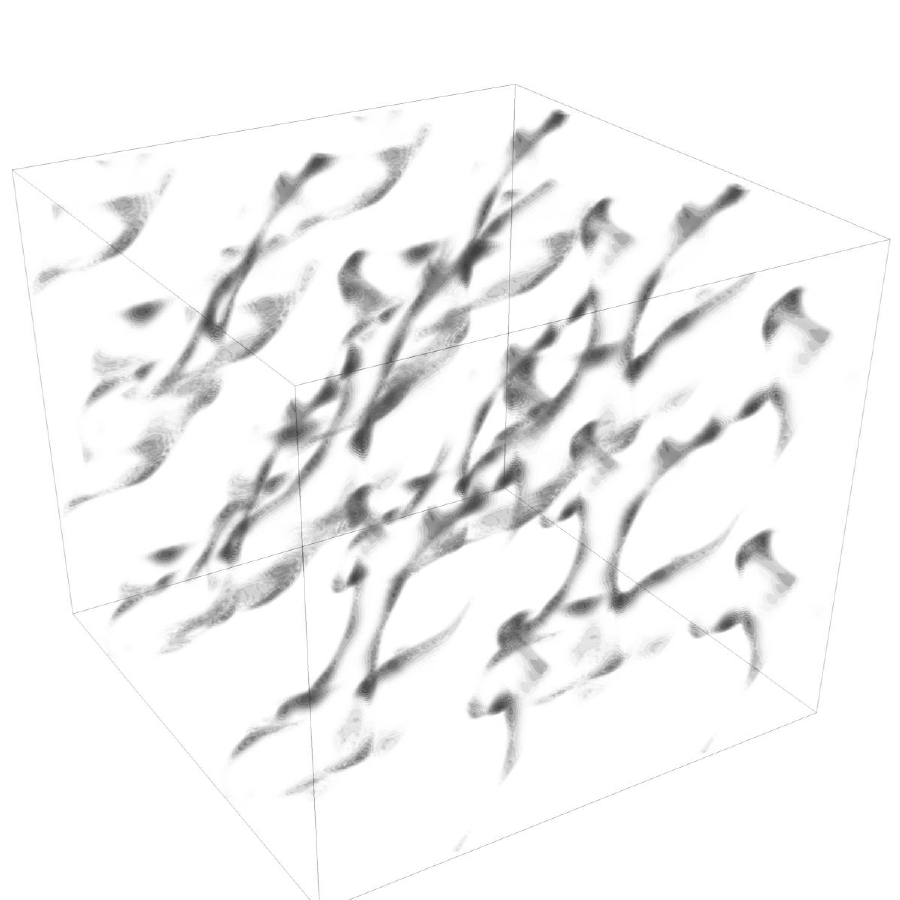}
        & \densfig{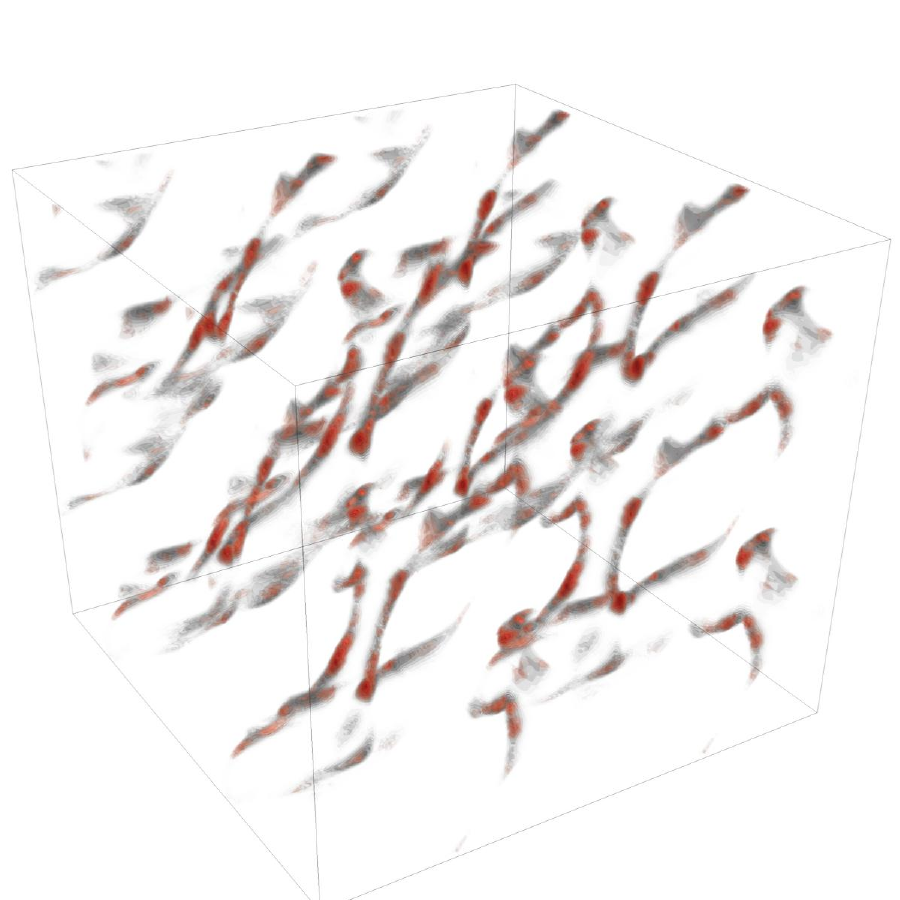}
        & \densfig{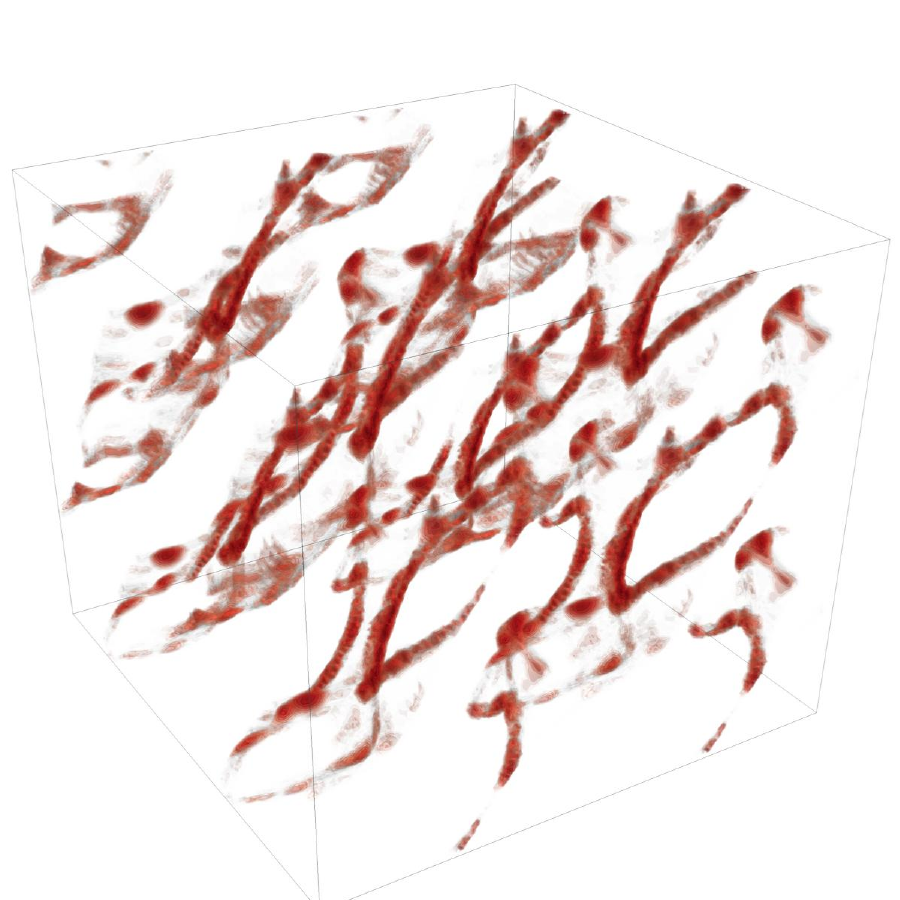}
        & \densfig{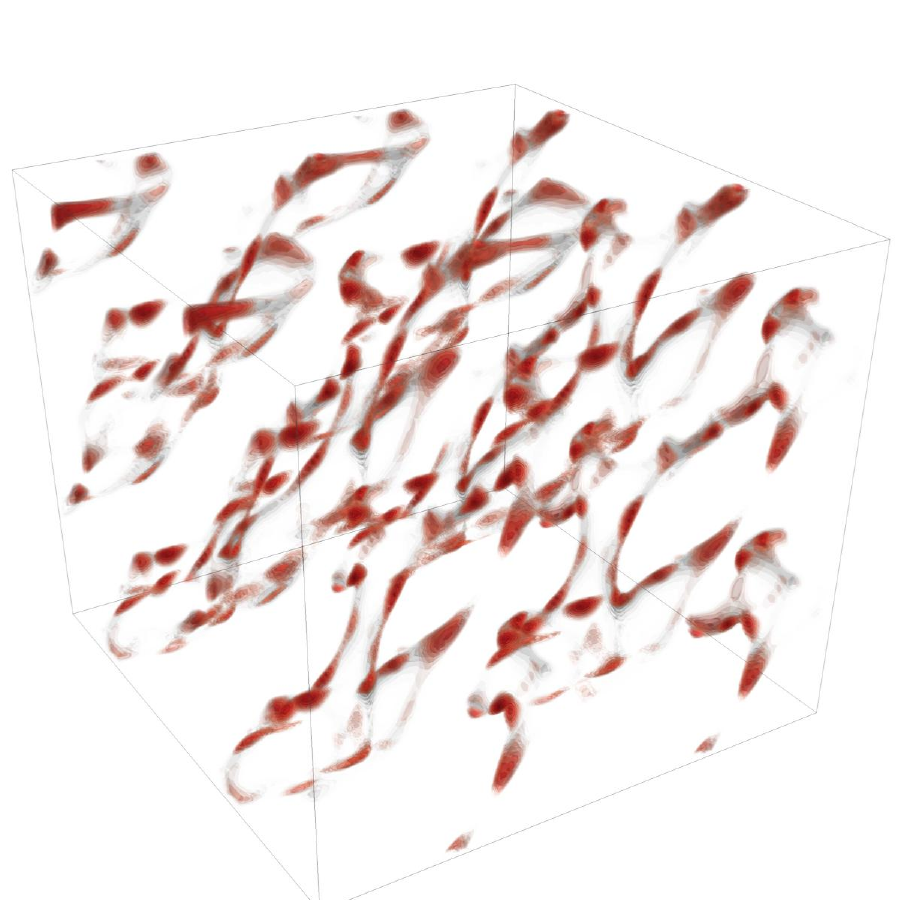} \\
        &
        &
        & \simtext{0.972}
        & \simtext{0.541}
        & \simtext{0.785} \\

        \bottomrule
    \end{tabular}%
    }
    }
    \caption{
    \textbf{Visualization of \ce{CH4} density field and error at cDFT-level.}
    Rows show an ID QMOF case and an OOD AC case.
    Columns show the framework, \hlcdft{\textbf{cDFT}} reference density, and density errors for MADField, DeepAPD, and SorbIIT.
    The reference density is shown in \grayhl{{gray}}, while \redhl{{red}} marks larger normalized absolute error under a shared per-row scale.
    Tanimoto similarity is reported below each model output.
    MADField remains near-white across both cases, indicating substantially lower spatial density error than the density modeling baselines. 
    }
    \vspace{-5pt}
    \label{fig:density_prediction_example}
\end{figure}

\subsection{Gas uptake prediction on MOFs}
\label{sec:uptake_bench}

\begin{table}[t]
\caption{\textbf{Gas uptake prediction on MOFs.}
Gas uptake mean absolute error $N_\text{err}$ in cm$^3$(STP)/g ($\downarrow$) on \hlcdft{\textbf{cDFT}} uptakes (nine adsorbates) and on predicting \hlgcmc{\textbf{GCMC}} uptakes (\ce{CH4}, \ce{Xe}). \textbf{Total} denotes the sample-weighted mean over all evaluated state points within each fidelity group. The conditioning columns indicate whether adsorbate identity (Ads.) and pressure ($P$) enter as conditioning inputs.}
\label{tab:uptake_bench_updated}
\footnotesize
\centering
\setlength{\tabcolsep}{3.0pt}
\renewcommand{\arraystretch}{1.1}
\resizebox{\textwidth}{!}{%
\begin{tabular}{cc l ccccccccc c cc c}
\toprule
\multicolumn{2}{c}{Cond.} & \multicolumn{1}{c}{\multirow{2}{*}{Model}} 
& \multicolumn{10}{c}{\hlcdft{\textbf{cDFT uptake}}} 
& \multicolumn{3}{c}{\hlgcmc{\textbf{GCMC uptake}}} \\
\cmidrule(lr){1-2} \cmidrule(lr){4-13} \cmidrule(lr){14-16}
Ads. & $P$ & & \ce{H2} & \ce{Ar} & \ce{Kr} & \ce{Xe} & \ce{N2} & \ce{CH4} & \ce{CO2} & \ce{C2H6} & \ce{C3H8} & \textbf{Total} & \ce{CH4} & \ce{Xe} & \textbf{Total} \\
\midrule
\multirow{2}{*}{\no} & \multirow{2}{*}{\no}
& RetNet    
& \ptz4.83 & \ptz3.93 & \ptz9.46 & 22.05 & \ptz3.31 & 17.44 & 17.91 & 27.70 & 46.62 & 12.50 & 10.72 & 10.65 & 10.70 \\
& & MOFTransformer     
& \ptz0.12 & \ptz1.87 & \ptz6.21 & 15.09 & \ptz1.32 & \ptz5.32 & 12.12 & 15.66 & 23.40 & \ptz5.76 & \ptz7.22 & 12.32 & \ptz8.92 \\
\arrayrulecolor{black!40}\midrule\arrayrulecolor{black}
\no & \umm
& DeepSorption
& \ptz4.07 & \ptz5.00 & 10.65 & 21.24 & \ptz3.66 & 12.71 & 21.19 & 19.19 & 45.37 & 11.43 & 12.73 & 20.19 & 15.22 \\
\arrayrulecolor{black!40}\midrule\arrayrulecolor{black}
\multirow{3}{*}{\yes} & \multirow{3}{*}{\yes}
& IsothermNet
& \ptz5.21 & \ptz7.53 & 18.71 & 40.25 & \ptz5.53 & 21.71 & 44.11 & 52.15 & 81.59 & 21.23 & 16.26 & 21.78 & 18.10 \\
& & Uni-MOF
& \ptz0.46 & \ptz1.81 & \ptz5.11 & 12.21 & \ptz1.22 & \ptz4.13 & 10.39 & 11.81 & 22.12 & \ptz4.93 & \ptz8.11 & 12.29 & \ptz9.50 \\

& &\textbf{MADField}
& \textbf{\ptz0.08} & \textbf{\ptz0.11} & \textbf{\ptz0.49} & \textbf{\ptz2.87} & \textbf{\ptz0.13} & \textbf{\ptz0.80} & \textbf{\ptz1.17} & \textbf{\ptz2.01} & \textbf{\ptz4.51} & \textbf{\ptz0.82} & \textbf{\ptz0.33} & \textbf{\ptz1.37} & \textbf{\ptz0.58}  \\
\bottomrule
\end{tabular}
}
\vspace{-5pt}
\end{table}


\newcommand{\figdebug}[2][\linewidth]{%
    \ifpaneldebug
        {\setlength{\fboxsep}{0pt}\fboxrule=0.5pt%
         \textcolor{red}{\fbox{\includegraphics[width=#1, trim=0pt 0pt 0pt 0pt, clip]{#2}}}}%
    \else
        \includegraphics[width=#1, trim=0pt 0pt 0pt 0pt, clip]{#2}%
    \fi
}

\newcommand{\figsixpair}[1]{%
    \begin{minipage}[t]{0.160\linewidth}
        \figdebug{figures/fig6_sub/#1.pdf}
    \end{minipage}\hfill
    \begin{minipage}[t]{0.160\linewidth}
        \figdebug{figures/fig6_sub/#1_iso.pdf}
    \end{minipage}%
}

\begin{figure}[t]
    \centering
    \figsixpair{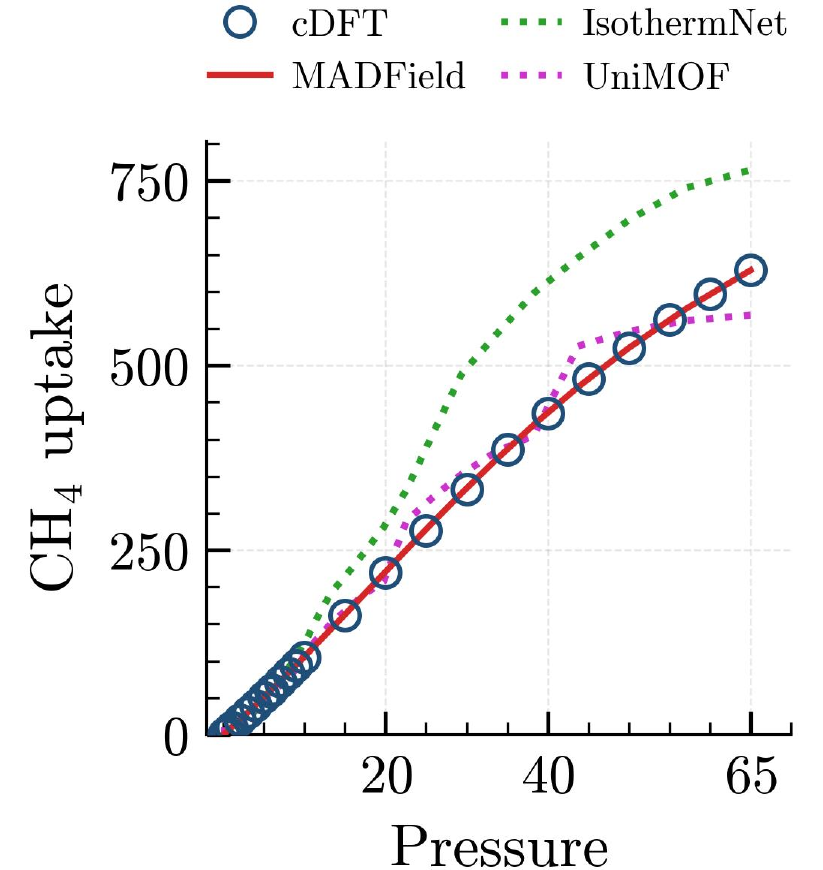}\hfill\figsixpair{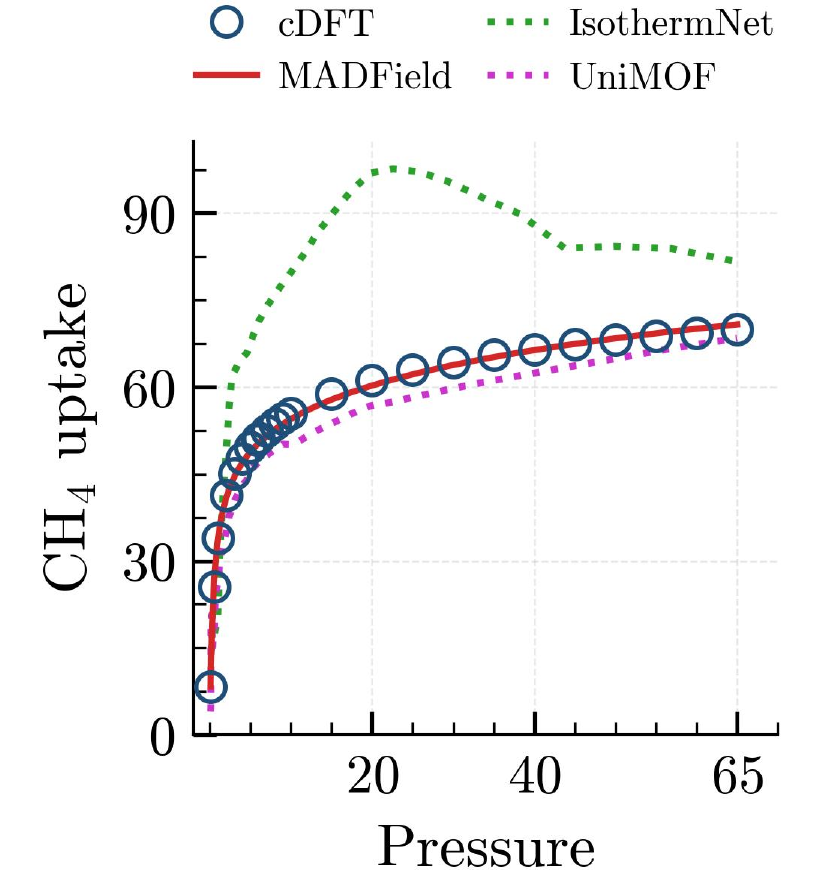}\hfill\figsixpair{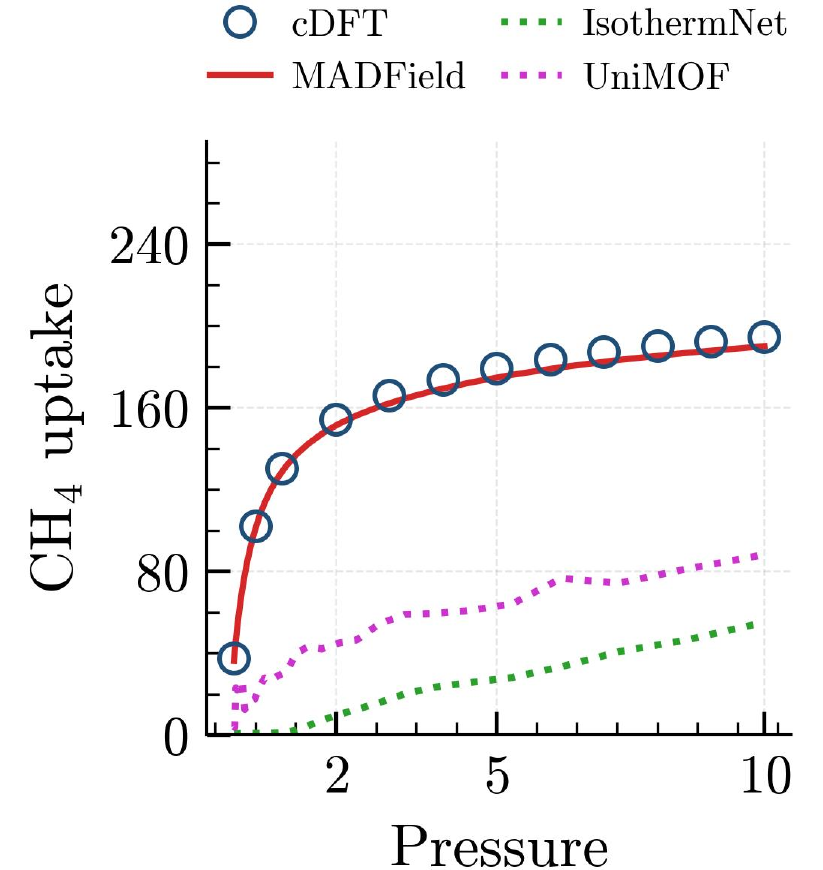}
    \figsixpair{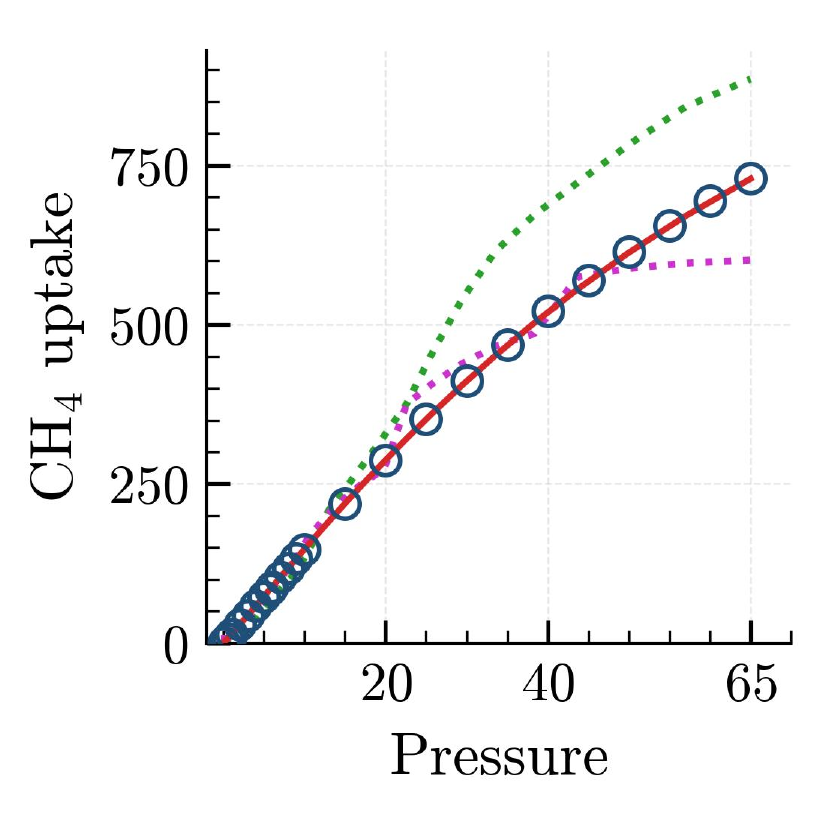}\hfill\figsixpair{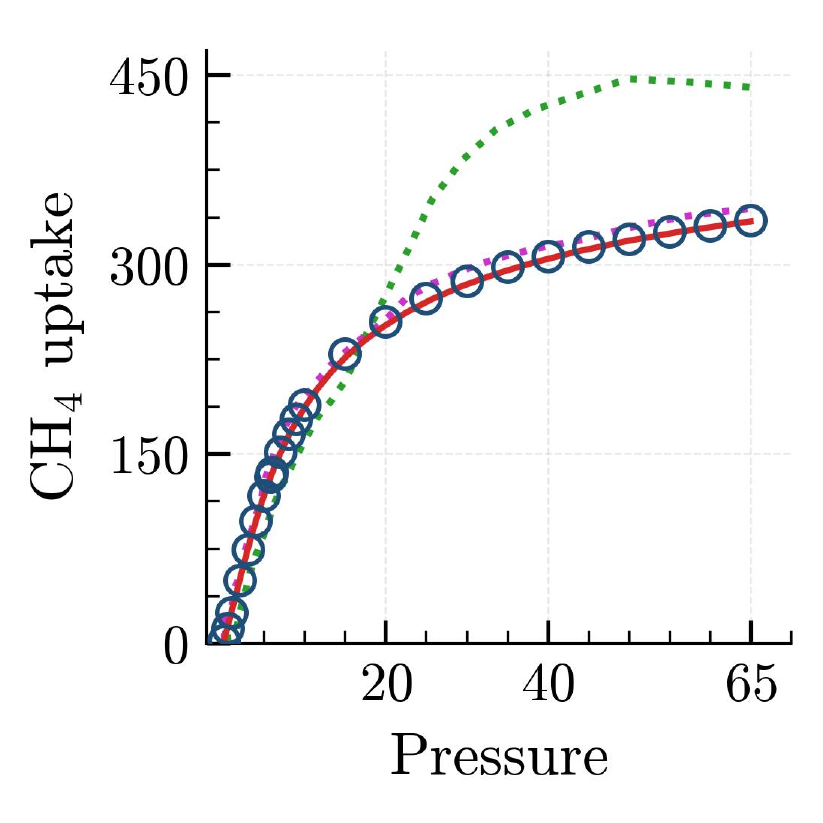}\hfill\figsixpair{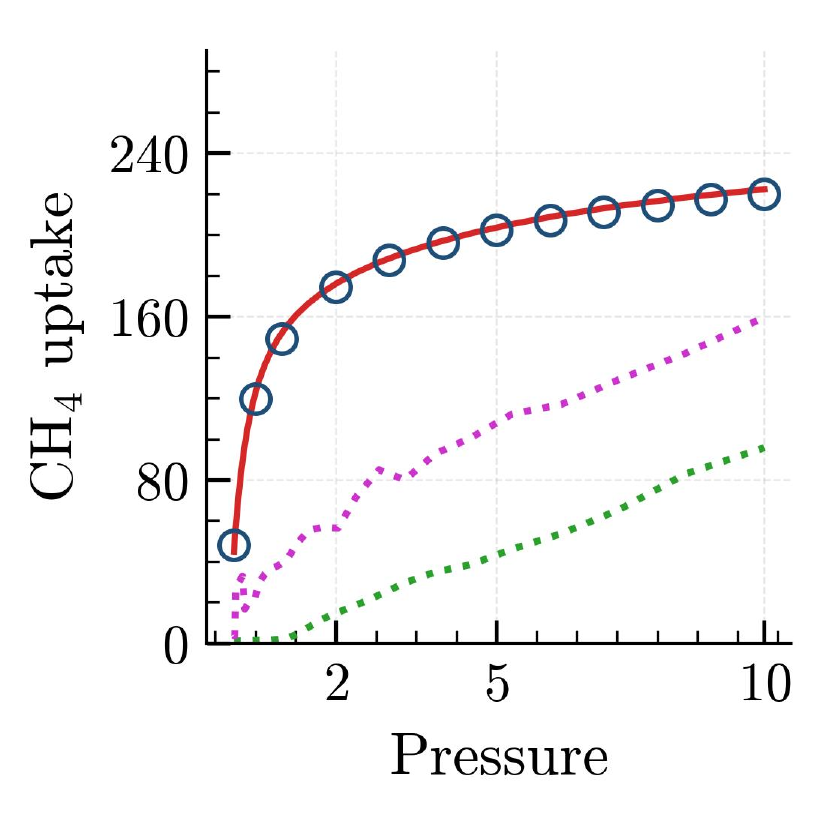}
    \figsixpair{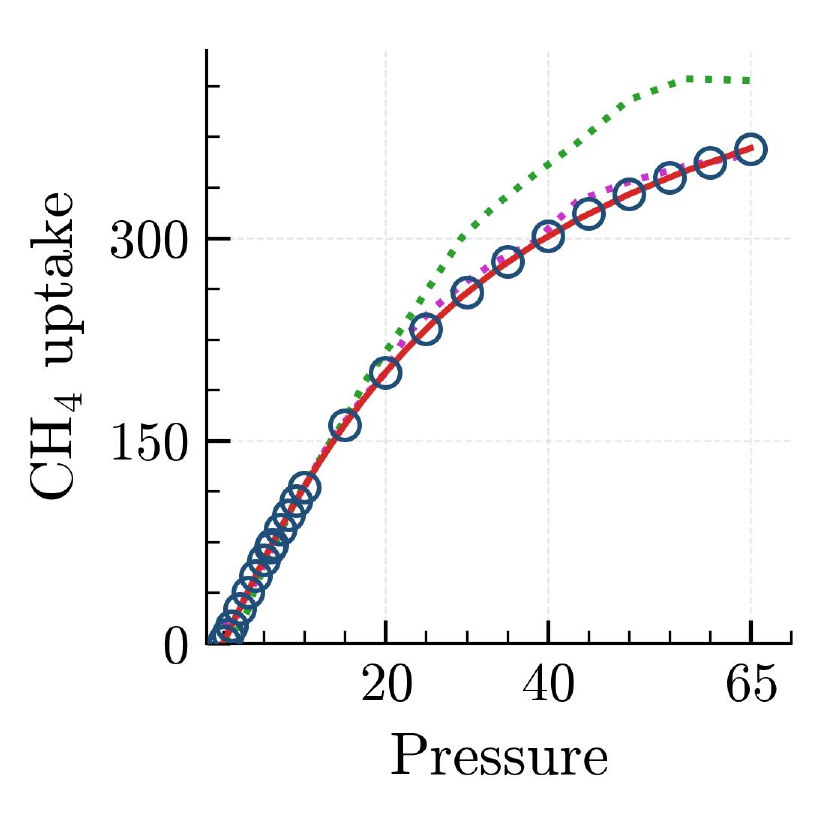}\hfill\figsixpair{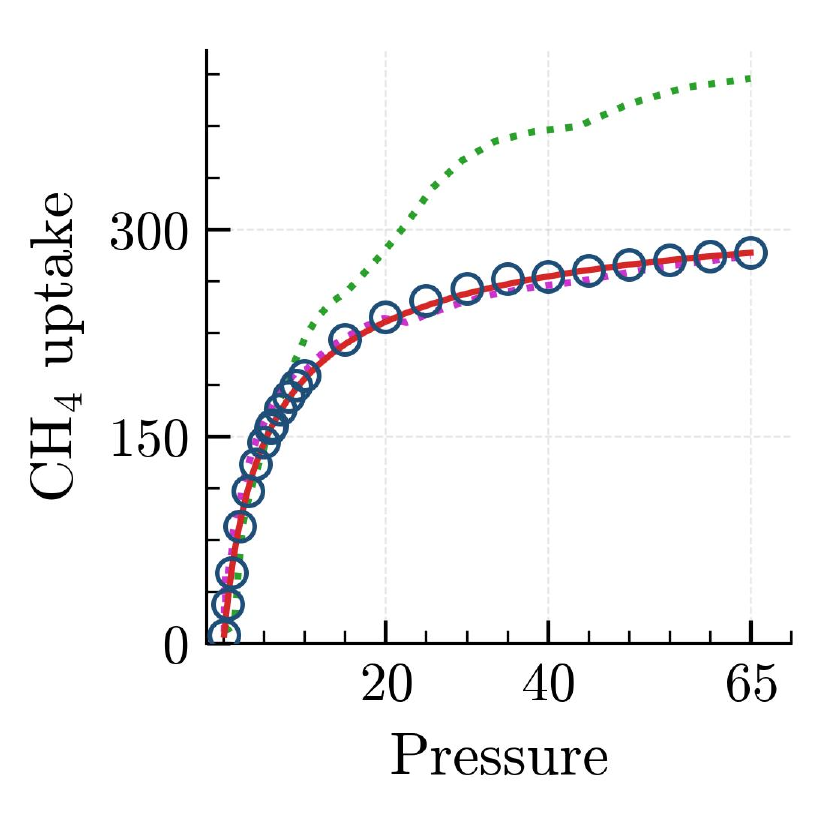}\hfill\figsixpair{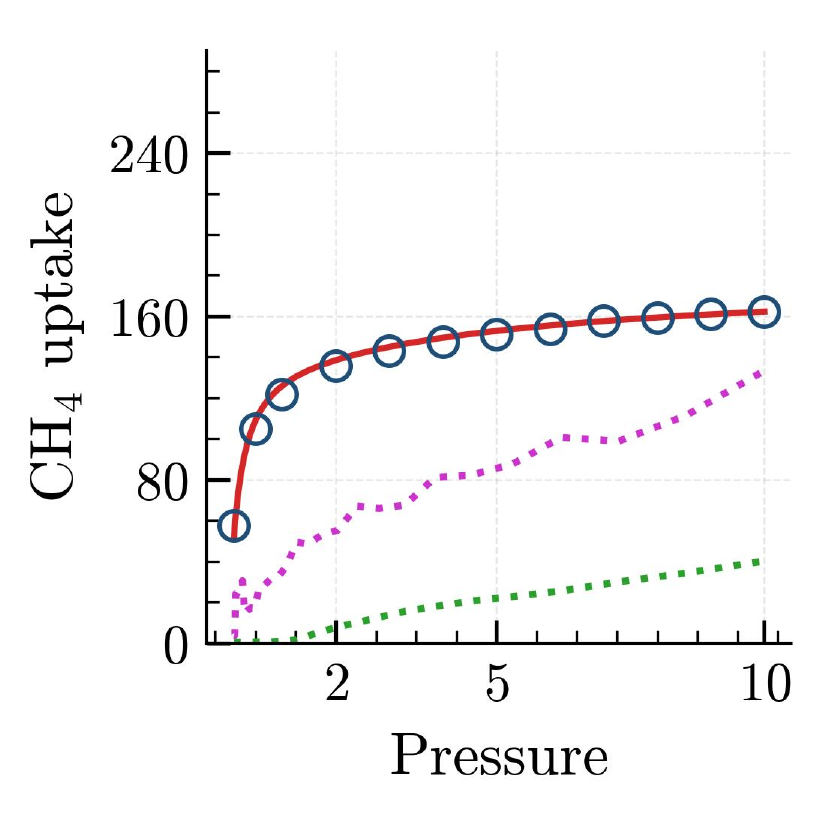}
    
    \caption{
    \textbf{Representative structures and adsorption isotherms.}
    Each pair of panels shows a framework structure and its corresponding \ce{CH4} adsorption isotherm, reported as uptake in $\mathrm{cm^3/g}$ versus pressure in bar.
    The first two rows show six in-distribution QMOF examples, and the last row shows three out-of-distribution JLA amorphous-carbon examples.
    }
    \label{fig:fig6}
    \vspace{-10pt}
\end{figure}

\cref{tab:uptake_bench_updated} reports uptake MAE ($N_{\mathrm{err}}$) on MOFs.
\method{} achieves $N_{\mathrm{err}}$ of $0.82\,\mathrm{cm^3/g}$ on the cDFT benchmark and $0.58\,\mathrm{cm^3/g}$ on the GCMC benchmark, outperforming all baselines by a wide margin.
On cDFT labels, \method{} improves over the strongest baseline Uni-MOF by $6.0\times$ and over MOFTransformer by $7.0\times$.
On GCMC labels, the gap widens further.
\method{} improves over MOFTransformer by $15.4\times$ and over Uni-MOF by $16.4\times$.
Notably, Uni-MOF and MOFTransformer are the two strongest baselines on this benchmark, yet \method{} surpasses both by an order of magnitude at both fidelity levels.
\cref{fig:fig6} shows representative isotherm comparisons on QMOF (ID) and JLA amorphous carbons (OOD).

To disentangle the contribution of the prediction target from the architecture, we evaluate a variant (\method{}-Scalar) that shares the same backbone but regresses uptake $N$ directly (see \cref{app:additional}).
\method{}-Scalar incurs $7.2\times$ higher $N_{\mathrm{err}}$ than \method{}, confirming that the primary accuracy gain stems from predicting the density field rather than a scalar.
Nevertheless, \methodscalar{} prediction is still comparable against all other baselines, indicating that the architecture itself provides additional gains.

Without cDFT pre-training, \method{} trained on GCMC data alone (see \cref{app:additional}) reaches $N_{\mathrm{err}}=1.68\,\mathrm{cm^3/g}$ on ID MOFs, $2.9\times$ higher than the multi-fidelity model ($0.58\,\mathrm{cm^3/g}$).
On OOD materials, the scratch model degrades to $13.60\,\mathrm{cm^3/g}$ versus $0.75\,\mathrm{cm^3/g}$ with multi-fidelity training, an $18.1\times$ gap that highlights the importance of the cDFT density prior for generalization.

\subsection{Transferability to other materials}
\label{sec:transfer_bench}

\setlength{\tabcolsep}{5.0pt}
\renewcommand{\arraystretch}{1}
\begin{table}[t]
\caption{\textbf{Transferability benchmark on disordered porous materials.}
Gas uptake mean absolute error $N_\text{err}$ in cm$^3$(STP)/g ($\downarrow$) on \hlcdft{\textbf{cDFT}} labels (AC at 1\,bar) and on \hlgcmc{\textbf{GCMC}} labels (\ce{CH4} on PIM, HCP, Kerogen, AC). \textbf{Total} denotes the sample-weighted mean over all GCMC families. The conditioning columns indicate whether adsorbate identity (Ads.) and pressure ($P$) enter as conditioning inputs.}
\label{tab:uptake_transfer}
\centering
\resizebox{0.9\textwidth}{!}{%
\footnotesize
\begin{tabular}{cc l c ccccc}
\toprule
\multicolumn{2}{c}{Cond.} & \multicolumn{1}{c}{\multirow{2}{*}{Model}} 
& \hlcdft{\textbf{cDFT uptake}} 
& \multicolumn{5}{c}{\hlgcmc{\textbf{GCMC uptake}}} \\
\cmidrule(lr){1-2} \cmidrule(lr){4-4} \cmidrule(lr){5-9}
Ads. & $P$ & & AC & PIM & HCP & Kerogen & AC & \textbf{Total} \\
\midrule
\multirow{2}{*}{\no} & \multirow{2}{*}{\no}
& RetNet           & 21.30 & \ptz2.28 & \ptz1.94 & \ptz6.36 & \ptz9.51 & \ptz2.61 \\
& & MOFTransformer & 56.40 & \ptz6.90 & 13.10 & 13.60 & 30.60 & 10.03 \\
\arrayrulecolor{black!40}\midrule\arrayrulecolor{black}
\no & \umm
& DeepSorption     & 52.50 & \ptz6.14 & \ptz6.72 & 26.80 & 23.10 & \ptz8.17 \\
\arrayrulecolor{black!40}\midrule\arrayrulecolor{black}
\multirow{3}{*}{\yes} & \multirow{3}{*}{\yes}
& IsothermNet      & 13.33 & \ptz9.63 & 11.41 & 44.23 & 46.03 & 13.48 \\
& & Uni-MOF        & 31.30 & \ptz8.20 & \ptz6.10 & 27.80 & 26.90 & \ptz9.48 \\
\cellcolor{white} & \cellcolor{white} & \textbf{MADField} & \textbf{\ptz2.71} & \textbf{\ptz0.57} & \textbf{\ptz0.40} & \textbf{\ptz1.42} & \textbf{\ptz0.67} & \textbf{\ptz0.62} \\
\bottomrule
\end{tabular}
}
\vspace{-8pt}
\end{table}



\cref{tab:uptake_transfer} reports $N_{\mathrm{err}}$ on disordered porous materials unseen during training, evaluating zero-shot transferability.
\method{} achieves $2.71\,\mathrm{cm^3/g}$ on the cDFT amorphous carbon benchmark, improving $4.9\times$ over the best baseline IsothermNet, and $0.62\,\mathrm{cm^3/g}$ on the GCMC disordered-material benchmark, improving $4.21\times$ over RetNet.
The GCMC benchmark on these OOD materials is particularly challenging since they comprise up to $18{,}019$ atoms with unit cell dimensions reaching about $56.7\,\text{\AA}$, yet \method{} maintains sub-$1\,\mathrm{cm^3/g}$ error across all four material families.
Notably, Uni-MOF and MOFTransformer, the strongest baselines on MOFs, degrade severely on these unseen materials (\textit{e.g.}, Uni-MOF rises from $7.70$ to $31.30\,\mathrm{cm^3/g}$ on cDFT ACs), while \method{} degrades only from $0.82$ to $2.71$, demonstrating that the learned density generalizes across material classes and scales.

\subsection{{High-throughput screening on ARC-MOF}}
\label{sec:wc_screening}

\begin{figure}[p]
    \centering
    \vspace*{\fill}
    \includegraphics[width=0.95\linewidth]{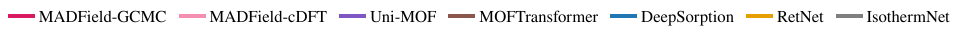}\\[-2pt]
    \begin{overpic}[width=0.325\linewidth]{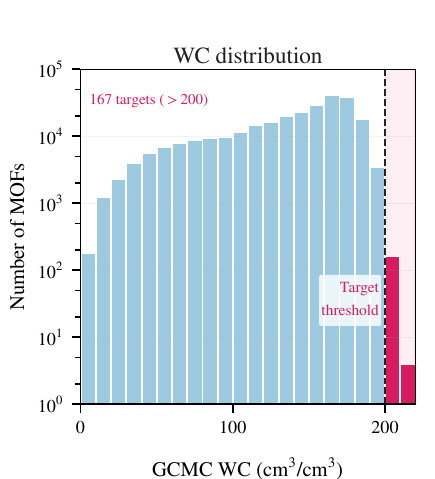}\put(1,95){\footnotesize\textbf{(a)}}\end{overpic}\hfill
    \begin{overpic}[width=0.325\linewidth]{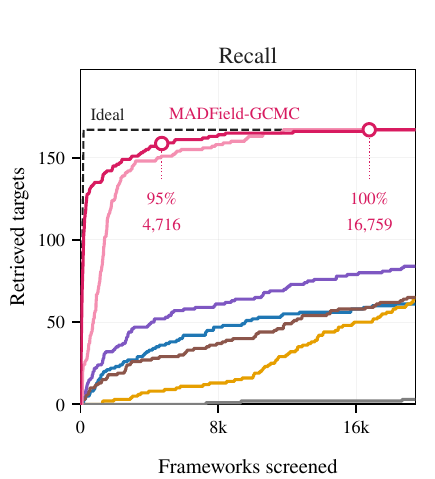}\put(1,95){\footnotesize\textbf{(b)}}\end{overpic}\hfill
    \begin{overpic}[width=0.325\linewidth]{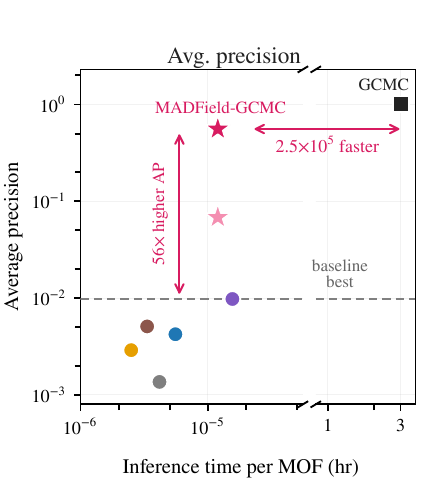}\put(1,95){\footnotesize\textbf{(c)}}\end{overpic}
    \caption{\textbf{\ce{CH4} working-capacity screening result on the ARC-MOF database} ($167$ high-WC targets, $\mathrm{WC}\ge 200~\mathrm{cm^3/cm^3}$, $0.06\%$ of $270{,}583$ frameworks).
    \textbf{(a)} GCMC WC distribution with the rare target regime beyond the threshold.
    \textbf{(b)} Targets retrieved versus the number of frameworks screened in predicted-WC order, where \methodgcmc{} reaches $95\%$ and $100\%$ recall after only $4{,}716$ and $16{,}759$ frameworks ($1.7\%$ and $6.2\%$).
    \textbf{(c)} Average precision versus per-MOF time, where \methodgcmc{} attains AP $=0.557$ ($56\times$ the best learned baseline) at five orders of magnitude lower cost than GCMC.}
    \label{fig:fig8_wc_screening}

    \vspace*{\fill}
    \includegraphics[width=\linewidth]{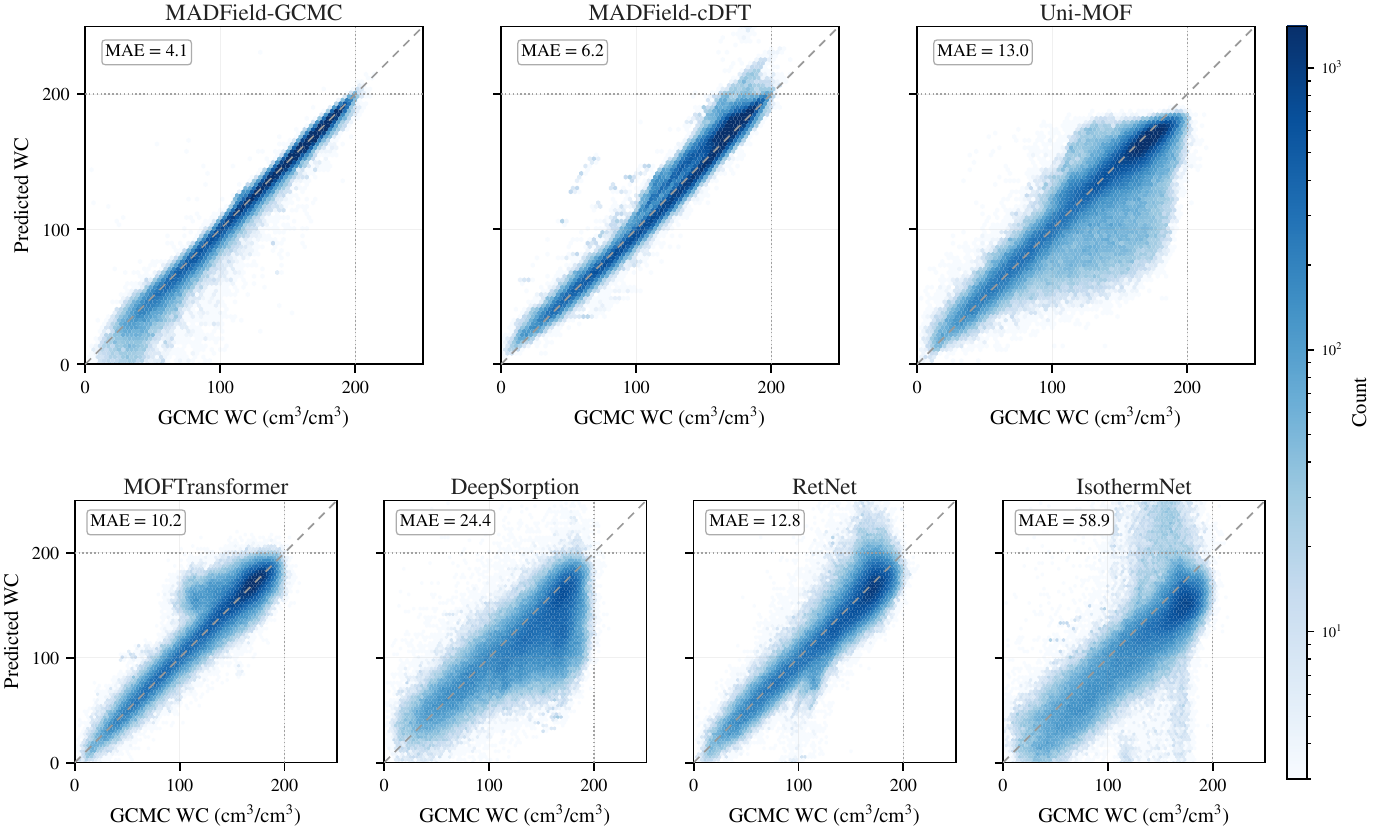}
    \caption{\textbf{\ce{CH4} working-capacity parity plot on the ARC-MOF database.}
    Each panel compares predicted and GCMC WC across all $270{,}583$ frameworks, with hexbin count on a shared log scale, the gray dashed line marking $y=x$, dotted lines marking the $\mathrm{WC}=200~\mathrm{cm^3/cm^3}$ threshold, and the per-panel WC MAE shown.
    Panels are ordered by screening average precision, with the top row showing \methodgcmc{}, \methodcdft{}, and the strongest baseline Uni-MOF, and the bottom row showing the remaining baselines.
    \methodgcmc{} resolves the rare high-WC regime, whereas the baselines show larger dispersion and systematic errors, especially among high-WC frameworks.}
    \label{fig:fig7_wc_parity}
    \vspace*{\fill}
\end{figure}

{Beyond per-condition accuracy, deployable screening must surface rare high-capacity frameworks from a highly imbalanced candidate pool.
We rank all $270{,}583$ ARC-MOF frameworks by predicted \ce{CH4} working capacity and evaluate retrieval of the $167$ high-WC targets defined in \cref{sec:setup}, which form only $0.06\%$ of the database (\cref{fig:fig8_wc_screening}(a)).
Ranking by \methodgcmc{} recovers $95\%$ of these targets after screening only $4{,}716$ frameworks and all $167$ after $16{,}759$ ($1.7\%$ and $6.2\%$ of the database), closely tracking the \emph{Ideal} upper bound (\cref{fig:fig8_wc_screening}(b)), whereas every baseline stays far behind.
In the precision versus speed comparison (\cref{fig:fig8_wc_screening}(c)), \methodgcmc{} attains AP $=0.557$, a $56\times$ improvement over the strongest learned baseline, Uni-MOF (AP $=0.010$), while MOFTransformer reaches AP $=0.005$ and the cDFT-only ablation \methodcdft{} drops to AP $=0.068$, isolating the contribution of GCMC fine-tuning, all at a per-MOF cost five orders of magnitude lower than fully converged GCMC.}

The per-MOF parity plots in \cref{fig:fig7_wc_parity} reveal the source of this advantage. \methodgcmc{} aligns tightly with the GCMC working-capacity reference across the full range, whereas the baselines show substantially larger dispersion and systematic errors, especially in the rare high-WC regime.

\clearpage
\subsection{Accelerating cDFT with \method{}}
\label{sec:warmstart}

\begin{wraptable}[8]{r}{0.4\textwidth}
\vspace{-12pt}
\caption{
\textbf{\method{}-initialized cDFT}
\method{} increases the convergence rate and reduces Picard iterations compared with the standard Boltzmann init.
}
\label{tab:warmstart}
\centering
\resizebox{\linewidth}{!}{%
\begin{tabular}{lcc}
\toprule
Initialization & Conv.\ rate (\%) & Mean iter. \\
\midrule
Standard init & 90.06 & 329.2 \\
\textbf{\methodcdft{}} & \textbf{94.21} & \textbf{166.5} \\
\bottomrule
\end{tabular}%
}
\vspace{-10pt}
\end{wraptable}

Accurate density predictions from \methodcdft{} can serve as high-quality initial fields for iterative cDFT solvers.
As shown in \cref{tab:warmstart}, \methodcdft{} initialization improves the convergence rate by recovering 42\% of cases that failed under standard settings without initialization.
Furthermore, it achieves a $2.0\times$ reduction in iteration steps.
This acceleration and convergence recovery enable our strategy of bootstrapping cDFT labels.

\FloatBarrier
\section{Conclusion}
\label{sec:conclusion}

We presented \method{}, a model that amortizes adsorption simulation by predicting the 3D equilibrium density field of gas molecules within nanoporous materials. By formulating gas uptake as an integral of the predicted field, \method{} unifies scalar uptake prediction, spatial density estimation, and cDFT solver acceleration within a single framework. Multi-fidelity training on cDFT and GCMC data enables strong in-distribution performance on MOFs and robust generalization to out-of-distribution disordered materials, surpassing existing baselines by up to an order of magnitude in uptake accuracy.

\paragraph{Limitations \& future work. }
The cDFT labels used for pre-training inherit the approximation error of the underlying free energy functional, though this is mitigated by their role as low-fidelity supervision that is subsequently corrected by GCMC fine-tuning, and by the close agreement between PC-SAFT cDFT and GCMC for the non-polar adsorbates considered here. Additionally, \method{} currently treats single-component adsorption; extending to mixture predictions, for instance by combining single-component isotherms through IAST, is a promising direction for future work.


\section*{Acknowledgment}
We thank Jake Burner for providing the GCMC dataset for ARC-MOF, and Seonghwan Kim for insightful and constructive suggestions that improved this work.

\bibliographystyle{unsrtnat}
\bibliography{references}

\begin{thebibliography}{48}
\providecommand{\natexlab}[1]{#1}
\providecommand{\url}[1]{\texttt{#1}}
\expandafter\ifx\csname urlstyle\endcsname\relax
  \providecommand{\doi}[1]{doi: #1}\else
  \providecommand{\doi}{doi: \begingroup \urlstyle{rm}\Url}\fi

\bibitem[Wilmer et~al.(2012)Wilmer, Leaf, Lee, Farha, Hauser, Hupp, and Snurr]{wilmer2012large}
Christopher~E Wilmer, Michael Leaf, Chang~Yeon Lee, Omar~K Farha, Brad~G Hauser, Joseph~T Hupp, and Randall~Q Snurr.
\newblock Large-scale screening of hypothetical metal--organic frameworks.
\newblock \emph{Nature chemistry}, 4\penalty0 (2):\penalty0 83--89, 2012.

\bibitem[Mason et~al.(2014)Mason, Veenstra, and Long]{mason2014evaluating}
Jarad~A Mason, Mike Veenstra, and Jeffrey~R Long.
\newblock Evaluating metal--organic frameworks for natural gas storage.
\newblock \emph{Chemical Science}, 5\penalty0 (1):\penalty0 32--51, 2014.

\bibitem[Lin et~al.(2012)Lin, Berger, Martin, Kim, Swisher, Jariwala, Rycroft, Bhown, Deem, Haranczyk, et~al.]{lin2012silico}
Li-Chiang Lin, Adam~H Berger, Richard~L Martin, Jihan Kim, Joseph~A Swisher, Kuldeep Jariwala, Chris~H Rycroft, Abhoyjit~S Bhown, Michael~W Deem, Maciej Haranczyk, et~al.
\newblock In silico screening of carbon-capture materials.
\newblock \emph{Nature materials}, 11\penalty0 (7):\penalty0 633--641, 2012.

\bibitem[Charalambous et~al.(2024)Charalambous, Moubarak, Schilling, Sanchez~Fernandez, Wang, Herraiz, Mcilwaine, Peh, Garvin, Jablonka, et~al.]{charalambous2024holistic}
Charithea Charalambous, Elias Moubarak, Johannes Schilling, Eva Sanchez~Fernandez, Jin-Yu Wang, Laura Herraiz, Fergus Mcilwaine, Shing~Bo Peh, Matthew Garvin, Kevin~Maik Jablonka, et~al.
\newblock A holistic platform for accelerating sorbent-based carbon capture.
\newblock \emph{Nature}, 632\penalty0 (8023):\penalty0 89--94, 2024.

\bibitem[Adams(1975)]{adams1975grand}
DJ~Adams.
\newblock Grand canonical ensemble monte carlo for a lennard-jones fluid.
\newblock \emph{Molecular Physics}, 29\penalty0 (1):\penalty0 307--311, 1975.

\bibitem[Dubbeldam et~al.(2016)Dubbeldam, Calero, Ellis, and Snurr]{dubbeldam2016raspa}
David Dubbeldam, Sof{\'\i}a Calero, Donald~E Ellis, and Randall~Q Snurr.
\newblock Raspa: molecular simulation software for adsorption and diffusion in flexible nanoporous materials.
\newblock \emph{Molecular Simulation}, 42\penalty0 (2):\penalty0 81--101, 2016.

\bibitem[Ran et~al.(2024)Ran, Sharma, Balestra, Li, Calero, Vlugt, Snurr, and Dubbeldam]{ran2024raspa3}
YA~Ran, S~Sharma, SRG Balestra, Z~Li, S~Calero, TJH Vlugt, RQ~Snurr, and D~Dubbeldam.
\newblock Raspa3: A monte carlo code for computing adsorption and diffusion in nanoporous materials and thermodynamics properties of fluids.
\newblock \emph{The Journal of Chemical Physics}, 161\penalty0 (11), 2024.

\bibitem[Evans(1979)]{evans1979nature}
Robert Evans.
\newblock The nature of the liquid-vapour interface and other topics in the statistical mechanics of non-uniform, classical fluids.
\newblock \emph{Advances in physics}, 28\penalty0 (2):\penalty0 143--200, 1979.

\bibitem[Roth(2010)]{roth2010fundamental}
Roland Roth.
\newblock Fundamental measure theory for hard-sphere mixtures: a review.
\newblock \emph{Journal of Physics: Condensed Matter}, 22\penalty0 (6):\penalty0 063102, 2010.

\bibitem[Dufour-D{\'e}cieux et~al.(2025)Dufour-D{\'e}cieux, Rehner, Schilling, Moubarak, Gross, and Bardow]{dufour2025classical}
Vincent Dufour-D{\'e}cieux, Philipp Rehner, Johannes Schilling, Elias Moubarak, Joachim Gross, and Andr{\'e} Bardow.
\newblock Classical density functional theory as a fast and accurate method for adsorption property prediction of porous materials.
\newblock \emph{AIChE Journal}, 71\penalty0 (6):\penalty0 e18779, 2025.

\bibitem[Thiele et~al.(2026)Thiele, Teh, Bursik, Granderath, Bauer, Dufour-D{\'e}cieux, Rehner, Stierle, Bardow, Hansen, et~al.]{thiele2026efficient}
Nadine Thiele, Tiong~Wei Teh, Benjamin Bursik, Marcel Granderath, Gernot Bauer, Vincent Dufour-D{\'e}cieux, Philipp Rehner, Rolf Stierle, Andr{\'e} Bardow, Niels Hansen, et~al.
\newblock Efficient prediction of multicomponent adsorption isotherms and enthalpies of adsorption in mofs using classical density functional theory.
\newblock \emph{The Journal of Physical Chemistry B}, 2026.

\bibitem[Sarikas et~al.(2024)Sarikas, Gkagkas, and Froudakis]{sarikas2024gas}
Antonios~P Sarikas, Konstantinos Gkagkas, and George~E Froudakis.
\newblock Gas adsorption meets deep learning: voxelizing the potential energy surface of metal-organic frameworks.
\newblock \emph{Scientific Reports}, 14\penalty0 (1):\penalty0 2242, 2024.

\bibitem[Chen et~al.(2022)Chen, Jiao, Liu, Liu, and Lu]{chen2022interpretable}
Pin Chen, Rui Jiao, Jinyu Liu, Yang Liu, and Yutong Lu.
\newblock Interpretable graph transformer network for predicting adsorption isotherms of metal--organic frameworks.
\newblock \emph{Journal of Chemical Information and Modeling}, 62\penalty0 (22):\penalty0 5446--5456, 2022.

\bibitem[Sun and Siepmann(2024)]{sun2024understanding}
Yangzesheng Sun and J~Ilja Siepmann.
\newblock Understanding and predicting the spatially resolved adsorption properties of nanoporous materials.
\newblock \emph{Journal of Chemical Theory and Computation}, 20\penalty0 (12):\penalty0 5259--5275, 2024.

\bibitem[Burner et~al.(2026)Burner, Marchand, Cicciarella, Gibaldi, and Woo]{burner2026rapid}
Jake Burner, Olivier Marchand, Rosa Cicciarella, Marco Gibaldi, and Tom~K Woo.
\newblock Rapid prediction of single-site adsorbate probability distributions in metal--organic frameworks using graph neural networks.
\newblock \emph{Digital Discovery}, 2026.

\bibitem[Col{\'o}n and Snurr(2014)]{colon2014high}
Yamil~J Col{\'o}n and Randall~Q Snurr.
\newblock High-throughput computational screening of metal--organic frameworks.
\newblock \emph{Chemical Society Reviews}, 43\penalty0 (16):\penalty0 5735--5749, 2014.

\bibitem[Simon et~al.(2015)Simon, Kim, Gomez-Gualdron, Camp, Chung, Martin, Mercado, Deem, Gunter, Haranczyk, et~al.]{simon2015materials}
Cory~M Simon, Jihan Kim, Diego~A Gomez-Gualdron, Jeffrey~S Camp, Yongchul~G Chung, Richard~L Martin, Rocio Mercado, Michael~W Deem, Dan Gunter, Maciej Haranczyk, et~al.
\newblock The materials genome in action: identifying the performance limits for methane storage.
\newblock \emph{Energy \& Environmental Science}, 8\penalty0 (4):\penalty0 1190--1199, 2015.

\bibitem[Sauer and Gross(2017)]{sauer2017classical}
Elmar Sauer and Joachim Gross.
\newblock Classical density functional theory for liquid--fluid interfaces and confined systems: A functional for the perturbed-chain polar statistical associating fluid theory equation of state.
\newblock \emph{Industrial \& Engineering Chemistry Research}, 56\penalty0 (14):\penalty0 4119--4135, 2017.

\bibitem[Lee et~al.(2021)Lee, Kim, Cho, Lee, Lee, Cho, and Kim]{lee2021computational}
Sangwon Lee, Baekjun Kim, Hyun Cho, Hooseung Lee, Sarah~Yunmi Lee, Eun~Seon Cho, and Jihan Kim.
\newblock Computational screening of trillions of metal--organic frameworks for high-performance methane storage.
\newblock \emph{ACS Applied Materials \& Interfaces}, 13\penalty0 (20):\penalty0 23647--23654, 2021.

\bibitem[Shi et~al.(2023)Shi, Li, Anstine, Tang, Colina, Sholl, Siepmann, and Snurr]{shi2023two}
Kaihang Shi, Zhao Li, Dylan~M Anstine, Dai Tang, Coray~M Colina, David~S Sholl, J~Ilja Siepmann, and Randall~Q Snurr.
\newblock Two-dimensional energy histograms as features for machine learning to predict adsorption in diverse nanoporous materials.
\newblock \emph{Journal of chemical theory and computation}, 19\penalty0 (14):\penalty0 4568--4583, 2023.

\bibitem[Lin et~al.(2025)Lin, Zhong, Chen, and Deng]{lin2025unified}
Emily Lin, Yang Zhong, Gang Chen, and Sili Deng.
\newblock Unified physio-thermodynamic descriptors via learned co2 adsorption properties in metal-organic frameworks.
\newblock \emph{npj Computational Materials}, 11\penalty0 (1):\penalty0 225, 2025.

\bibitem[Cui et~al.(2023)Cui, Wu, Zhang, Yang, Hu, Fang, Ye, Zhang, Suo, Mo, et~al.]{cui2023direct}
Jiyu Cui, Fang Wu, Wen Zhang, Lifeng Yang, Jianbo Hu, Yin Fang, Peng Ye, Qiang Zhang, Xian Suo, Yiming Mo, et~al.
\newblock Direct prediction of gas adsorption via spatial atom interaction learning.
\newblock \emph{Nature Communications}, 14\penalty0 (1):\penalty0 7043, 2023.

\bibitem[Kang et~al.(2023)Kang, Park, Smit, and Kim]{kang2023multi}
Yeonghun Kang, Hyunsoo Park, Berend Smit, and Jihan Kim.
\newblock A multi-modal pre-training transformer for universal transfer learning in metal--organic frameworks.
\newblock \emph{Nature Machine Intelligence}, 5\penalty0 (3):\penalty0 309--318, 2023.

\bibitem[Wang et~al.(2024)Wang, Liu, Wang, Zhou, Ke, Zhang, Wu, Gao, and Lu]{wang2024comprehensive}
Jingqi Wang, Jiapeng Liu, Hongshuai Wang, Musen Zhou, Guolin Ke, Linfeng Zhang, Jianzhong Wu, Zhifeng Gao, and Diannan Lu.
\newblock A comprehensive transformer-based approach for high-accuracy gas adsorption predictions in metal-organic frameworks.
\newblock \emph{Nature Communications}, 15\penalty0 (1):\penalty0 1904, 2024.

\bibitem[J{\o}rgensen and Bhowmik(2020)]{jorgensen2020deepdft}
Peter~Bj{\o}rn J{\o}rgensen and Arghya Bhowmik.
\newblock Deepdft: Neural message passing network for accurate charge density prediction.
\newblock In \emph{Machine Learning for Molecules Workshop@ NeurIPS 2020}, 2020.

\bibitem[J{\o}rgensen and Bhowmik(2022)]{jorgensen2022equivariant}
Peter~Bj{\o}rn J{\o}rgensen and Arghya Bhowmik.
\newblock Equivariant graph neural networks for fast electron density estimation of molecules, liquids, and solids.
\newblock \emph{npj Computational Materials}, 8\penalty0 (1):\penalty0 183, 2022.

\bibitem[Liu et~al.(2021)Liu, Lin, Cao, Hu, Wei, Zhang, Lin, and Guo]{liu2021swin}
Ze~Liu, Yutong Lin, Yue Cao, Han Hu, Yixuan Wei, Zheng Zhang, Stephen Lin, and Baining Guo.
\newblock Swin transformer: Hierarchical vision transformer using shifted windows.
\newblock In \emph{Proceedings of the IEEE/CVF international conference on computer vision}, pages 10012--10022, 2021.

\bibitem[Rosen et~al.(2021)Rosen, Iyer, Ray, Yao, Aspuru-Guzik, Gagliardi, Notestein, and Snurr]{rosen2021machine}
Andrew~S Rosen, Shaelyn~M Iyer, Debmalya Ray, Zhenpeng Yao, Alan Aspuru-Guzik, Laura Gagliardi, Justin~M Notestein, and Randall~Q Snurr.
\newblock Machine learning the quantum-chemical properties of metal--organic frameworks for accelerated materials discovery.
\newblock \emph{Matter}, 4\penalty0 (5):\penalty0 1578--1597, 2021.

\bibitem[Gardner et~al.(2023)Gardner, Beaulieu, and Deringer]{gardner2023synthetic}
John~LA Gardner, Zo{\'e}~Faure Beaulieu, and Volker~L Deringer.
\newblock Synthetic data enable experiments in atomistic machine learning.
\newblock \emph{Digital Discovery}, 2\penalty0 (3):\penalty0 651--662, 2023.

\bibitem[Burner et~al.(2022)Burner, Luo, White, Mirmiran, Kwon, Boyd, Maley, Gibaldi, Simrod, and Woo]{burner_2023arcmof}
Jake Burner, Jun Luo, Andrew White, Adam Mirmiran, Ohmin Kwon, Peter~G. Boyd, Stephen Maley, Marco Gibaldi, Scott Simrod, and Tom~K. Woo.
\newblock ab initio repeat charge mof database (arc-mof), July 2022.
\newblock URL \url{https://doi.org/10.5281/zenodo.6908728}.

\bibitem[Burner(2025{\natexlab{a}})]{burner_2025_CH4_1bar}
Jake Burner.
\newblock Metal-organic framework ch4@1bar adsorbate probability distributions, August 2025{\natexlab{a}}.
\newblock URL \url{https://doi.org/10.5281/zenodo.16800893}.

\bibitem[Burner(2025{\natexlab{b}})]{burner_2025_CH4_65bar}
Jake Burner.
\newblock Metal-organic framework ch4@65bar adsorbate probability distributions, August 2025{\natexlab{b}}.
\newblock URL \url{https://doi.org/10.5281/zenodo.16801034}.

\bibitem[Burner(2025{\natexlab{c}})]{burner_2025_Xe_1bar}
Jake Burner.
\newblock Metal-organic framework xe@1bar adsorbate probability distributions, August 2025{\natexlab{c}}.
\newblock URL \url{https://doi.org/10.5281/zenodo.16801181}.

\bibitem[Thyagarajan and Sholl(2020)]{thyagarajan2020database}
Raghuram Thyagarajan and David~S Sholl.
\newblock A database of porous rigid amorphous materials.
\newblock \emph{Chemistry of Materials}, 32\penalty0 (18):\penalty0 8020--8033, 2020.

\bibitem[Martin and Siepmann(1998)]{martin1998transferable}
Marcus~G Martin and J~Ilja Siepmann.
\newblock Transferable potentials for phase equilibria. 1. united-atom description of n-alkanes.
\newblock \emph{The Journal of Physical Chemistry B}, 102\penalty0 (14):\penalty0 2569--2577, 1998.

\bibitem[Gross and Sadowski(2001)]{gross2001perturbed}
Joachim Gross and Gabriele Sadowski.
\newblock Perturbed-chain saft: An equation of state based on a perturbation theory for chain molecules.
\newblock \emph{Industrial \& engineering chemistry research}, 40\penalty0 (4):\penalty0 1244--1260, 2001.

\bibitem[Stierle et~al.(2024)Stierle, Bauer, Thiele, Bursik, Rehner, and Gross]{stierle2024classical}
Rolf Stierle, Gernot Bauer, Nadine Thiele, Benjamin Bursik, Philipp Rehner, and Joachim Gross.
\newblock Classical density functional theory in three dimensions with gpu-accelerated automatic differentiation: Computational performance analysis using the example of adsorption in covalent-organic frameworks.
\newblock \emph{Chemical Engineering Science}, 298:\penalty0 120380, 2024.

\bibitem[Rapp{\'e} et~al.(1992)Rapp{\'e}, Casewit, Colwell, Goddard~III, and Skiff]{rappe1992uff}
Anthony~K Rapp{\'e}, Carla~J Casewit, Kent~S Colwell, William~A Goddard~III, and W~Mason Skiff.
\newblock Uff, a full periodic table force field for molecular mechanics and molecular dynamics simulations.
\newblock \emph{Journal of the American chemical society}, 114\penalty0 (25):\penalty0 10024--10035, 1992.

\bibitem[Loshchilov and Hutter(2017)]{loshchilov2017decoupled}
Ilya Loshchilov and Frank Hutter.
\newblock Decoupled weight decay regularization.
\newblock \emph{arXiv preprint arXiv:1711.05101}, 2017.

\bibitem[Hu et~al.(2022)Hu, Shen, Wallis, Allen-Zhu, Li, Wang, Wang, Chen, et~al.]{hu2022lora}
Edward~J Hu, Yelong Shen, Phillip Wallis, Zeyuan Allen-Zhu, Yuanzhi Li, Shean Wang, Liang Wang, Weizhu Chen, et~al.
\newblock Lora: Low-rank adaptation of large language models.
\newblock \emph{Iclr}, 1\penalty0 (2):\penalty0 3, 2022.

\bibitem[Dosovitskiy et~al.(2020)Dosovitskiy, Beyer, Kolesnikov, Weissenborn, Zhai, Unterthiner, Dehghani, Minderer, Heigold, Gelly, et~al.]{dosovitskiy2020image}
Alexey Dosovitskiy, Lucas Beyer, Alexander Kolesnikov, Dirk Weissenborn, Xiaohua Zhai, Thomas Unterthiner, Mostafa Dehghani, Matthias Minderer, Georg Heigold, Sylvain Gelly, et~al.
\newblock An image is worth 16x16 words: Transformers for image recognition at scale.
\newblock \emph{arXiv preprint arXiv:2010.11929}, 2020.

\bibitem[Dao(2023)]{dao2023flashattention}
Tri Dao.
\newblock Flashattention-2: Faster attention with better parallelism and work partitioning.
\newblock \emph{arXiv preprint arXiv:2307.08691}, 2023.

\bibitem[Shazeer(2020)]{shazeer2020glu}
Noam Shazeer.
\newblock Glu variants improve transformer.
\newblock \emph{arXiv preprint arXiv:2002.05202}, 2020.

\bibitem[Peebles and Xie(2023)]{peebles2023scalable}
William Peebles and Saining Xie.
\newblock Scalable diffusion models with transformers.
\newblock In \emph{Proceedings of the IEEE/CVF international conference on computer vision}, pages 4195--4205, 2023.

\bibitem[Esper et~al.(2023)Esper, Bauer, Rehner, and Gross]{esper2023pcp}
Timm Esper, Gernot Bauer, Philipp Rehner, and Joachim Gross.
\newblock Pcp-saft parameters of pure substances using large experimental databases.
\newblock \emph{Industrial \& Engineering Chemistry Research}, 62\penalty0 (37):\penalty0 15300--15310, 2023.

\bibitem[Paszke et~al.(2019)Paszke, Gross, Massa, Lerer, Bradbury, Chanan, Killeen, Lin, Gimelshein, Antiga, et~al.]{paszke2019pytorch}
Adam Paszke, Sam Gross, Francisco Massa, Adam Lerer, James Bradbury, Gregory Chanan, Trevor Killeen, Zeming Lin, Natalia Gimelshein, Luca Antiga, et~al.
\newblock Pytorch: An imperative style, high-performance deep learning library.
\newblock \emph{Advances in neural information processing systems}, 32, 2019.

\bibitem[Rehner et~al.(2023)Rehner, Bauer, and Gross]{rehner2023feos}
Philipp Rehner, Gernot Bauer, and Joachim Gross.
\newblock Feos: an open-source framework for equations of state and classical density functional theory.
\newblock \emph{Industrial \& Engineering Chemistry Research}, 62\penalty0 (12):\penalty0 5347--5357, 2023.

\bibitem[Hjorth~Larsen et~al.(2017)Hjorth~Larsen, J{\o}rgen~Mortensen, Blomqvist, Castelli, Christensen, Du{\l}ak, Friis, Groves, Hammer, Hargus, et~al.]{hjorth2017atomic}
Ask Hjorth~Larsen, Jens J{\o}rgen~Mortensen, Jakob Blomqvist, Ivano~E Castelli, Rune Christensen, Marcin Du{\l}ak, Jesper Friis, Michael~N Groves, Bj{\o}rk Hammer, Cory Hargus, et~al.
\newblock The atomic simulation environment—a python library for working with atoms.
\newblock \emph{Journal of Physics: Condensed Matter}, 29\penalty0 (27):\penalty0 273002, 2017.

\end{thebibliography}
\clearpage


\newpage
\appendix

\section{Additional Background on Adsorption Simulation}
\label{app:cdft_gcmc}

This appendix provides the theoretical and numerical background behind the density-field labels used in MADField.
The main text introduces GCMC and cDFT as particle-level and density-level routes to the same equilibrium adsorption state.
Here we give the corresponding ensemble definitions, density-functional equations, Boltzmann reference density, Picard iteration, PC-SAFT functional structure, and external-potential construction.

\subsection{Grand Canonical Monte Carlo and Published GCMC References}
\label{app:gcmc}

Grand Canonical Monte Carlo samples adsorption equilibrium in the grand-canonical ensemble at fixed temperature \(T\), chemical potential \(\mu\), and framework structure~\citep{dubbeldam2016raspa,ran2024raspa3}.
For an adsorbate configuration with \(N\) molecule centers \(\mathbf r_{1:N}\), the force-field energy can be written as
\begin{equation}
    U_N(\mathbf r_{1:N})
    =
    \sum_{i=1}^{N}
    V_{\mathrm{ext}}^{\mathrm{GCMC}}(\mathbf r_i)
    +
    \sum_{i<j}
    u_{ss}(\mathbf r_i,\mathbf r_j).
    \label{eq:app_gcmc_energy}
\end{equation}
Here \(V_{\mathrm{ext}}^{\mathrm{GCMC}}\) is the framework-fluid interaction under the GCMC force field.
The term \(u_{ss}\) is the adsorbate-adsorbate interaction.
A GCMC trajectory samples configurations with different particle numbers using moves such as insertion, deletion, translation, and rotation.
The Markov chain is constructed so that the stationary distribution is proportional to the grand-canonical Boltzmann weight.

The equilibrium one-body density is the ensemble average of the microscopic particle density,
\begin{equation}
    \rho_{\mathrm{GCMC}}(\mathbf r)
    =
    \left\langle
    \sum_{i=1}^{N}
    \delta(\mathbf r-\mathbf r_i)
    \right\rangle_{T,\mu}.
    \label{eq:app_gcmc_density_continuous}
\end{equation}
In practice, this density is estimated by binning molecule centers onto a voxel grid and averaging over production samples,
\begin{equation}
    \rho_{\mathrm{GCMC}}(v)
    =
    \frac{1}{V_v}
    \left\langle
    n_v
    \right\rangle_{T,\mu},
    \label{eq:app_gcmc_density_voxel}
\end{equation}
where \(v\) is a voxel, \(V_v\) is the voxel volume, and \(n_v\) is the number of adsorbate centers falling in that voxel.

MADField-GCMC uses published GCMC references rather than newly generated GCMC trajectories.
We use ARC-MOF GCMC density fields and uptake values from Burner et al.~\citep{burner_2025_CH4_1bar,burner_2025_CH4_65bar,burner_2025_Xe_1bar,burner_2023arcmof} for in-domain GCMC-fidelity training and evaluation.
We use PRAM methane uptake values for zero-shot evaluation on disordered porous materials~\citep{thyagarajan2020database}.
The force-field choices, simulation settings, and convergence protocols therefore follow the original ARC-MOF density dataset and PRAM dataset.

The MADField-GCMC molecular-conditioning slot is populated with the TraPPE-style Lennard-Jones registry\citep{martin1998transferable} associated with the published GCMC references~\citep{burner_2025_CH4_1bar,burner_2025_CH4_65bar,burner_2025_Xe_1bar,burner_2023arcmof}.
This registry is fidelity-specific and differs from the PC-SAFT registry used for MADField-cDFT.
The detailed force-field parameters and simulation settings should be read from the original ARC-MOF density release and the PRAM database.

\subsection{Density-Grid Convergence and Tanimoto Similarity}
\label{app:gcmc_density_convergence}

For density-field supervision, convergence of scalar uptake alone is not sufficient.
A GCMC density field must also have a stable spatial distribution.
The adsorption-probability-distribution literature uses a Tanimoto-type spatial similarity to compare two accumulated density grids~\citep{burner2026rapid}.
For two non-negative density tensors \(A\) and \(B\) defined on the same grid, the continuous Tanimoto similarity is
\begin{equation}
    T(A,B)
    =
    \frac{
        \sum_i A_i B_i
    }{
        \sum_i A_i^2
        +
        \sum_i B_i^2
        -
        \sum_i A_i B_i
        +
        \epsilon
    }.
    \label{eq:app_tanimoto}
\end{equation}
The small \(\epsilon\) prevents numerical division by zero.
A value of \(T=1\) indicates identical density tensors up to the grid representation.
In the published GCMC density references, this type of overlap criterion is used to ensure that the spatial adsorption distribution has stabilized, not only the integrated loading.

\subsection{Classical Density Functional Theory}
\label{app:cdft}

Classical density functional theory describes adsorption equilibrium through the one-body number density field \(\rho(\mathbf r)\)~\citep{evans1979nature}.
For a rigid framework, adsorbate species \(s\), temperature \(T\), and chemical potential \(\mu\), the grand-potential functional is
\begin{equation}
    \Omega[\rho]
    =
    F_{\mathrm{id}}[\rho]
    +
    F_{\mathrm{exc}}[\rho;s]
    +
    \int_{\mathcal V}
    \rho(\mathbf r)
    \left[
        V_{\mathrm{ext}}(\mathbf r;s)
        -
        \mu
    \right]
    d\mathbf r .
    \label{eq:app_grand_potential}
\end{equation}
Here \(F_{\mathrm{id}}\) is the ideal-gas free-energy functional.
\(F_{\mathrm{exc}}\) is an approximate excess free-energy functional that accounts for intermolecular correlations.
\(V_{\mathrm{ext}}\) is the framework-fluid external potential.
The equilibrium density is a stationary point of \(\Omega[\rho]\),
\begin{equation}
    \left.
    \frac{\delta \Omega[\rho]}{\delta \rho(\mathbf r)}
    \right|_{\rho=\rho_{\mathrm{eq}}}
    =
    0 .
    \label{eq:app_stationarity}
\end{equation}

The ideal-gas functional is
\begin{equation}
    F_{\mathrm{id}}[\rho]
    =
    k_B T
    \int_{\mathcal V}
    \rho(\mathbf r)
    \left[
        \log\left(\rho(\mathbf r)\Lambda^3\right)-1
    \right]
    d\mathbf r ,
    \label{eq:app_fid}
\end{equation}
where \(\Lambda\) is the thermal de Broglie wavelength.
Its functional derivative is
\begin{equation}
    \frac{\delta F_{\mathrm{id}}}{\delta \rho(\mathbf r)}
    =
    k_B T
    \log\left(\rho(\mathbf r)\Lambda^3\right).
    \label{eq:app_fid_derivative}
\end{equation}
Combining \cref{eq:app_stationarity} and \cref{eq:app_fid_derivative} gives the Euler-Lagrange equation
\begin{equation}
    k_B T
    \log\left(\rho_{\mathrm{eq}}(\mathbf r)\Lambda^3\right)
    +
    \mu_{\mathrm{exc}}(\mathbf r;[\rho_{\mathrm{eq}}])
    +
    V_{\mathrm{ext}}(\mathbf r)
    -
    \mu
    =
    0 ,
    \label{eq:app_euler_lagrange}
\end{equation}
where
\begin{equation}
    \mu_{\mathrm{exc}}(\mathbf r;[\rho])
    =
    \frac{\delta F_{\mathrm{exc}}[\rho]}{\delta \rho(\mathbf r)}
    \label{eq:app_mu_exc}
\end{equation}
is the local excess chemical-potential field.

The chemical potential is fixed by the bulk fluid at the same \(T\) and pressure \(P\).
Writing the bulk chemical potential as
\begin{equation}
    \mu
    =
    k_B T
    \log\left(\rho_{\mathrm{bulk}}\Lambda^3\right)
    +
    \mu_{\mathrm{exc}}^{\mathrm{bulk}},
    \label{eq:app_bulk_mu}
\end{equation}
the Euler-Lagrange equation can be rearranged into the fixed-point form
\begin{equation}
    \rho_{\mathrm{eq}}(\mathbf r)
    =
    \rho_{\mathrm{bulk}}
    \exp\left[
        -\beta V_{\mathrm{ext}}(\mathbf r)
        -
        \beta \mu_{\mathrm{exc}}(\mathbf r;[\rho_{\mathrm{eq}}])
        +
        \beta \mu_{\mathrm{exc}}^{\mathrm{bulk}}
    \right],
    \label{eq:app_cdft_fixed_point}
\end{equation}
where \(\beta=(k_B T)^{-1}\).

\subsection{Boltzmann Reference Density}
\label{app:boltzmann_density}

The Boltzmann reference density is the non-interacting solution obtained by dropping the excess term in \cref{eq:app_cdft_fixed_point}.
Equivalently, it is the equilibrium density of an ideal gas in the same external potential.
This gives
\begin{equation}
    \rho_{\mathrm{Boltz}}(\mathbf r)
    =
    \rho_{\mathrm{bulk}}
    \exp\left[
        -\beta V_{\mathrm{ext}}(\mathbf r)
    \right].
    \label{eq:app_boltzmann}
\end{equation}
This density captures the dominant one-body response to the framework.
It is high in attractive wells, low in repulsive wall regions, and independent of fluid-fluid correlations.
MADField uses this field as a physical reference and predicts a log-density residual relative to it.
This makes the neural target closer to the many-body correction than to the full adsorption field.

\subsection{Picard Iteration and Damped Mixing}
\label{app:picard}

The fixed-point equation in \cref{eq:app_cdft_fixed_point} defines a nonlinear map from a current density to an updated density.
For PC-SAFT parameters \(\boldsymbol{\xi}\), define
\begin{equation}
    \mathcal T_{\boldsymbol{\xi}}[\rho](\mathbf r)
    =
    \rho_{\mathrm{bulk}}
    \exp\left[
        -\beta V_{\mathrm{ext}}(\mathbf r)
        -
        \beta \mu_{\mathrm{exc}}(\mathbf r;[\rho])
        +
        \beta \mu_{\mathrm{exc}}^{\mathrm{bulk}}
    \right].
    \label{eq:app_picard_map}
\end{equation}
A Picard solver repeatedly evaluates this map,
\begin{equation}
    \tilde{\rho}^{(n+1)}(\mathbf r)
    =
    \mathcal T_{\boldsymbol{\xi}}[\rho^{(n)}](\mathbf r).
    \label{eq:app_raw_picard}
\end{equation}
Because the raw update can be unstable in strongly attractive or high-loading regimes, we use damped mixing,
\begin{equation}
    \rho^{(n+1)}(\mathbf r)
    =
    (1-\alpha)
    \rho^{(n)}(\mathbf r)
    +
    \alpha
    \tilde{\rho}^{(n+1)}(\mathbf r),
    \label{eq:app_damped_picard}
\end{equation}
where \(\alpha\in(0,1]\) is the damping factor.
The converged solution satisfies
\begin{equation}
    \rho_{\mathrm{eq}}
    =
    \mathcal T_{\boldsymbol{\xi}}[\rho_{\mathrm{eq}}].
    \label{eq:app_picard_fixed_point}
\end{equation}
The conventional cold-start initialization is
\begin{equation}
    \rho^{(0)}_{\mathrm{cold}}(\mathbf r)
    =
    \rho_{\mathrm{Boltz}}(\mathbf r).
    \label{eq:app_cold_start}
\end{equation}
MADField warm-starting replaces this initial density by a learned approximation to the interacting equilibrium density.
The detailed warm-start schedule, clipping, and fallback protocol are given in \cref{app:warmstart_protocol}.

\subsection{PC-SAFT Excess Functional}
\label{app:pcsaft}

MADField-cDFT labels are generated with an inhomogeneous PC-SAFT cDFT functional~\citep{gross2001perturbed,sauer2017classical}.
The PC-SAFT excess free energy is decomposed as
\begin{equation}
    F_{\mathrm{exc}}[\rho]
    =
    F_{\mathrm{hs}}[\rho]
    +
    F_{\mathrm{chain}}[\rho]
    +
    F_{\mathrm{disp}}[\rho]
    +
    F_{\mathrm{assoc}}[\rho]
    +
    F_{\mathrm{polar}}[\rho].
    \label{eq:app_pcsaft_decomp}
\end{equation}
The hard-sphere term \(F_{\mathrm{hs}}\) represents excluded-volume packing and is evaluated with fundamental measure theory~\citep{roth2010fundamental}.
The chain term \(F_{\mathrm{chain}}\) accounts for connectivity of PC-SAFT segments.
The dispersion term \(F_{\mathrm{disp}}\) accounts for attractive fluid-fluid interactions.
The association term \(F_{\mathrm{assoc}}\) is not used because the reported \method adsorbates are treated as non-associating fluids.
The polar term \(F_{\mathrm{polar}}\) is not used in the reported labels because we use an effective non-polar parameterization.

Each \method{} adsorbate is represented by the PC-SAFT triplet
\begin{equation}
    \boldsymbol{\xi}_{\mathrm{cDFT}}
    =
    (m,\sigma,\epsilon).
    \label{eq:app_pcsaft_triplet}
\end{equation}
The segment number \(m\), segment diameter \(\sigma\), and dispersion energy \(\epsilon\) determine the bulk equation of state and the inhomogeneous excess functional.
The same triplet is also used in the cDFT external-potential construction.
The numerical parameter values are listed in \cref{tab:app_pcsaft_params}.

The FMT weighted densities and PC-SAFT convolutional quantities are evaluated on the three-dimensional periodic grid.
These convolutional evaluations are the main reason FFT-based implementations are natural for three-dimensional cDFT solvers~\citep{stierle2024classical,dufour2025classical}.
This grid-based functional evaluation makes cDFT much faster than particle sampling for many adsorption state points, but the nonlinear fixed-point problem still requires iterative convergence.

\subsection{cDFT External Potential}
\label{app:vext_cdft}

The cDFT external potential is the one-body interaction energy between an adsorbate molecule and the rigid framework.
In the reported cDFT labels, it is computed from Lennard-Jones interactions between adsorbate segments and framework atoms.
For a grid point \(\mathbf r\), we use
\begin{equation}
    V_{\mathrm{ext}}^{\mathrm{cDFT}}(\mathbf r)
    =
    m
    \sum_{j\in\mathcal N(\mathbf r)}
    4\epsilon_{fj}
    \left[
        \left(
            \frac{\sigma_{fj}}{r_j(\mathbf r)}
        \right)^{12}
        -
        \left(
            \frac{\sigma_{fj}}{r_j(\mathbf r)}
        \right)^6
    \right].
    \label{eq:app_vext_cdft}
\end{equation}
Here \(j\) indexes framework atoms within the cutoff neighborhood \(\mathcal N(\mathbf r)\).
The distance \(r_j(\mathbf r)\) is computed under periodic boundary conditions.
For non-orthorhombic cells, we use fractional-coordinate wrapping and expand each structure consistently with the cutoff radius \(r_c=12.8\) \AA.
The fluid-framework mixed parameters are obtained with Lorentz-Berthelot mixing,
\begin{equation}
    \sigma_{fj}
    =
    \frac{\sigma_f+\sigma_j}{2},
    \qquad
    \epsilon_{fj}
    =
    \sqrt{\epsilon_f\epsilon_j}.
    \label{eq:app_lorentz_berthelot}
\end{equation}
The adsorbate parameters \((m,\sigma_f,\epsilon_f)\) come from the PC-SAFT registry.
The framework parameters \((\sigma_j,\epsilon_j)\) come from UFF~\citep{rappe1992uff}, with selected values listed in \cref{tab:app_params_uff}.
The prefactor \(m\) scales the one-body framework interaction by the PC-SAFT segment count.
For numerical stability, the reduced potential is clipped before being passed to the neural model and before defining pore masks.

\subsection{Relation Between Theory and MADField Inputs}
\label{app:theory_to_inputs}

The theory above defines the physical inputs used by MADField.
The framework geometry and adsorbate parameters determine \(V_{\mathrm{ext}}(\mathbf r)\).
The pressure and temperature determine \(\rho_{\mathrm{bulk}}\) through the bulk equation of state.
Together these define the Boltzmann reference density in \cref{eq:app_boltzmann}.
MADField receives the reduced external potential, the log-Boltzmann density, unit-cell parameters, thermodynamic state, and molecular-conditioning vector.
The model then predicts the interacting equilibrium density field by learning the residual correction that cDFT would otherwise obtain through Picard iteration or that GCMC would estimate through particle sampling.

\section{Training}
\label{app:training}

This appendix gives the training details for the released cDFT and GCMC models.
The cDFT model is trained with supervised density labels only.
The GCMC model is obtained by LoRA adaptation of the released cDFT model.
Exploratory Euler--Lagrange pre-training runs were not used for the reported checkpoint and are therefore omitted.

\subsection{Supervised Density Loss}
\label{app:loss_full}

Both fidelity stages use the same supervised density objective.
The target density \(\rho^*\) is the solver-converged PC-SAFT cDFT density for \method{} cDFT training and the GCMC density for \methodgcmc{}.
The pore mask is
\begin{equation}
    \mathcal P
    =
    \{
    \mathbf r
    \mid
    \beta V_{\mathrm{ext}}(\mathbf r)<50
    \}.
\end{equation}
The complement \(\bar{\mathcal P}\) is treated as the wall region.
The predicted and target log-residuals are
\begin{equation}
    \Delta_\theta(\mathbf r)
    =
    \log \hat\rho_\theta(\mathbf r)
    -
    \log\rho_{\mathrm{Boltz}}(\mathbf r),
    \qquad
    \Delta^*(\mathbf r)
    =
    \log \rho^*(\mathbf r)
    -
    \log\rho_{\mathrm{Boltz}}(\mathbf r).
\end{equation}
For numerical stability, log-domain targets are computed only inside the pore mask and are explicitly bounded.
Framework voxels are excluded from log-residual losses by the pore mask \(\mathcal P=\{\mathbf r\mid \beta V_{\mathrm{ext}}(\mathbf r)<50\}\).
Within \(\mathcal P\), target densities are floored before taking logarithms,
\[
    \log\rho^*_{\epsilon}(\mathbf r)
    =
    \log\bigl(\max(\rho^*(\mathbf r),\epsilon_{\log})\bigr),
\]
with \(\epsilon_{\log}=10^{-12}\) for the residual loss and \(10^{-30}\) for the hierarchical low-pass target.
The resulting log-density values are hard-clamped to \([-30,5]\).
This makes the log-residual target finite for every voxel, including zero-count GCMC bins, without modifying the original density histogram used by the density and uptake losses.
The log-residual target makes the residual head learn the correction on top of the Boltzmann baseline.
The prediction itself still depends explicitly on \(\rho_{\mathrm{Boltz}}\) through the output parameterization.

The full loss is
\begin{align}
\mathcal L
=&
\eta_{\mathrm{res}}
\mathcal L_{\mathrm{res}}
+
\eta_{\rho}
\mathcal L_{\rho}
+
\eta_{\rho,\mathrm{top}}
\mathcal L_{\rho,\mathrm{top}}
+
\eta_N
\mathcal L_N
+
\eta_{\mathrm{hier}}
\mathcal L_{\mathrm{hier}} .
\label{eq:app_training_loss}
\end{align}
The residual loss is
\begin{equation}
    \mathcal L_{\mathrm{res}}
    =
    \frac{1}{|\mathcal P|}
    \sum_{\mathbf r\in\mathcal P}
    \left(
        \Delta_\theta(\mathbf r)-\Delta^*(\mathbf r)
    \right)^2 .
\end{equation}
The pore density loss is the relative \(L_1\) error inside the pore region,
\begin{equation}
    \mathcal L_{\rho}
    =
    \frac{
        \sum_{\mathbf r\in\mathcal P}
        \left|
            \hat\rho_\theta(\mathbf r)-\rho^*(\mathbf r)
        \right|
    }{
        \sum_{\mathbf r\in\mathcal P}\rho^*(\mathbf r)+\epsilon_\rho
    } .
\end{equation}
The top-density loss applies the same relative \(L_1\) error to the highest-density pore voxels,
\begin{equation}
    \mathcal L_{\rho,\mathrm{top}}
    =
    \frac{
        \sum_{\mathbf r\in\mathcal P_{\mathrm{top}}}
        \left|
            \hat\rho_\theta(\mathbf r)-\rho^*(\mathbf r)
        \right|
    }{
        \sum_{\mathbf r\in\mathcal P_{\mathrm{top}}}\rho^*(\mathbf r)+\epsilon_\rho
    } .
\end{equation}
Here \(\mathcal P_{\mathrm{top}}\) is the top \(10\%\) of pore voxels by target density.
The loading loss is the absolute relative error in integrated uptake,
\begin{equation}
    \mathcal L_N
    =
    \left|
    \frac{
        N[\hat\rho_\theta]-N[\rho^*]
    }{
        N[\rho^*]+\epsilon_N
    }
    \right| .
\end{equation}
The hierarchical loss supervises the coarse output branch against a low-pass residual target,
\begin{equation}
    \mathcal L_{\mathrm{hier}}
    =
    \frac{1}{|\mathcal P|}
    \sum_{\mathbf r\in\mathcal P}
    \left(
        c_\theta(\mathbf r)-\mathrm{LP}(\Delta^*)(\mathbf r)
    \right)^2 .
\end{equation}
The operator \(\mathrm{LP}\) is 3D average pooling with kernel size \(4\), followed by trilinear upsampling to the \(128^3\) grid.
We use
\begin{equation}
    (
    \eta_{\mathrm{res}},
    \eta_\rho,
    \eta_{\rho,\mathrm{top}},
    \eta_N,
    \eta_{\mathrm{hier}}
    )
    =
    (
    0.5,
    2.0,
    1.5,
    0.25,
    0.05
    ).
\end{equation}
The sum-rule loss used in earlier internal experiments is disabled in the released model.

\subsection{\method{} cDFT Training}
\label{app:stage1_training}

The released cDFT model is trained in two supervised phases.
The first phase trains from scratch for \(50{,}000\) optimizer steps.
The peak learning rate is \(6\times10^{-4}\).
The warmup length is \(2{,}000\) steps.
The second phase resumes from the best checkpoint of the first phase and runs for \(30{,}000\) additional steps.
The peak learning rate is \(2\times10^{-4}\).
The warmup length is \(1{,}000\) steps.
The released checkpoint is selected at step \(26{,}000\) of the second phase.

Both phases use AdamW~\citep{loshchilov2017decoupled}.
Weight decay is \(0.01\), excluding bias, normalization, and positional-encoding parameters.
The Adam coefficients are \(\beta_1=0.9\) and \(\beta_2=0.999\).
The schedule is linear warmup followed by cosine decay.
Training uses \(8\) H200 GPUs with batch size \(8\) per GPU.
The effective batch size is \(64\).
Gradient accumulation is \(1\).
Gradient clipping uses max norm \(1.0\).
Mixed precision uses bf16 autocast.

\subsection{\methodgcmc{} LoRA Adaptation}
\label{app:stage2_training}

MADField-GCMC adapts the released cDFT model to GCMC density fidelity with LoRA~\citep{hu2022lora}.
The base cDFT weights are frozen.
LoRA adapters are inserted into \texttt{attn.qkv}, \texttt{attn.proj}, \texttt{ffn.w1}, \texttt{ffn.w2}, and \texttt{ffn.w3} in each Swin block.
The LoRA rank is \(r=4\).
The LoRA scale is \(\alpha_{\mathrm{LoRA}}=2r=8\).
The scaling mode is standard \(\alpha_{\mathrm{LoRA}}/r\).
The adapter dropout is \(0\).

The released \methodgcmc{} run trains on paired GCMC density data for \ce{CH4} and Xe.
The training set contains \(50{,}621\) paired state points.
This consists of \(32{,}852\) \ce{CH4} state points and \(17{,}769\) Xe state points.
The validation set contains \(5{,}796\) paired state points.
This consists of \(3{,}864\) \ce{CH4} state points and \(1{,}932\) Xe state points.

\methodgcmc{} uses the same density loss as \method{} cDFT training, with \(\rho^*\) replaced by the GCMC density. 
The molecular-conditioning vector switches from the PC-SAFT registry to the TraPPE-style registry used by the GCMC reference data.
For \ce{CH4}, the \methodgcmc{} conditioning triplet is \((m,\sigma,\epsilon/k_B)=(1.0,3.73,148.0)\).
For Xe, the \methodgcmc{} conditioning triplet is \((1.0,4.10,221.0)\).
The bulk density \(\rho_{\mathrm{bulk}}(P,T)\) is still computed from the PC-SAFT equation of state.
This registry switch is part of the released \methodgcmc{} checkpoint.

The \methodgcmc{} run uses AdamW with peak learning rate \(5\times10^{-4}\).
The warmup length is \(2{,}500\) steps.
The total training length is \(50{,}000\) steps.
The best checkpoint is selected at step \(45{,}000\).
Training uses \(8\) H200 GPUs with batch size \(8\) per GPU.
The effective batch size is \(64\).
The released LoRA checkpoint adds \(89{,}856\) adapter parameters.
The total number of trainable parameters in the \methodgcmc{} run is \(1{,}572{,}898\) out of \(6{,}443{,}410\), because output-side and conditioning modulation parameters are also trained.

\section{cDFT Warm-Start Protocol}
\label{app:warmstart_protocol}

This appendix describes how MADField predictions are converted into initial conditions for the iterative cDFT solver.
The cold-start baseline initializes the solver with the Boltzmann density,
\begin{equation}
    \rho_{\mathrm{cold}}^{(0)}(\mathbf r)
    =
    \rho_{\mathrm{bulk}}
    \exp[-\beta V_{\mathrm{ext}}(\mathbf r)].
\end{equation}
The warm-start run first computes the same \(128^3\) external-potential grid and the same PC-SAFT bulk density as the cold-start run.
It then evaluates the \method{} checkpoint to obtain a predicted density field \(\hat\rho_\theta(\mathbf r)\).
Before passing this prediction to the solver, we rescale it by the pore-region mean density,
\begin{equation}
    s
    =
    \mathrm{clip}
    \left(
        \frac{\rho_{\mathrm{bulk}}}
        {
        \frac{1}{|\mathcal P|}
        \sum_{\mathbf r\in\mathcal P}
        \hat\rho_\theta(\mathbf r)
        },
        0.3,
        5.0
    \right).
\end{equation}
The warm-start initial condition is then
\begin{equation}
    \rho_{\mathrm{warm}}^{(0)}(\mathbf r)
    =
    \mathrm{clip}
    \left(
        s\hat\rho_\theta(\mathbf r),
        10^{-30},
        0.01
    \right).
\end{equation}
The scaling prevents large global mismatch between the predicted field and the bulk density from destabilizing the first Picard steps.
The upper density clip is used only for numerical initialization stability.

Both cold and warm runs solve the same PC-SAFT cDFT fixed-point equation.
They differ only in the initial density and the Picard damping schedule.
The cold-start schedule is conservative,
\begin{equation}
    (0.005,500),
    \quad
    (0.01,2000),
    \quad
    (0.05,5000),
\end{equation}
where each pair denotes damping factor and maximum iteration count.
The warm-start schedule uses larger damping factors and a lower iteration cap,
\begin{equation}
    (0.01,200),
    \quad
    (0.05,1000),
    \quad
    (0.10,3000).
\end{equation}
The cold-start cap is \(7500\) iterations.
The warm-start cap is \(4200\) iterations.
Both use the same convergence tolerance of \(10^{-5}\).

If the warm-start run diverges or collapses to an unphysical low-density solution, the solver falls back to the cold-start schedule.
This fallback is used only to make the warm-start protocol robust.
The iteration-count speedups reported in the main text are computed on state points that converge under both initializations.
Thus the reported speedup measures the reduction in Picard iterations when the same cDFT functional is solved from a better initial density field.

For multi-pressure isotherms, pressure continuation is separate from the MADField initialization.
When a lower-pressure cDFT solution is already available, it can initialize the next pressure step through a pressure ladder.
The MADField warm-start analysis in the main text isolates the effect of replacing the Boltzmann initialization by the learned density field.

\section{Architecture Details}
\label{app:architecture}

This appendix provides implementation details for the MADField volumetric density operator.
The main text describes the architecture at the level needed to understand the method.
Here we report the input construction, tokenization, periodic attention, conditioning, decoder, output head, and parameter counts used in the reported model.

\subsection{Input Channels}
\label{app:input_channels}

MADField takes a \(128^3\) periodic voxel grid with two volumetric input channels.
The first channel is the reduced external potential,
\begin{equation}
    \tilde V(\mathbf r)
    =
    \frac{
    \mathrm{clamp}\!\left(\beta V_{\mathrm{ext}}(\mathbf r),-15,60\right)
    }{15}.
\end{equation}
The second channel is the clipped log-Boltzmann density,
\begin{equation}
    \tilde \rho_B(\mathbf r)
    =
    \frac{
    \mathrm{clamp}\!\left(\log\rho_{\mathrm{bulk}}-\beta V_{\mathrm{ext}}(\mathbf r),-30,5\right)
    }{15}.
\end{equation}
The upper clamp on \(\beta V_{\mathrm{ext}}\) makes strongly repulsive wall voxels saturate to the same value, which gives the network an implicit wall signal without adding a separate binary mask.
The pore mask threshold \(\beta V_{\mathrm{ext}}<50\) is used only for loss and evaluation and is described in \cref{app:training}.
Although the two channels are analytically related before clipping, they play different numerical roles after normalization: \(\beta V_{\mathrm{ext}}\) provides a pressure-independent wall and well geometry, while \(\log\rho_B\) provides the pressure-dependent ideal-gas density scale used by the residual reconstruction.

\subsection{Fixed-Grid Representation}
\label{app:fixed_grid}

All reported MADField models represent each periodic simulation cell on a \(128^3\) voxel grid.
This gives a common tensor interface for batching heterogeneous unit cells and matches the grid-based cDFT and processed GCMC density labels used in the paper.
For a cell with maximum side length \(L_{\max}\), the largest voxel spacing is approximately \(L_{\max}/128\).
In our datasets, even large cells with \(L_{\max}\lesssim 60\,\text{\AA}\) therefore have spacing below \(0.47\,\text{\AA}\), which is comparable to or finer than the \(0.5\,\text{\AA}\)-scale resolution commonly used for adsorption energy or density grids.
For the QMOF structures used in the main cDFT benchmark, cell dimensions are typically below this upper bound, giving finer effective spacings than the worst-case \(60\,\text{\AA}\) cell.
The lattice lengths and angles are also provided in the non-volumetric conditioning vector, allowing the model to condition on the physical scale and shape represented by the fixed voxel grid.

\subsection{ConvStem}
\label{app:convstem}

A two-stage 3D convolutional stem tokenizes the \(128^3\) grid into a \(32^3\) token grid.
Both convolutions use circular padding to respect periodic boundary conditions.
The stem is
\begin{align}
    &\mathrm{Conv3d}(2\to48,\ k=3,\ s=2)
    \rightarrow
    \mathrm{GroupNorm}(1,48)
    \rightarrow
    \mathrm{GELU},
    \nonumber\\
    &\mathrm{Conv3d}(48\to288,\ k=3,\ s=2)
    \rightarrow
    \mathrm{GroupNorm}(1,288)
    \rightarrow
    \mathrm{GELU}.
\end{align}
The resulting token grid has \(32^3=32{,}768\) tokens.
Each token corresponds to a \(4^3\) voxel patch in the original density grid, following the patch-tokenization paradigm used in vision transformers~\citep{dosovitskiy2020image}.

\subsection{Periodic Fourier Positional Encoding}
\label{app:pe}

We use a periodic Fourier positional encoding on the \(32^3\) token grid.
The encoding satisfies \(\mathrm{PE}(\mathbf x)=\mathrm{PE}(\mathbf x+\mathbf L)\) by construction.
For a token index \(p\) along one axis with \(n=32\) tokens, integer harmonics are encoded as
\begin{equation}
    \mathrm{PE}_{\mathrm{axis}}[p,2k]
    =
    \sin\!\left(\frac{2\pi h_k p}{n}\right)\mathrm{decay}_k,
    \qquad
    \mathrm{PE}_{\mathrm{axis}}[p,2k+1]
    =
    \cos\!\left(\frac{2\pi h_k p}{n}\right)\mathrm{decay}_k.
\end{equation}
Integer frequencies guarantee exact periodicity across opposite faces of the unit cell.
The three axis-wise encodings are concatenated to match \(d_{\mathrm{model}}=288\).

\subsection{Conditioning Vector}
\label{app:conditioning}

The non-volumetric conditioning vector $\mathbf{c} \in \mathbb{R}^{11}$ concatenates the bulk thermodynamic state, the unit-cell axes, the unit-cell angle deviations from $90^\circ$, and the adsorbate triplet,
\begin{equation*}
    \mathbf{c}
    = \mathrm{standardize}\bigl(
    \rho'_{\mathrm{bulk}},\, P',\;
    a',\, b',\, c',\;
    \alpha',\, \beta',\, \gamma',\;
    m',\, \sigma',\, \varepsilon'
    \bigr),
\end{equation*}
where $x' = \log x$ for the eight scale variables $\{\rho_{\mathrm{bulk}}, P, a, b, c, m, \sigma, \varepsilon\}$ and $\theta' = \theta - 90^\circ$ for the three cell angles $\{\alpha, \beta, \gamma\}$.
Standardization statistics are computed once on the training split and stored as fixed buffers.
The angle-deviation parameterization places an orthorhombic cell at the origin of the angular axes. For cDFT training, $(m, \sigma, \varepsilon)$ is the PC-SAFT parameter triplet. For GCMC fine-tuning, the same three slots are populated by the TraPPE-style registry of \cref{app:cdft_gcmc}.

\subsection{Periodic Swin Blocks}
\label{app:swin_attn}

The backbone contains \(L=6\) Swin-style blocks with \(d_{\mathrm{model}}=288\)~\citep{liu2021swin}.
Each attention layer uses 8 heads, giving head dimension 36.
Window attention uses windows of size \(8^3\) on the \(32^3\) token grid.
This gives \(4^3=64\) windows with 512 tokens per window.
Attention is evaluated by \texttt{scaled\_dot\_product\_attention}, using the FlashAttention-2 backend on H100 and H200 GPUs when available~\citep{dao2023flashattention}.
Odd-indexed blocks shift the token grid by \(w/2=4\) tokens along all three axes using \texttt{torch.roll}.
Because the grid is periodic, rolled tokens are physically adjacent across unit-cell boundaries.
No attention mask is used.

Each block uses a SwiGLU feed-forward network~\citep{shazeer2020glu}.
The nominal FFN width is \(d_{\mathrm{ff}}=2d_{\mathrm{model}}=576\).
The SwiGLU inner width is
\begin{equation}
    d_{\mathrm{inner}}
    =
    \mathrm{round\_up\_to\_8}
    \left(
        \frac{2d_{\mathrm{ff}}}{3}
    \right)
    =
    384.
\end{equation}
This keeps the SwiGLU parameter count comparable to a standard FFN with expansion ratio 2.

\subsection{AdaLN-Zero Conditioning}
\label{app:adaln}

Each Swin block is modulated by AdaLN-Zero~\citep{peebles2023scalable}.
The conditioning vector \(\mathbf c\in\mathbb R^{11}\) is mapped by a three-layer MLP,
\begin{equation}
    11
    \rightarrow
    128
    \rightarrow
    128
    \rightarrow
    6d_{\mathrm{model}}.
\end{equation}
The output dimension is \(6d_{\mathrm{model}}=1728\).
It provides \((\gamma_1,\beta_1,\alpha_1)\) for attention and \((\gamma_2,\beta_2,\alpha_2)\) for the FFN.
For the attention sublayer, the update is
\begin{equation}
    y
    =
    (1+\gamma_1)\odot\mathrm{LN}(x)+\beta_1,
    \qquad
    x
    \leftarrow
    x+\alpha_1\odot\mathrm{Attn}(y).
\end{equation}
The FFN sublayer uses the same form with \((\gamma_2,\beta_2,\alpha_2)\).
The final linear layer of the AdaLN MLP is zero-initialized.
Thus \(\gamma\), \(\beta\), and \(\alpha\) start at zero, and each block is initialized as an identity residual block.

\subsection{U-Shape Decoder}
\label{app:ushape}

The six Swin blocks are arranged as a shallow U-shape.
Blocks \(0,1,2\) form the encoder.
Blocks \(3,4,5\) form the decoder.
Encoder hidden states are cached and concatenated with the corresponding decoder states.
Each skip connection uses a linear projection
\begin{equation}
    \mathrm{Linear}(2d_{\mathrm{model}},d_{\mathrm{model}}).
\end{equation}
There are three such projections.
Together they contain \(498{,}528\) parameters.

\subsection{Output Head}
\label{app:output_head}

The output head reconstructs a \(128^3\) log-density residual from the final \(32^3\) token grid.
It contains a fine residual head, a gate head, and a coarse low-frequency head.
The full output equation is
\begin{equation}
    \log \hat\rho_\theta(\mathbf r)
    =
    \log \rho_{\mathrm{Boltz}}(\mathbf r)
    +
    \alpha
    \,
    \sigma\!\left(g_\theta(\mathbf r)\right)
    \left[
        r_\theta(\mathbf r)
        +
        \lambda
        \,
        \mathrm{Up}
        \left(
            r^{\mathrm{coarse}}_\theta
        \right)(\mathbf r)
    \right].
    \label{eq:app_output_head}
\end{equation}
Here \(\alpha=1.0\) is a fixed registered buffer and is not learned.
The fine residual head is \(\mathrm{Linear}(d_{\mathrm{model}},64)\), producing one \(4^3\) patch per token.
The gate head is \(\mathrm{Linear}(d_{\mathrm{model}},64)\), also producing one \(4^3\) patch per token.
The coarse head is \(\mathrm{Linear}(d_{\mathrm{model}},1)\), producing one scalar per token on the \(32^3\) token grid.
The coarse output is trilinearly upsampled to the \(128^3\) voxel grid before being added to the fine residual.
The scalar \(\lambda\) is an unconstrained learned parameter initialized to \(0.3\).
It is not passed through a sigmoid and is not constrained to \([0,1]\).

The residual head uses Gaussian initialization with standard deviation \(0.02\).
The gate head has zero weights and bias \(-2\), so the initial gate value is \(\sigma(-2)\approx0.12\).
The coarse head is zero-initialized.
This makes the initial prediction close to the Boltzmann baseline while still allowing a small learned correction from the residual branch.

After reconstruction, the log-density is clamped per sample to
\begin{equation}
    \log \hat\rho_\theta
    \in
    [
    \log\rho_{\mathrm{bulk}}-20,
    \log\rho_{\mathrm{bulk}}+10
    ].
\end{equation}
This numerical clamp prevents extreme densities while preserving the orders-of-magnitude variation across pressure conditions.
\clearpage

\subsection{Parameter Summary}
\label{app:param_summary}

The final model has \(6.35\)M parameters.
Most parameters are in the Swin blocks.
The approximate module-level breakdown is shown in \cref{tab:app_param_breakdown}, and model hyperparameters are shown in \cref{tab:app_hparams}.

\begin{table}[h]
\caption{\textbf{Approximate parameter breakdown of MADField.}}
\label{tab:app_param_breakdown}
\centering
\footnotesize
\setlength{\tabcolsep}{6pt}
\renewcommand{\arraystretch}{1.15}
\begin{tabular}{l c}
\toprule
Module & Share of parameters \\
\midrule
Swin blocks & \(86\%\) \\
U-shape skip projections & \(8\%\) \\
ConvStem and input processing & \(6\%\) \\
Output heads & \(<1\%\) \\
Coarse blend parameter \(\lambda\) & \(1\) parameter \\
\bottomrule
\end{tabular}
\end{table}

\begin{table}[h]
\caption{\textbf{\method{} architecture hyperparameters.}}
\label{tab:app_hparams}
\centering
\footnotesize
\setlength{\tabcolsep}{6pt}
\renewcommand{\arraystretch}{1.15}
\begin{tabular}{l l}
\toprule
Component & Value \\
\midrule
Input grid & \(128^3\) \\
Token grid & \(32^3\) \\
Patch size & \(4^3\) voxels \\
\(d_{\mathrm{model}}\) & 288 \\
Number of Swin blocks & 6 \\
Encoder blocks & 3 \\
Decoder blocks & 3 \\
Attention heads & 8 \\
Head dimension & 36 \\
Window size & \(8^3\) tokens \\
FFN type & SwiGLU \\
Nominal FFN width & 576 \\
SwiGLU inner width & 384 \\
Conditioning dimension & 11 \\
AdaLN MLP & \(11\to128\to128\to1728\) \\
Drop-path rate & 0.1 linear schedule \\
Residual head & \(\mathrm{Linear}(288,64)\) \\
Gate head & \(\mathrm{Linear}(288,64)\) \\
Coarse head & \(\mathrm{Linear}(288,1)\) \\
Gate bias & \(-2\) \\
Output \(\alpha\) & fixed \(1.0\) \\
Coarse blend \(\lambda\) & raw learned scalar initialized to \(0.3\) \\
Output clamp & \([\log\rho_{\mathrm{bulk}}-20,\log\rho_{\mathrm{bulk}}+10]\) \\
Total parameters & \(6.35\)M \\
\bottomrule
\end{tabular}
\end{table}

\clearpage
\section{PC-SAFT Parameters and Selected Constants}
\label{app:parameters}
We show our chosen PC-SAFT parameters and selected constants here. 

\begin{table}[h]
\caption{PC-SAFT non-polar parameters for the nine adsorbates. All species use the Esper non-polar set~\citep{esper2023pcp}.}
\label{tab:app_pcsaft_params}
\centering
\footnotesize
\begin{tabular}{lcccl}
\toprule
Adsorbate & $m$ & $\sigma$ (\AA) & $\varepsilon/k_B$ (K)
  & $M_w$ (g/mol) \\
\midrule
H$_2$       & 1.000 & 2.960 &  34.20 &   2.016 \\
Ar          & 1.000 & 3.378 & 117.81 &  39.95 \\
Kr          & 1.000 & 3.608 & 164.02 &  83.80 \\
Xe          & 1.000 & 3.927 & 227.70 & 131.29 \\
CH$_4$      & 1.000 & 3.701 & 150.07 &  16.04 \\
N$_2$       & 1.238 & 3.300 &  89.41 &  28.01 \\
C$_2$H$_6$  & 1.607 & 3.517 & 191.45 &  30.07 \\
C$_3$H$_8$  & 1.986 & 3.624 & 209.09 &  44.10 \\
CO$_2$      & 2.531 & 2.579 & 153.32 &  44.01 \\
\bottomrule
\end{tabular}
\end{table}

\begin{table}[h]
\caption{Selected Universal Force Field~\citep{rappe1992uff}
Lennard-Jones parameters for framework atoms. The full UFF table
covers all 102 elements; values for non-listed elements follow
directly from~\citet{rappe1992uff}.}
\label{tab:app_params_uff}
\centering
\footnotesize
\begin{tabular}{lcc lcc}
\toprule
Element & $\sigma$ (\AA) & $\varepsilon/k_B$ (K)
  & Element & $\sigma$ (\AA) & $\varepsilon/k_B$ (K) \\
\midrule
C  & 3.431 &  52.84 & O  & 3.118 &  30.19 \\
H  & 2.571 &  22.14 & N  & 3.261 &  34.72 \\
S  & 3.595 & 137.88 & F  & 2.997 &  25.16 \\
Cl & 3.516 & 114.23 & Si & 3.826 & 202.29 \\
Zn & 2.462 &  62.40 & Cu & 3.114 &   2.52 \\
Zr & 2.783 &  34.72 & Fe & 2.594 &   6.54 \\
Co & 2.559 &   7.05 & Ni & 2.525 &   7.55 \\
Al & 4.008 & 254.13 & Mg & 2.691 &  55.86 \\
\bottomrule
\end{tabular}
\end{table}


\section{Data Generation Pipeline}
\label{app:data_pipeline}

This appendix summarizes the data sources and label-generation pipeline used for the cDFT and GCMC experiments.
MADField-cDFT uses cDFT density fields generated by our PC-SAFT cDFT solver.
MADField-GCMC uses published GCMC references from ARC-MOF and PRAM.
All density fields are represented on \(128^3\) periodic voxel grids unless otherwise stated.

\subsection{Dataset Sources}
\label{app:dataset_sources}

We use four material sources in the paper.
QMOF provides the main crystalline-MOF dataset for cDFT training and in-domain evaluation~\citep{rosen2021machine}.
ARC-MOF provides published GCMC density and uptake references for the GCMC-fidelity benchmark~\citep{burner_2023arcmof,burner_2025_CH4_1bar,burner_2025_CH4_65bar,burner_2025_Xe_1bar}.
The JLA-Gardner amorphous-carbon ensemble provides an out-of-domain cDFT transfer set~\citep{gardner2023synthetic}.
PRAM provides out-of-domain GCMC uptake references for disordered nanoporous materials~\citep{thyagarajan2020database}.

\textbf{QMOF.}
We use the QMOF database, whose structures are provided as DFT-relaxed MOF geometries.
After structure parsing, pore filtering, and simulation filtering, the cDFT dataset contains \(5{,}770\) unique MOFs.
This prevents leakage across different adsorbates or pressures of the same material.

\paragraph{ARC-MOF.}
The GCMC in-domain benchmark uses ARC-MOF structures with published GCMC adsorption references.
For the present benchmark, we use the available GCMC density and uptake data for \ce{CH4} and Xe.
These labels define the target fidelity for MADField-GCMC.

\paragraph{JLA-Gardner amorphous carbon.}
The cDFT out-of-domain transfer benchmark uses amorphous carbon structures from the JLA-Gardner ensemble.
These structures are non-crystalline carbon networks and do not share the metal-linker motifs present in QMOF.
We generate cDFT references for this set using the same cDFT pipeline as QMOF.

\paragraph{PRAM.}
The GCMC out-of-domain transfer benchmark uses the PRAM disordered nanoporous dataset.
We use PRAM methane uptake references across PIMs, HCPs, kerogens, and amorphous carbons.
PRAM is used only for uptake evaluation in the main paper, not for density-field supervision.

\subsection{cDFT Reference Generation}
\label{app:cdft_reference_data}

MADField-cDFT data labels are generated by solving PC-SAFT cDFT on a periodic \(128^3\) grid~\citep{gross2001perturbed,sauer2017classical,stierle2024classical,dufour2025classical}.
The solver uses a non-polar PC-SAFT decomposition with FMT hard-sphere, hard-chain, and dispersion contributions.

Each calculation starts from the Boltzmann density
\begin{equation}
    \rho_{\mathrm{Boltz}}(\mathbf r)
    =
    \rho_{\mathrm{bulk}}
    \exp[-\beta V_{\mathrm{ext}}(\mathbf r)].
\end{equation}
The Picard iteration then solves the cDFT fixed-point equation until convergence or until the iteration cap is reached.
Pressure points within the same isotherm are evaluated from low to high pressure, using the nearest lower-pressure converged solution as a warm start when available.
Failed pressure points are excluded from supervised training but retained for the warm-start recovery analysis in \cref{app:warmstart_protocol}.

\paragraph{External potential.}
For each framework and adsorbate pair, we precomputed one external-potential grid and reuse it across all pressures.
The potential is a Lennard-Jones 12-6 framework-fluid interaction with Lorentz-Berthelot mixing.
Framework atoms use UFF parameters.
Adsorbates use the cDFT PC-SAFT parameter registry described in \cref{app:cdft_gcmc}.
The cutoff is \(r_c=12.8\) \AA.
For triclinic and non-orthorhombic cells, the simulation cell is expanded consistently with this cutoff and periodic distances are computed under the minimum-image convention.
The resulting \(V_{\mathrm{ext}}\) is stored in Kelvin, and \(\beta V_{\mathrm{ext}}\) is stored as a normalized model input.

\paragraph{Pressure grids.}
All cDFT data are generated at \(T=298.15\) K, and the base grid is shown in \cref{tab:app_pressure_grid}.

\begin{table}[t]
\caption{\textbf{cDFT pressure grids at \(T=298.15\) K.}}
\label{tab:app_pressure_grid}
\centering
\footnotesize
\setlength{\tabcolsep}{6pt}
\renewcommand{\arraystretch}{1.25}
\begin{tabular}{l c l}
\toprule
Fluid & \# pressures & Pressure range \\
\midrule
\begin{tabular}[t]{@{}l@{}}Ar, \ce{C2H6}, \ce{C3H8}, \ce{CO2}, Kr, \ce{N2}, Xe\end{tabular}
& \begin{tabular}[t]{@{}c@{}}12\end{tabular}
& \begin{tabular}[t]{@{}l@{}}\(0.1,0.5,1,2,3,4,5,6,7,8,9,10\) bar\end{tabular} \\

\begin{tabular}[t]{@{}l@{}}\ce{CH4}\end{tabular}
& \begin{tabular}[t]{@{}c@{}}24\end{tabular}
& \begin{tabular}[t]{@{}l@{}}\(0.1,0.5,1,2,3,4,5,6,7,8,9,10,\)\\\(15,20,25,30,35,40,45,50,55,60,65\) bar\end{tabular} \\

\begin{tabular}[t]{@{}l@{}}\ce{H2}\end{tabular}
& \begin{tabular}[t]{@{}c@{}}30\end{tabular}
& \begin{tabular}[t]{@{}l@{}}\(0.1,0.5,1,2,3,4,5,6,7,8,9,10,\)\\\(15,20,25,30,35,40,45,50,\)\\\(55,60,65,70,75,80,85,90,95,100\) bar\end{tabular} \\
\bottomrule
\end{tabular}
\end{table}

\subsection{GCMC Reference Data}
\label{app:gcmc_reference_data}

We use the released GCMC density from ARC-MOF and released GCMC uptake from PRAM in \methodgcmc{}.
The ARC-MOF references provide density grids and uptake values for in-domain GCMC-fidelity training and evaluation.
The PRAM references provide methane uptake values for out-of-domain evaluation on disordered nanoporous materials.

\paragraph{ARC-MOF GCMC references.}
The in-domain GCMC benchmark uses ARC-MOF structures with published GCMC density and uptake references.
We use the available \ce{CH4} and Xe GCMC data for \methodgcmc{} training and evaluation.
The density fields are mapped to the same \(128^3\) grid convention used by the model.
The corresponding uptake values are used for scalar GCMC benchmark comparisons.

\paragraph{PRAM GCMC references.}
The PRAM benchmark is used only for zero-shot uptake evaluation.
It contains methane GCMC uptake references across PIMs, HCPs, kerogens, and amorphous carbons.
No PRAM structures are used for \methodgcmc{} training.
Because PRAM methane loadings are smaller than the high-pressure ARC-MOF MOF loadings, we report both absolute MAE and mean relative error in the main text.

\section{Baseline Implementations}
\label{app:baselines}

This appendix describes how scalar and density adsorption baselines are adapted to our cDFT and GCMC benchmarks.
All baselines are trained and evaluated on the same train, validation, and test splits as MADField whenever the model interface allows it.
We keep each baseline architecture unchanged unless an implementation fix is required for training or inference.
Because the original scalar models differ in whether they accept adsorbate and pressure as inputs, we use the most faithful task formulation supported by each model.

\subsection{Scalar Uptake Baselines}
\label{app:scalar_baselines}

\paragraph{RetNet.}
RetNet is used as a voxel-CNN scalar baseline.
We keep the original architecture and training recipe unchanged.
The model consists of five 3D convolutional layers, two max-pooling layers, and three fully connected layers, with BatchNorm and LeakyReLU activations.
The first convolution uses circular padding to respect periodic boundary conditions.
The model has \(0.367\)M parameters.

RetNet does not condition on adsorbate identity or pressure.
We therefore train a separate RetNet model for each \((\mathrm{adsorbate},P)\) condition.
For the cDFT benchmark, this produces one model per gas and pressure state.
For the GCMC benchmark, we train separate models for the available \ce{CH4} and Xe conditions.
The input voxel transform follows the original RetNet setup:
\[
    X=\mathrm{clip}(-U/T_{\mathrm{room}},0,50),
\]
followed by standardization using the training-set mean and standard deviation.
We use the original random rotation, flip, reflection, and identity augmentations.
We also keep the original optimizer and schedule, using Adam with learning rate \(10^{-3}\), StepLR with step size \(10\) and decay factor \(0.5\), MSE loss, and \(30\) training epochs.
The final-layer bias is initialized to the training-label mean, as in the original implementation.
Labels are trained in raw uptake units without target normalization.
We store the training-set voxel normalization statistics with each checkpoint and use them at evaluation time.

\paragraph{DeepSorption.}
DeepSorption is used through its Matformer implementation.
We keep the published architecture unchanged, including the \(N=8\) encoder layers, embedding dimension \(512\), feed-forward dimension \(2048\), \(4\) attention heads, constrained attention distance bins, pre-trained RotatE embeddings, 3D positional encoding, and generator head.
The resulting model has \(26.06\)M parameters.

DeepSorption predicts a vector of scalar targets for a given structure.
We therefore train one DeepSorption model per adsorbate, with the output dimension set to the number of pressure points for that adsorbate.
This allows the model to predict a pressure grid for a fixed gas, while avoiding an unsupported pressure-conditioned regression interface.
For cDFT, the output dimension varies across gases because the pressure grid differs by adsorbate.
For GCMC, the output dimension matches the available pressure points in the published GCMC benchmark.

The main changes are limited to the data and training pipeline.
Targets are pivoted from the long-format \((\mathrm{MOF},\mathrm{adsorbate},P)\) table into per-gas pressure vectors.
Missing or unconverged cDFT state points are represented by a binary mask and excluded from the loss.
The training loss is masked MSE,
\[
    \mathcal L_{\mathrm{DS}}
    =
    \frac{
        \sum_k m_k
        \left(
            \hat y_k-y_k
        \right)^2
    }{
        \sum_k m_k+\epsilon
    },
\]
where \(m_k=1\) for valid pressure points and \(m_k=0\) for missing or unconverged labels.
Target normalization follows the original per-target z-score strategy, but statistics are computed only over valid training labels.
We use the same ReduceLROnPlateau scheduler as the original code, with factor \(0.6\), patience \(10\), and minimum learning rate \(10^{-7}\).
The batch size is reduced to \(16\) because the maximum sequence length is increased for our MOF structures.
We select checkpoints using a robust validation uptake error computed only on state points with reference loading above \(0.05\) mol kg\(^{-1}\), which avoids unstable relative errors in the near-zero loading regime.
We remove the evaluation-time ReLU applied to the first output column in the original script and use the raw regression outputs for all pressure points.

\paragraph{MOFTransformer.}
We use the pip-installable \texttt{moftransformer==2.1.4} package without modifying the source code.
The transformer backbone, data modules, featurization utilities, and regression head are kept unchanged.
We add wrapper scripts only for data preparation, launching fine-tuning runs, and saving per-structure predictions.

MOFTransformer does not take pressure, temperature, or adsorbate identity as regression inputs in its scalar head.
We therefore fine-tune a separate checkpoint for each \((\mathrm{adsorbate},P)\) condition.
Each checkpoint starts from the same pre-trained MOFTransformer backbone and predicts one scalar uptake target per MOF.
The target is normalized with the z-score statistics computed from the corresponding training split.
For cDFT, we train separate checkpoints for the gas and pressure conditions evaluated in the benchmark.
For GCMC, we train separate checkpoints for the available \ce{CH4} and Xe conditions.

Input preparation follows the upstream MOFTransformer pipeline.
Each structure is converted into the CGCNN graph representation and GRIDAY Lennard-Jones probe grid used by the original package.
We parallelize the upstream data-preparation routine across CPU workers, but do not change the generated features.
For test evaluation, we wrap the upstream test-only path to save per-MOF predictions to CSV, since the original script logs aggregate metrics but does not export all predictions.
For amorphous-carbon OOD inference, we use an inference-only CIF parsing bypass to avoid failures on large P1 cells.
This bypass is not used for the crystalline QMOF or ARC-MOF training data.

\paragraph{Uni-MOF.}
We use the released Uni-MOF code with minimal task-registration changes.
The transformer backbone, UniMat atom-pair encoder, data loader, loss function, and target normalization scheme are kept unchanged.
We only register new task identifiers for our cDFT and GCMC datasets by adding target normalization statistics and pressure-temperature normalization ranges to the existing registry dictionaries.

Unlike MOFTransformer and RetNet, Uni-MOF natively conditions on adsorbate and thermodynamic variables.
Its input includes a gas identifier, gas attribute vector, \(\log_{10}P\), and temperature.
We therefore train one checkpoint for the multi-gas cDFT benchmark and one checkpoint for each GCMC training setup.
The training LMDB stores atoms, coordinates, lattice matrix, gas identifier, gas attributes, temperature, pressure, target uptake, and task name for each state point.
Target values use the upstream log1p-standardization transform.
At inference time, the upstream inverse transform is used to recover uptake in physical units.

We add wrapper scripts to build LMDB files from our CSV splits, compute target statistics, register tasks idempotently, launch distributed training, and merge shard-level inference outputs.
No architectural or algorithmic change is made to the released Uni-MOF model.

\paragraph{IsothermNet.}
We adapt the published IsothermNet implementation for multi-gas adsorption prediction.
The original implementation is designed around a single-gas CO\(_2\) isotherm setting with hard-coded gas constants in the structural global features.
We replace these constants with per-sample molecular parameters so that the same architecture can receive the appropriate gas information for each state point.
For our cDFT benchmark, the molecular slots are populated with the nine-gas PC-SAFT registry used elsewhere in the paper.
For the GCMC benchmark, the slots are populated with the corresponding GCMC molecular registry.

We make several implementation-level fixes needed for stable training.
Modules that were instantiated inside the forward pass are moved into \texttt{\_\_init\_\_} so that their parameters and batch-normalization statistics are registered and updated during training.
We also use float32 training, vectorize repeated tensor operations in the forward pass, and reconstruct batch indices inside the model for reliable multi-GPU execution.
These changes do not alter the intended model architecture but make training deterministic and reproducible.

The original feature-generation pipeline relies on molecular files and RDKit-based bond perception, which is not robust for our MOF, amorphous-carbon, and PRAM structures.
We therefore rewrite the featurization pipeline using pymatgen structures, periodic distance matrices, covalent-radius bond detection, graph-based ring detection, and parallel structure processing.
The Zeo++ textural feature pipeline is also automated to produce the textural inputs required by IsothermNet.
We train IsothermNet with log-space MSE on \(\log(q+10^{-3})\), use per-gas and per-pressure validation metrics, and convert predictions back to cm\(^3\)(STP) g\(^{-1}\) for the benchmark tables.

\paragraph{Summary.}
The scalar baselines differ in the conditioning axes they support.
RetNet and MOFTransformer do not natively condition on pressure and adsorbate identity, so we train separate checkpoints for unsupported axes.
DeepSorption supports multi-pressure output for a fixed gas, so we train one checkpoint per adsorbate.
Uni-MOF and IsothermNet support molecular and thermodynamic conditioning after task registration or molecular-parameter injection, so they are trained as conditional scalar predictors.
These choices give each scalar baseline the strongest task formulation supported by its released architecture while preserving a fair comparison to MADField.

\subsection{Density-Field Baselines}
\label{app:density_baselines}

\paragraph{DeepAPD.}
DeepAPD predicts normalized adsorbate probability distributions for binding-site localization in MOFs~\citep{burner2026rapid}.
We retain the released PaiNN-Charge backbone, including the atom and probe encoders, hidden dimension, radial basis cutoff, Softplus output, and disconnected-probe masking policy.
We adapt only the training target, loss, and inference normalization so that the model can be evaluated on our physically scaled density benchmarks.

The original DeepAPD objective normalizes both prediction and target before computing a Tanimoto-style loss.
This is appropriate for adsorption probability distributions, but removes the physical density scale needed for uptake integration.
We therefore remove target normalization in the LMDB construction and train on physically scaled density,
\[
    \tilde\rho(\mathbf r)
    =
    \rho(\mathbf r) / \rho_{\mathrm{norm}},
    \qquad
    \rho_{\mathrm{norm}}=5\times10^4~\mathrm{mol\,m^{-3}} .
\]
The predicted field is multiplied by \(\rho_{\mathrm{norm}}\) to recover density in physical units.

For GCMC evaluation, we use public checkpoint where DeepAPD is trained on the same published Burner et al. ARC-MOF density references used by MADField-GCMC~\citep{burner_2023arcmof,burner_2025_CH4_1bar,burner_2025_CH4_65bar,burner_2025_Xe_1bar}.
The released DeepAPD checkpoint contains task heads for \ce{CH4} at \(1\) bar, \ce{CH4} at \(65\) bar, and Xe at \(1\) bar.
We use the heads matching our GCMC benchmark and retrain the probe heads for our supervised target.
This gives a field directly comparable to MADField outputs.

\paragraph{SorbIIT.}
SorbIIT predicts spatially resolved adsorption information using a periodic-aware 3D U-Net architecture~\citep{sun2024understanding}.
We use the released \texttt{CellUNet} architecture with periodic unit-cell convolutions and the same channel configuration.
Since the released code does not provide pre-trained checkpoints or public training data for our benchmark setting, we retrain SorbIIT from scratch on our splits.

The original SorbIIT pipeline takes proprietary zeolite energy-grid inputs and predicts voxel-wise isotherm coefficients.
We replace the ZTB energy grid with our cDFT external-potential grid \(V_{\mathrm{ext}}/k_B\).
Following the released preprocessing convention, we clamp the input to \([-10^4,10^4]\) K and normalize it as
\[
      x(\mathbf r) = -\frac{V_{\mathrm{ext}}(\mathbf r)/k_B - 6400}{6400}.
\]
For the cDFT benchmark, the volumetric fields are resampled to a fixed physical resolution following the SorbIIT convention, with density grids rescaled to preserve the cell integral.

Because our benchmark is fixed-temperature adsorption, we use the fixed-temperature quadratic isotherm parameterization.
The network predicts voxel-wise isotherm coefficient fields, from which density fields are reconstructed at the target pressures.
We train with a hybrid objective combining coefficient reconstruction, direct density supervision, integrated-loading supervision, non-negativity regularization, and the SorbIIT amplitude regularizer.
In our main cDFT runs, the coefficient loss is down-weighted and direct density supervision is dominant
(\(\lambda_X=0.1\), \(\lambda_\rho=1.0\), \(\lambda_N=0.1\), \(\lambda_{\mathrm{neg}}=0.05\), and regularization weight \(0.05\)).

For cDFT comparison, we train one SorbIIT model per fluid on the same train, validation, and test split used by MADField-cDFT.
For GCMC comparison, we train on the same ARC-MOF density references used by the other GCMC-fidelity models.
For amorphous-carbon evaluation, crystallographic symmetry augmentation is disabled and only the identity transform is used, since these cells do not have the crystallographic symmetry operations assumed in the original zeolite setting.

\paragraph{Faithfulness of density baselines.}
Both density baselines preserve their released model architectures.
DeepAPD keeps the PaiNN-Charge architecture, and SorbIIT keeps the \texttt{CellUNet} architecture.
The modifications are restricted to data formatting, target scaling, loss definitions, and inference reconstruction so that both models can be evaluated on the same physically scaled density task as MADField.
s

\section{Evaluation Metrics}
\label{app:metrics}

We use two primary evaluation metrics.
Uptake accuracy is measured by mean absolute error in gas loading.
Density-field quality is measured by Tanimoto similarity after normalizing each density field to unit mass.

\paragraph{Uptake MAE.}
For each state point \(i\), the reference loading is \(q_i\) and the predicted loading is \(\hat q_i\), both reported in cm\(^3\)(STP) g\(^{-1}\).
For density-prediction models, \(\hat q_i\) is obtained by integrating the predicted density field over the accessible pore region and converting the result using the sample-specific unit-cell volume and framework mass.
For scalar baselines, \(\hat q_i\) is the directly predicted uptake after unit conversion when needed.
We report
\begin{equation}
    \mathrm{MAE}_{q}
    =
    \frac{1}{N}
    \sum_{i=1}^{N}
    \left|
        \hat q_i - q_i
    \right| .
    \label{eq:app_uptake_mae}
\end{equation}
Lower values indicate more accurate uptake prediction.

\paragraph{Density Tanimoto similarity.}
We evaluate spatial density-field quality using the continuous Tanimoto coefficient.
For each sample \(i\), we restrict the predicted and reference density fields to the evaluation voxels and normalize each field to unit mass,
\begin{equation}
    \tilde{\rho}_i(v)
    =
    \frac{
        \rho_i(v)
    }{
        \sum_{u}\rho_i(u)+\epsilon
    },
    \qquad
    \tilde{\hat\rho}_i(v)
    =
    \frac{
        \hat\rho_i(v)
    }{
        \sum_{u}\hat\rho_i(u)+\epsilon
    } .
    \label{eq:app_density_normalization}
\end{equation}
The per-sample Tanimoto similarity is
\begin{equation}
    T_i
    =
    \frac{
        \sum_v
        \tilde{\rho}_i(v)
        \tilde{\hat\rho}_i(v)
    }{
        \sum_v
        \tilde{\rho}_i(v)^2
        +
        \sum_v
        \tilde{\hat\rho}_i(v)^2
        -
        \sum_v
        \tilde{\rho}_i(v)
        \tilde{\hat\rho}_i(v)
        +
        \epsilon
    } .
    \label{eq:app_tanimoto_metric}
\end{equation}
We report the mean over the test set,
\begin{equation}
    T_{\rho}
    =
    \frac{1}{N}
    \sum_{i=1}^{N}
    T_i .
    \label{eq:app_mean_tanimoto}
\end{equation}
A value of \(1\) indicates identical normalized density fields, while lower values indicate weaker spatial overlap.
Because each field is normalized before comparison, \(T_{\rho}\) measures density-shape agreement independently of total uptake magnitude.

\paragraph{Aggregation.}
Per-adsorbate and per-family columns are computed by averaging over the corresponding state points.
Unless otherwise stated, \textbf{Total} denotes the sample-weighted mean over all evaluated state points in the listed group.
For MOF benchmarks, this aggregates over the evaluated adsorbate-pressure-structure state points.
For PRAM transfer, this aggregates over all evaluated structures across the listed material families.
We additionally report mean relative error for PRAM because methane uptake values are smaller than in the high-pressure MOF benchmark.

\section{Additional Experiments}
\label{app:additional}

This appendix provides additional analyses for the cDFT uptake benchmark, OOD density transfer, and warm-start experiment.
We include controlled ablations, zero-shot density transfer to amorphous carbon, external-potential well-depth stratification, and inference-time measurements.

\subsection{Ablation Study}
\label{app:ablation}

We report controlled comparisons that isolate three design choices.
The scalar-head variant tests whether field supervision is more effective than direct uptake regression.
The no-Boltzmann-residual variant tests the contribution of the Boltzmann residual parameterization.
The FNO baseline tests whether the Swin volumetric backbone improves the density-prediction route under comparable supervision.
\cref{tab:app_ablation_individual} reports uptake MAE on the held-out cDFT test set.
We also compare direct GCMC training with cDFT-pre-trained LoRA adaptation\cref{tab:app_gcmc_pre-training_ablation}.
MADField-GCMC-scratch is trained directly on the GCMC split without cDFT pre-training.
MADField-GCMC starts from MADField-cDFT and adapts to GCMC density fields with LoRA.

\begin{table}[t]
\caption{
\textbf{Ablation study on cDFT uptake prediction.}
Errors are MAE of gas uptake in cm$^3$(STP)/g on the held-out cDFT test set.
\textbf{Density Head} is the full MADField-cDFT model that predicts \(\rho(\mathbf r)\) and integrates it to obtain uptake.
\textbf{Scalar Head} uses the same backbone but directly predicts uptake.
\textbf{w/o Boltzmann residual} removes the Boltzmann-residual output parameterization.
\textbf{FNO baseline} replaces the Swin backbone with a Fourier neural operator baseline. \textbf{Total} denotes the sample-weighted mean over all evaluated cDFT state points.
}
\label{tab:app_ablation_individual}
\centering
\footnotesize
\setlength{\tabcolsep}{4.5pt}
\renewcommand{\arraystretch}{1.15}
\resizebox{\textwidth}{!}{%
\begin{tabular}{l cccccccccc}
\toprule
Configuration
& \ce{H2}& \ce{Ar}& \ce{Kr}& \ce{Xe}& \ce{N2}& \ce{CH4}& \ce{CO2}& \ce{C2H6}& \ce{C3H8}& \textbf{Total} \\
\midrule
Scalar Head
& 2.46& 2.17& 4.71& 11.93& 1.72& 9.06& 8.41& 10.85& 17.66& 5.91 \\
w/o Boltzmann residual
& 0.16& 0.21& 0.72& 3.28& 0.17& 1.26& 1.46& 2.15& 4.55& 1.01 \\
FNO baseline
& 0.24& 0.66& 2.98& 9.74& 0.44& 5.31& 6.44& 8.03& 16.07& 3.66 \\
\textbf{Density Head}
& \textbf{0.08}& \textbf{0.11}& \textbf{0.49}& \textbf{2.87}& \textbf{0.13}& \textbf{0.80}& \textbf{1.17}& \textbf{2.01}& \textbf{4.51}& \textbf{0.82} \\
\bottomrule
\end{tabular}
}
\end{table}

\begin{table}[t]
\caption{
\textbf{cDFT pre-training ablation for GCMC uptake prediction.}
Errors are MAE of gas uptake in cm$^3$(STP)/g.
MOF evaluates in-distribution GCMC uptake prediction on crystalline MOFs.
PRAM evaluates zero-shot transfer to disordered porous materials for \ce{CH4}. \textbf{Tot.} denotes the sample-weighted mean.
}
\label{tab:app_gcmc_pre-training_ablation}
\centering
\footnotesize
\setlength{\tabcolsep}{5.5pt}
\renewcommand{\arraystretch}{1.15}
\resizebox{\textwidth}{!}{%
\begin{tabular}{l ccc ccccc}
\toprule
\multirow{2}{*}{Model} & \multicolumn{3}{c}{MOF (ID)} & \multicolumn{5}{c}{PRAM (OOD, \ce{CH4})} \\
\cmidrule(lr){2-4}\cmidrule(l){5-9}
& \ce{CH4} & \ce{Xe} & \textbf{Tot.} & PIM & HCP & Kerogen & Bhatia & \textbf{Tot.} \\
\midrule
MADField-GCMC-scratch & 0.89 & 3.14 & 1.68 & 10.06 & 10.39 & 22.35 & 102.65 & 13.60 \\
\textbf{MADField-GCMC} & \textbf{0.33} & \textbf{1.37} & \textbf{0.58} & \textbf{0.57} & \textbf{0.40} & \textbf{1.42} & \textbf{0.67} & \textbf{0.62} \\
\bottomrule
\end{tabular}
}
\end{table}

\subsection{Zero-shot cDFT Density Transfer to Amorphous Carbon}
\label{app:jla_density_transfer}

We evaluate zero-shot density-field transfer on the JLA-Gardner amorphous-carbon set, shown in \cref{tab:app_jla_density_transfer}.
Models are trained on crystalline MOF cDFT labels and evaluated on \ce{CH4} density fields without fine-tuning.
Density quality is measured by Tanimoto similarity between normalized predicted and reference density fields, so the metric evaluates spatial adsorption-pattern agreement.

\begin{table}[t]
\caption{
\textbf{Zero-shot cDFT density transfer to amorphous carbon.}
Tanimoto similarity is computed on \ce{CH4} density fields from the JLA-Gardner amorphous-carbon set.
Higher is better.
}
\label{tab:app_jla_density_transfer}
\centering
\footnotesize
\setlength{\tabcolsep}{7pt}
\renewcommand{\arraystretch}{1.15}
\begin{tabular}{l c}
\toprule
Model & \ce{CH4} Tanimoto \(\uparrow\) \\
\midrule
DeepAPD & 0.418 \\
SorbIIT & 0.492 \\
\textbf{MADField-cDFT} & \textbf{0.885} \\
\bottomrule
\end{tabular}
\end{table}

MADField-cDFT transfers substantially better than the density-field baselines, indicating that the learned cDFT density operator generalizes beyond crystalline MOFs to disordered carbon geometries.

\subsection{Inference Time Analysis}
\label{app:inference_time}

We measure inference time on 100 held-out samples using the same evaluation scripts used for benchmark prediction, where table is shown in \cref{tab:app_inference_time}.
The reported time is wall-clock time divided by the number of evaluated samples.
For MADField models, the timing includes model loading, data loading, preprocessing, neural forward pass, and output evaluation inside the benchmark script.
For IsothermNet, the timing is cached-feature forward-only and is therefore not directly comparable to end-to-end volumetric evaluation.
For DeepAPD, we report full-grid probe evaluation because the model produces normalized density fields.

\begin{table}[t]
\caption{
\textbf{Inference time comparison.}
Times are measured over 100 samples and reported as wall-clock seconds per sample.
MADField models and the density baselines are evaluated with their benchmark scripts.
IsothermNet uses cached features and is included only as a forward-only scalar baseline.
}
\label{tab:app_inference_time}
\centering
\footnotesize
\setlength{\tabcolsep}{6pt}
\renewcommand{\arraystretch}{1.15}
\begin{tabular}{l c c c}
\toprule
Model & Evaluation mode & sec/sample \(\downarrow\) & samples/sec \(\uparrow\) \\
\midrule
MADField-cDFT & end-to-end eval& 0.253& 3.95 \\
MADField-GCMC LoRA& end-to-end eval& 0.097& 10.26 \\
SorbIIT& end-to-end eval& 0.197& 5.08 \\
DeepAPD& full-grid probe eval& 1.872& 0.53 \\
DeepSorption & CIF parse + DS forward & 0.0191& 52.43 \\
RetNet & cached voxel load/transform + RN forward & 0.0096 & 104.44 \\
IsothermNet & cached-feature forward-only & 0.00018 & 5655.15 \\
\bottomrule
\end{tabular}
\end{table}

MADField-GCMC LoRA runs in \(0.097\) seconds per sample, and MADField-cDFT runs in \(0.253\) seconds per sample.
The GCMC-adapted LoRA model is faster than the \method{} cDFT model in this benchmark because the two timings are end-to-end script timings and use different checkpoint wrappers and evaluation heads.
Both MADField variants are sub-second volumetric predictors.
Compared with DeepAPD, MADField-GCMC LoRA is \(19.2\times\) faster while predicting physically scaled density rather than normalized density.
SorbIIT has similar end-to-end runtime but requires uptake information as input, so it is not an independent uptake predictor.
The scalar baselines are faster when evaluated from cached or compact features, but they do not produce three-dimensional density fields and cannot be used for cDFT warm-starting.

\subsection{Experiment specs}
\label{app:repro}

\paragraph{Software.}
We use PyTorch 2.3.1 with CUDA 12.1 and bf16 autocast~\citep{paszke2019pytorch}.
The Python environment uses Python 3.11, NumPy 1.26, and SciPy 1.17.
The PC-SAFT bulk equation of state uses FeOs 0.6.1~\citep{rehner2023feos}.
CIF parsing and structure handling use ASE~\citep{hjorth2017atomic}.
The cDFT solver is a custom PyTorch implementation for three-dimensional PC-SAFT cDFT with periodic and non-orthorhombic cells.
\methodgcmc{} uses published GCMC reference data from ARC-MOF and PRAM rather than newly generated GCMC trajectories.
Experiment tracking uses Weights \& Biases.

\paragraph{Hardware and training runs.}
\method{} cDFT training is run on a single node with \(8\times\) NVIDIA H200 GPUs using DDP.
The effective batch size is \(64\), with batch size \(8\) per GPU and gradient accumulation \(1\).
The released \method{} checkpoint is obtained from a supervised \(50{,}000\)-step run followed by a supervised \(30{,}000\)-step resume run.
The released checkpoint is selected at step \(26{,}000\) of the second run.
\methodgcmc{} GCMC adaptation is run on the same \(8\times\) H200 setup for \(50{,}000\) LoRA fine-tuning steps.
Inference timings are measured on a single H200 and are reported in \cref{app:inference_time}.

\section{Broader impacts}

This work studies learned surrogates for adsorption simulation in porous materials. 
By predicting equilibrium adsorbate density fields, the proposed approach may reduce the computational cost of material screening and provide spatial information that is not available from scalar property predictors. 
Such tools could support research on gas storage, separation, and carbon capture by helping prioritize candidates for more expensive simulations or experiments.
Responsible use requires careful benchmarking, uncertainty awareness, and clear reporting of the model's domain of applicability.

\end{document}